\documentclass[12pt,a4paper]{article}

\usepackage{amsmath,amssymb}
\usepackage{graphicx}
\usepackage{pst-all}
\usepackage{bbm}

\makeatletter \@addtoreset{equation}{section} \makeatother

\makeatletter
\let\old@startsection=\@startsection
\let\oldl@section=\l@section
\renewcommand{\@startsection}[6]{\old@startsection{#1}{#2}{#3}{#4}{#5}{#6\mathversion{bold}}}
\renewcommand{\l@section}[2]{\oldl@section{\mathversion{bold}#1}{#2}}
\makeatother

\makeatletter
\let\old@makecaption=\@makecaption
\def\@makecaption{\small\old@makecaption}
\makeatother

\let\oldPhi=\Phi
\let\oldPsi=\Psi
\let\oldGamma=\Gamma
\let\oldDelta=\Delta
\let\oldSigma=\Sigma
\let\oldTheta=\Theta
\let\oldPi=\Pi
\let\oldUpsilon=\Upsilon
\renewcommand{\Phi}{\mathnormal{\oldPhi}}
\renewcommand{\Psi}{\mathnormal{\oldPsi}}
\renewcommand{\Gamma}{\mathnormal{\oldGamma}}
\renewcommand{\Sigma}{\mathnormal{\oldSigma}}
\renewcommand{\Delta}{\mathnormal{\oldDelta}}
\renewcommand{\Theta}{\mathnormal{\oldTheta}}
\renewcommand{\Pi}{\mathnormal{\oldPi}}
\renewcommand{\Upsilon}{\mathnormal{\oldUpsilon}}

\newcommand{\Action}{\mathcal{S}}
\newcommand{\Lagr}{\mathcal{L}}
\newcommand{\Ham}{\mathcal{H}}

\newcommand{\tr}{\mathop{\mathrm{tr}}}
\newcommand{\str}{\mathop{\mathrm{str}}}

\newcommand{\order}{\mathcal{O}}

\newcommand{\Integers}{\mathbbm{Z}}
\newcommand{\Reals}{\mathbbm{R}}

\newcommand{\Sphere}{S}  
\newcommand{\AdS}{\mathrm{AdS}}

\ifx\genfrac\sdflkaj

\else

\fi
\newcommand{\sfrac}[2]{{\textstyle\frac{#1}{#2}}}
\newcommand{\half}{\sfrac{1}{2}}
\newcommand{\ihalf}{\sfrac{i}{2}}
\newcommand{\quarter}{\sfrac{1}{4}}
\newcommand{\Half}{\frac{1}{2}}

\newcommand{\Quarter}{\frac{1}{4}}

\newcommand{\rep}[1]{{\mathbf{#1}}}
\newcommand{\matr}[2]{\left(\begin{array}{#1}#2\end{array}\right)}
\newcommand{\alg}[1]{\mathfrak{#1}}
\newcommand{\grp}[1]{\mathrm{#1}}

\newcommand{\grSU}{\grp{SU}}
\newcommand{\grSO}{\grp{SO}}
\newcommand{\grPSU}{\grp{PSU}}
\newcommand{\algSU}{\alg{su}}
\newcommand{\algSO}{\alg{so}}
\newcommand{\algPSU}{\alg{psu}}

\newcommand{\brk}[1]{(#1)}
\newcommand{\lrbrk}[1]{\left(#1\right)}
\newcommand{\bigbrk}[1]{\bigl(#1\bigr)}
\newcommand{\Bigbrk}[1]{\Bigl(#1\Bigr)}

\newcommand{\lrsbrk}[1]{\left[#1\right]}

\newcommand{\Bigsbrk}[1]{\Bigl[#1\Bigr]}
\newcommand{\biggsbrk}[1]{\biggl[#1\biggr]}
\newcommand{\Biggsbrk}[1]{\Biggl[#1\Biggr]}
\newcommand{\ket}[1]{\mathopen{|}#1\mathclose{\rangle}}
\newcommand{\bra}[1]{\mathopen{\langle}#1\mathclose{|}}

\newcommand{\comm}[2]{[#1,#2]}

\newcommand{\acomm}[2]{\{#1,#2\}}

\newcommand{\abs}[1]{{|#1|}}

\newcommand{\grade}[1]{[#1]}

\newcommand{\nn}{\nonumber}

\def\[{\begin{equation}}
\def\]{\end{equation}}

\makeatletter
\def\mr@ignsp#1 {\ifx\:#1\@empty\else #1\expandafter\mr@ignsp\fi}%
\newcommand{\multiref}[1]{\begingroup
\xdef\mr@no@sparg{\expandafter\mr@ignsp#1 \: }%
\def\mr@comma{}%
\@for\mr@refs:=\mr@no@sparg\do{\mr@comma\def\mr@comma{,}\ref{\mr@refs}}%
\endgroup}
\makeatother

\newcommand{\hypref}[2]{\ifx\href\asklfhas #2\else\href{#1}{#2}\fi}

\newcommand{\secref}[1]{Sec.~\multiref{#1}}

\newcommand{\appref}[1]{App.~\multiref{#1}}

\newcommand{\tabref}[1]{Tab.~\multiref{#1}}

\newcommand{\figref}[1]{Fig.~\multiref{#1}}
\renewcommand{\eqref}[1]{(\multiref{#1})}

\newcommand{\bibtitle}[1]{\emph{#1}}

\ifx\href\asklfhas\newcommand{\href}[2]{#2}\fi

\newcommand{\comma}{\quad,\quad}
\newcommand{\unit}{\mathbbm{1}}

\newcommand{\tim}[1]{\dot{#1}}
\newcommand{\spa}[1]{\acute{#1}}

\newcommand{\tl}{\tilde{\lambda}}

\newcommand{\Bsi}{\Upsilon}

\newcommand{\lAA}{{a}}
\newcommand{\rAA}{{\dot{a}}}
\newcommand{\laa}{{\alpha}}
\newcommand{\raa}{{\dot{\alpha}}}
\newcommand{\lBB}{{b}}
\newcommand{\rBB}{{\dot{b}}}
\newcommand{\lbb}{{\beta}}
\newcommand{\rbb}{{\dot{\beta}}}
\newcommand{\lCC}{{c}}
\newcommand{\rCC}{{\dot{c}}}
\newcommand{\lcc}{{\gamma}}
\newcommand{\rcc}{{\dot{\gamma}}}
\newcommand{\lDD}{{d}}

\newcommand{\ldd}{{\delta}}

\newcommand{\lx}{{1}}
\newcommand{\ly}{{2}}
\newcommand{\lz}{{3}}
\newcommand{\lw}{{4}}
\newcommand{\rx}{{\dot{1}}}
\newcommand{\ry}{{\dot{2}}}
\newcommand{\rz}{{\dot{3}}}
\newcommand{\rw}{{\dot{4}}}

\newcommand{\conj}{\natural}

\newcommand{\vx}{\vec{x}}
\newcommand{\vp}{\vec{p}}

\newcommand{\cpp}{\energy' p - \energy p'}

\newcommand{\Smatrix}{\mathbbm{S}}  
\newcommand{\smatrix}{\mathbf{S}}   
\newcommand{\Tmatrix}{\mathbbm{T}}  
\newcommand{\tmatrix}{\mathrm{T}}   

\newcommand{\Asmatrix}{\mathbf{A}}
\newcommand{\Bsmatrix}{\mathbf{B}}
\newcommand{\Csmatrix}{\mathbf{C}}
\newcommand{\Dsmatrix}{\mathbf{D}}
\newcommand{\Esmatrix}{\mathbf{E}}
\newcommand{\Fsmatrix}{\mathbf{F}}
\newcommand{\Gsmatrix}{\mathbf{G}}
\newcommand{\Hsmatrix}{\mathbf{H}}
\newcommand{\Ksmatrix}{\mathbf{K}}
\newcommand{\Lsmatrix}{\mathbf{L}}

\newcommand{\Atmatrix}{\mathrm{A}}
\newcommand{\Btmatrix}{\mathrm{B}}
\newcommand{\Ctmatrix}{\mathrm{C}}
\newcommand{\Dtmatrix}{\mathrm{D}}
\newcommand{\Etmatrix}{\mathrm{E}}
\newcommand{\Ftmatrix}{\mathrm{F}}
\newcommand{\Gtmatrix}{\mathrm{G}}
\newcommand{\Htmatrix}{\mathrm{H}}
\newcommand{\Ktmatrix}{\mathrm{K}}
\newcommand{\Ltmatrix}{\mathrm{L}}

\newcommand{\levi}{\epsilon}
\newcommand{\energy}{\varepsilon}

\newcommand{\be}{\begin{eqnarray}}
\newcommand{\ee}{\end{eqnarray}}

\newcommand{\genR}[2]{\mathfrak{R}_{#1}{}^{#2}}
\newcommand{\genL}[2]{\mathfrak{L}_{#1}{}^{#2}}
\newcommand{\genQ}[2]{\mathfrak{Q}_{#1}{}^{#2}}
\newcommand{\genS}[2]{\mathfrak{S}_{#1}{}^{#2}}
\newcommand{\genRd}[2]{\dot{\mathfrak{R}}_{#1}{}^{#2}}
\newcommand{\genLd}[2]{\dot{\mathfrak{L}}_{#1}{}^{#2}}
\newcommand{\genQd}[2]{\dot{\mathfrak{Q}}_{#1}{}^{#2}}
\newcommand{\genSd}[2]{\dot{\mathfrak{S}}_{#1}{}^{#2}}
\newcommand{\genP}{\mathfrak{P}}
\newcommand{\genH}{\mathfrak{H}}
\newcommand{\genJ}{\mathfrak{J}}
\newcommand{\genF}{\mathfrak{F}}
\newcommand{\genK}{\mathfrak{K}}

\def\pic #1#2{\hbox{\lower#1pt\hbox{~\mbox{\epsfxsize=20truemm \epsffile{#2}}}}}

\def\pic #1#2#3{\hbox{\lower#1pt\hbox{~\mbox{\includegraphics[scale=#3]{#2}}}}}

\begin{document}

\thispagestyle{empty}
\begin{flushright}\footnotesize
\texttt{hep-th/0611169}\\
\texttt{ITEP-TH-61/06}\\
\texttt{UUITP-15/06} \vspace{0.8cm}
\end{flushright}

\renewcommand{\thefootnote}{\fnsymbol{footnote}}
\setcounter{footnote}{0}

\begin{center}
{\Large\textbf{\mathversion{bold} Worldsheet Scattering in
$AdS_5\times S^5$ }\par}

\vspace{1.5cm}

\textrm{T.~Klose$^{1}$, T.~McLoughlin$^{2}$, R.~Roiban$^{2}$ and
K.~Zarembo$^{1}$\footnote{Also at ITEP, Moscow, Russia}}
\vspace{8mm}

\textit{$^{1}$ Department of Theoretical Physics, Uppsala University\\
SE-751 08 Uppsala, Sweden}\\
\texttt{Thomas.Klose,Konstantin.Zarembo@teorfys.uu.se} \vspace{3mm}

\textit{$^{2}$ Department of Physics, The Pennsylvania State
University\\ University Park, PA 16802, USA}\\
\texttt{tmclough,radu@phys.psu.edu} \vspace{3mm}

\par\vspace{1cm}

\textbf{Abstract} \vspace{5mm}

\begin{minipage}{14cm}
We calculate the S-matrix in the gauge-fixed sigma-model on
$\AdS_5\times \Sphere^5$ to the leading order in perturbation theory,
and analyze how supersymmetry is realized on the scattering states. A
mild nonlocality of the supercharges implies that their action on
multi-particle states does not follow the Leibniz rule, which is
replaced by a nontrivial coproduct. The plane wave symmetry algebra is
thus naturally enhanced to a Hopf algebra.
This structure mirrors that of the large 't~Hooft coupling
expansion of the S-matrix for the spin chain in the dual
super-Yang-Mills theory.
\end{minipage}

\end{center}

\newpage
\setcounter{page}{1}
\renewcommand{\thefootnote}{\arabic{footnote}}
\setcounter{footnote}{0}

\tableofcontents


\section{Introduction}

According to the AdS/CFT duality, type IIB string theory in the
$\AdS_5\times \Sphere^5$ background is equivalent to $\mathcal{N}=4$
super-Yang-Mills theory in four dimensions \cite{Maldacena:1998re}.
Understanding and proving the AdS/CFT correspondence requires however
solving both the planar limit of $\mathcal{N}=4$
super-Yang-Mills theory and the $\AdS_5\times \Sphere^5$
worldsheet string theory at finite
values of their coupling constants.

While this remains a formidable task, questions of a kinematical nature
-- such as the determination of their spectra -- may be answered by
making use of the special properties of these two theories. For
$\mathcal{N}=4$ super-Yang-Mills the spectrum of anomalous dimensions
of gauge-invariant operators is determined by an auxiliary spin chain
whose Hamiltonian is the dilatation operator of the theory. In the
appropriate variables, the AdS/CFT correspondence implies that the
anomalous dimensions of gauge-invariant operators should equal the
worldsheet energies of the corresponding closed string states.
Even though both the worldsheet sigma-model \cite{Bena:2003wd,
Kazakov:2004qf, Berk2}
and the spin chain that describes the
spectrum of the super-Yang-Mills theory \cite{Minahan:2002ve,longrange} are
integrable (see \cite{Beisert:2004ry} for a review), explicitly
solving them is a daunting task.

In the flat space limit the worldsheet theory is free and its
spectrum is built from noninteracting oscillators. The curvature and
RR flux of $\AdS_5\times \Sphere^5$ introduce nontrivial
interactions such that the spectrum is expected to be a complicated
collection of discrete levels.
However, the integrability of the theory guarantees that the
spectrum retains a Fock space structure. Indeed, one of the many
definitions of integrability is that one can globally separate
action-angle variables and thus define a set of independent
oscillators \cite{Faddeev's_book}.

Although explicitly separating variables is not easy in the quantum
theory \cite{Sklyanin:1995bm}, the features of the outcome of this
procedure are quite universal. The spectrum is determined by
quantization conditions for a set of particle's momenta which
typically constitute a set of coupled functional equations (the Bethe
equations \cite{Bethe:1931hc}). The $2\rightarrow 2$ S-matrix is of
central importance for this construction.  The S-matrix usually
determines the spectrum in an asymptotically large volume and with
some additional input the generalization to the exact finite-size
spectrum is possible in many cases.

The S-matrix for the super-Yang-Mills spin chain was introduced in
\cite{Staudacher:2004tk}. As discussed in
\cite{Beisert:su22-S-matrix,Beisert:2006qh} the non-perturbative
S-matrix is almost completely determined by the global symmetries
unbroken by the choice of vacuum state for the spin chain Hamiltonian. An
overall abelian phase remains undetermined by symmetries. It was suggested
\cite{Janik:2006dc} that it should obey a constraint of a similar
nature to the crossing symmetry in relativistic quantum field
theories.

The first two terms in the large 't~Hooft coupling expansion of the
abelian phase have been found in \cite{Arutyunov:2004vx} and
\cite{Hernandez:2006tk}, respectively. Subsequently an asymptotic
series solution
to the crossing condition was constructed in \cite{BHL}. An analytic
continuation to weak coupling, which reproduces the explicit
calculation \cite{Bern:2006ew} of the four-loop anomalous dimensions
of twist-two large spin operators was put forward in a recent paper
\cite{Beisert:2006ez} and further discussed in
\cite{GoHe_dressing}.

The aim of our work is to initiate the perturbative study of the
S-matrix of the entire worldsheet sigma-model.
Earlier studies, discussing special truncations of the field content
of the worldsheet theory, have appeared in \cite{KZ, Roiban:2006yc}.
Such calculations have the potential of checking the validity of
algebraic considerations for both the tensor structure and the abelian
phase of the S-matrix while providing insight into the realization of
the symmetries in the interacting theory as well as further confirming
its integrability.

Our starting point is the light-cone gauge-fixed worldsheet theory in
$\AdS_5\times \Sphere^5$. The Lagrangian has terms with arbitrary
numbers of fields of which the quadratic part is that of a free massive
theory.%
\footnote{This theory is also the light-cone gauge-fixed string
theory in a plane wave which was shown in \cite{Blau} to be a
Penrose limit of $\AdS_5\times \Sphere^5$.  It
was quantized in \cite{Metsaev:2001bj} and its complete spectrum was
constructed in \cite{Metsaev:2002re}. The relation between the string
theory spectrum and gauge-invariant super-Yang-Mills operators was
described in detail in \cite{Berenstein:2002jq}.}
The closed string spectrum is the Fock space of massive modes
with quantized momenta (BMN modes). The interactions are generated by
the geometric curvature and RR-flux; they induce corrections to the
free massive spectrum, which have been calculated to leading order in
\cite{Callan:2003xr} (see also \cite{Parnachev:2002kk,Frolov:2006cc}).
In the infinite-volume regime the spectrum is continuous and
interactions cause a non-trivial scattering of asymptotic states.

We will calculate the worldsheet scattering amplitudes%
\footnote{They can only be defined on an infinite string worldsheet
and should not be confused with the more familiar target-space
amplitudes.}
in the light-cone gauge to leading order in perturbation theory.
The residual symmetry of the sigma-model in that
gauge, the centrally extended $\algPSU(2|2)\otimes \algPSU(2|2)$, is
the same as the symmetry of the spin chain S-matrix
\cite{Beisert:su22-S-matrix}. On the worldsheet the central
charges arise once the level matching
condition is relaxed \cite{Arutyunov:2006ak}. As we will show,
a mild nonlocality of the supersymmetry generators enhances the
symmetry algebra to a Hopf algebra. We will argue that the main
consequences of this algebra hold also at the quantum level.

While rigorously proving (quantum) integrability is probably as hard
as solving the model exactly, the additional conservation laws
present in an integrable theory have testable consequences. In
particular, they kinematically forbid particle production in the
scattering processes and require factorization of the many-body
S-matrix. We will check these properties at tree level for the
gauge-fixed sigma-model in $\AdS_5\times \Sphere^5$ by explicit
calculations of scattering amplitudes. We should mention that
classical integrability (well established for the AdS string) does
not automatically guarantee that the corresponding quantum theory is
integrable, because conservation laws of higher charges may suffer
from quantum anomalies
\cite{Abdalla:1980jt}. For the case of the
strings in $\AdS_5\times \Sphere^5$  arguments in favor of quantum
integrability and the absence of anomalies have been formulated in
\cite{Berk2}.

We begin in \secref{sec:results} by describing the field content of
the gauge-fixed worldsheet theory and certain puzzling facts about
the interplay between its  Lagrangian
and its expected symmetries. We also summarize our results for the classical
S-matrix.  In \secref{sec:Hopf} we derive the action of
the symmetry generators on the S-matrix and thus solve the issues
raised in the previous section. The two-body S-matrix is
calculated to the leading order in perturbation theory in
\secref{sec:scalars}. There we also show that $2\rightarrow 4$
scattering amplitudes vanish for bosonic in- and out-states. In
\secref{sec:fermions} we calculate the complete tree-level S-matrix, which
we compare with the strong-coupling limit of the spin chain S-matrix
in \secref{sec:SpinChainSmatrices}. We conclude with the discussion
of the results in \secref{sec:conclusions}.

{\bf Note added:}  Arutyunov, Frolov and Zamaklar
\cite{Arutyunov:2006yd} constructed the S-matrix matrix that
satisfies the quantum Yang-Baxter equation and yields in the
weak-coupling limit the tree-level scattering matrix found here. As
a consequence, the tree-level scattering matrix should obey the
classical Yang-Baxter equation. We refer the reader to
\cite{Arutyunov:2006yd} for the detailed discussion of this
important property of the world-sheet S-matrix.

\section{Summary of results}
\label{sec:results}

The quantization of the Green-Schwarz string is a longstanding problem and over
time various solutions have been proposed, each preserving various
parts of the original symmetries of the theory; the more symmetry is
preserved the larger the number of unphysical fields appearing in
the worldsheet theory. The AdS/CFT correspondence relates gauge
theory observables to string theory observables. Consequently, for
the purpose of string theory calculations, one is tempted to explicitly
eliminate all unphysical degrees of freedom by fixing a unitary
gauge.  With this motivation in mind we will use the light-cone
gauge%
\footnote{There are essentially two ways to fix the light-cone
gauge in $\AdS_5\times \Sphere^5$, which differ by picking inequivalent
light-cone geodesics. In one case, the light-cone directions lie in
$\AdS_5$ \cite{MTT}; this gauge choice is possible only in the
Poincar\'e patch of $\AdS_5$.
In the other case the light cone is shared
between
$\AdS_5$ and  $\Sphere^5$
\cite{Arutyunov:2005hd,Frolov:2006cc}. We
consider the latter case.}
\cite{Frolov:2006cc}, the fixed-$J$ gauge
\cite{Callan:2003xr,Arutyunov:2004yx}
as well as a one-parameter superposition \cite{Arutyunov:2006gs}.%
\footnote{Since such gauges (which, incidentally, preserve the least
amount of symmetry) typically involve solving the classical
constraints of the theory, it is not immediately clear whether
any gauge in this class
is justified at the quantum level. We are however interested
in the classical theory where no subtleties can arise.}

\paragraph{The fields.} For our purpose it is most convenient to
choose the global coordinatization of $\AdS_5\times \Sphere^5$; we will
choose the metric
\[ \label{eqn:adsxs}
 ds^2 = -G_{tt}(z) dt^2
        +G_{zz}(z) dz^2
        +G_{\varphi\varphi}(y) d\varphi^2
        +G_{yy}(y) dy^2
\]
where
\[
       G_{tt} = \lrbrk{\frac{1+\frac{z^2}{4}}{1-\frac{z^2}{4}}}^2 \; ,
 \quad G_{zz} = \frac{1}{\lrbrk{1-\frac{z^2}{4}}^2} \; ,
 \quad G_{\varphi\varphi} = \lrbrk{\frac{1-\frac{y^2}{4}}{1+\frac{y^2}{4}}}^2 \; ,
 \quad G_{yy} = \frac{1}{\lrbrk{1+\frac{y^2}{4}}^2} \; .
\]
$y^m$ and $z^\mu $ are four-component vectors. $y^2$ and $z^2$ stand
for their Euclidean scalar squares. The corresponding worldsheet
fields are denoted by capital letters $T,Z,\Phi,Y$. One combination
of the longitudinal fields $T$ and $\Phi$ will be used in our gauge
choice while the derivatives of the other (independent) combination
are determined by the Virasoro constraints.  As usual in light-cone
gauge, its zero-mode is however undetermined.

The $\grSO(8)\subset\grSO(6)\times\grSO(4,2)$ preserved by the
gauge choice at the quadratic level is broken by interactions to
$\grSO(4)\times\grSO(4)$. The transverse bosonic fields, $Y^m$ and
$Z^\mu$, form the defining representation of this group. A more
efficient parametrization in the presence of fermions is provided by
the isomorphism $\grSO(4)\simeq (\grSU(2)\times \grSU(2))/\Integers_2$. Its
explicit realization -- in terms of the Pauli matrices
$\sigma_m =(\unit,i\vec\sigma)$ and $\sigma_\mu=(\unit,i\vec\sigma)$
for the two copies of $\grSO(4)$ -- represents $Y$ and $Z$ as
bispinors of the relevant $\grSO(4)$:
\[\label{indexiana}
 Y_{\lAA\rAA} = (\sigma_m)_{\lAA\rAA} Y^m
 \comma
 Z_{\laa\raa} = (\sigma_\mu)_{\laa\raa} Z^\mu \; .
\]
The fermions also transform as bi-spinors of $\grSO(4)\times
\grSO(4)$, but they are charged with respect to different
$\grSU(2)$ factors. The worldsheet fermions that remain after
fixing the $\kappa $-symmetry gauge will be denoted by
\[
\Psi_{\lAA\raa}\qquad\mbox{and}\qquad\Bsi_{\laa\rAA} \; .
\]
The quantum numbers of all fields with respect to $\grSU(2)^4$ are
summarized in \tabref{fig:fields-indices}.
\begin{table}
\begin{center}
\begin{tabular}{|c|c|c|c|c|} \cline{2-5}
\multicolumn{1}{c|}{} & \multicolumn{2}{c|}{$\Sphere^5$} &
\multicolumn{2}{c|}{$\AdS_5$ \vphantom{\rule[-2mm]{0mm}{7mm}}} \\
\hline
                  & $\grSU(2)$  & $\grSU(2)$  & $\grSU(2)$  & $\grSU(2)$  \vphantom{\rule[-2mm]{0mm}{7mm}} \\ \hline
``Spin''          & $J$         & $\dot{J}$   & $\dot{S}$   & $S$         \vphantom{\rule{0mm}{5mm}} \\
Index             & $\lAA = 1,2$
                  & $\rAA = \dot{1},\dot{2}$
                  & $\raa = \dot{3},\dot{4}$
                  & $\laa = 3,4$                                          \\ \hline
$Y_{\lAA\rAA}$    & $\rep{2}$   & $\rep{2}$   & $\rep{1}$   & $\rep{1}$   \vphantom{\rule{0mm}{5mm}} \\
$Z_{\laa\raa}$    & $\rep{1}$   & $\rep{1}$   & $\rep{2}$   & $\rep{2}$   \\
$\Psi_{\lAA\raa}$ & $\rep{2}$   & $\rep{1}$   & $\rep{2}$   & $\rep{1}$   \\
$\Bsi_{\laa\rAA}$ & $\rep{1}$   & $\rep{2}$   & $\rep{1}$   &
$\rep{2}$   \\ \hline
\end{tabular}
\end{center}
\caption{{$\grSU(2)^4$ quantum numbers of the physical degrees of freedom.}
We use different values for the $\Sphere^5$ part
and the $\AdS_5$ part, such that an index can be identified from its
value without giving the index symbol. Representations of
$\grSU(2)^4$ will be denoted by $(\rep{2J+1 , 2\dot{J}+1 , 2\dot{S}+1, 2S+1})$.}
\label{fig:fields-indices}
\end{table}
This description does not fix the action of the supercharges on
fields. It turns out \cite{Callan:2003xr} that bosons and fermions together form
the bi-fundamental representation $(\rep{(2|2),(2|2)})$ of
$\grPSU(2|2)_L\times \grPSU(2|2)_R$. The bosonic subgroup of each
$\grPSU(2|2)$ factor consists of two $\grSU(2)$ groups, one from each
of the original $\grSO(4)$ factors. The supercharges relate bosons and
fermions following the edges of the diagram:
\[
 \begin{array}{ccc}
 Y_{\lAA\rAA}    & \leftrightarrow & \Psi_{\lAA\raa} \\[1mm]
 \updownarrow    &                 & \updownarrow    \\[1mm]
 \Bsi_{\laa\rAA} & \leftrightarrow & Z_{\laa\raa}
 \end{array}
\]
The odd generators of $\grPSU(2|2)_L$ act vertically and the ones of $\grPSU(2|2)_R$ act horizontally.

Even though the complete supergroup symmetry is not manifest, one may
formally define superindices $A=(\lAA|\laa)$ and ${\dot
A}=(\rAA|\raa)$, where the lower-case latin indices are
Gra{\ss}mann-even and the greek indices are Gra{\ss}mann-odd. Thus,
all fields combine into a single bi-fundamental supermultiplet of
$\grPSU(2|2)_L\times
\grPSU(2|2)_R$ which we will denote by $\Phi_{A{\dot A}}$.

\paragraph{The S-matrix.} The two-particle S-matrix is an operator
between two copies of the tensor product of the module $W_{p}$,
generated by $\Phi_{A{\dot A}}(p)$, with itself for different
momenta:
\[
\Smatrix:{W}_{p}\otimes W_{p'}\rightarrow W_{p}\otimes W_{p'} \; .
\]
In the basis provided by $\Phi_{A{\dot A}}(p)$, its matrix representation is
\[\label{smatrix_components}
 \Smatrix \, \ket{\Phi_{A\dot{A}}(p) \Phi_{B\dot{B}}(p')}
 = \ket{\Phi_{C\dot{C}}(p) \Phi_{D\dot{D}}(p')} \,
\Smatrix_{A\dot{A}B\dot{B}}^{C\dot{C}D\dot{D}}(p,p')
\]
and barring anomalies, the S-matrix respects the symmetries of the
theory. In an integrable theory the S-matrix satisfies a number of
additional kinematic constraints: there should be no particle
production and the many-body S-matrix should factorize into the
products of the two-particle S-matrices. Consistency of the
factorization requires that the latter S-matrix satisfies the
quantum Yang-Baxter equation (YBE). The YBE is very constraining and
in particular a factorizable S-matrix invariant under a non-simple
group, such as $\grPSU(2|2)\times \grPSU(2|2)$, must be a tensor
product of S-matrices for each of the factors (see e.g. \cite{OWR})%
\footnote{This can be understood as a requirement that the
Faddeev-Zamolodchikov algebra is also a direct product: the field
$\Phi_{A{\dot A}}$ is represented by a bilinear in oscillators:
$\Phi_{A{\dot A}}\sim z_Az_{\dot{A}}$ each transforming under one of
the $\grPSU(2|2)$ factors. The two sets of oscillators mutually
commute. The braiding relations for each of these sets are
determined by an $\grPSU(2|2)$-invariant S-matrix $\smatrix$
consistent with the Lagrangian of the theory.}
:
\be
\Smatrix=\smatrix\otimes \smatrix \comma
\Smatrix_{A\dot{A}B\dot{B}}^{C\dot{C}D\dot{D}}(p,p')=
\smatrix_{AB}^{CD}(p,p')\smatrix_{\dot{A}\dot{B}}^{\dot{C}\dot{D}}(p,p')
\; .
\label{factorize_0}
\ee
It is important to note that a factorized tensor structure does not
follow solely from the $\grPSU(2|2)\times\grPSU(2|2)$ symmetry
considerations. For example, it is in principle possible to scatter a
pair of excitations uncharged under the first $\grPSU(2|2)$ in a
singlet combination under the second $\grPSU(2|2)$ into a pair of
excitations uncharged under the second $\grPSU(2|2)$ in a singlet
combination under the first $\grPSU(2|2)$. In fact, simple inspection
of the gauge-fixed Lagrangian yields no hint of the factorized structure
\eqref{factorize_0}. Confirming group factorization is thus an
important test of integrability.

Since only $\grSU(2)\times\grSU(2)\subset\grPSU(2|2)$
is a manifest symmetry of the gauge-fixed worldsheet theory,
$\smatrix$ may be parametrized in terms of ten unknown functions
of the momenta $p$ and $p'$ of the two incoming particles:%
\footnote{These definitions are similar but not identical to those
of \cite{Beisert:su22-S-matrix}. The relationship between the two
definitions is given in equation \eqref{relation} below.}.
\psset{unit=1mm}\psset{xunit=0.8mm,yunit=0.8mm}
\begin{align}
\quad\smatrix_{\lAA\lBB}^{\lCC\lDD} & = \Asmatrix \,\delta_\lAA^\lCC \delta_\lBB^\lDD + \Bsmatrix \,\delta_\lAA^\lDD \delta_\lBB^\lCC &&
\raisebox{-1mm}{\begin{pspicture}(5,5)
  \psline[linewidth=0.2]{-}(0,0)(0,5)
  \psline[linewidth=0.2]{-}(5,0)(5,5)
\end{pspicture}} \;\quad
\raisebox{-1mm}{\begin{pspicture}(5,5)
  \psline[linewidth=0.2]{-}(0,0)(5,5)
  \psline[linewidth=0.2]{-}(5,0)(0,5)
\end{pspicture}} \;\; , \qquad\quad &
\smatrix_{\lAA\lBB}^{\lcc\ldd} & = \Csmatrix \,\levi_{\lAA\lBB} \levi^{\lcc\ldd} &&
\raisebox{-1mm}{\begin{pspicture}(5,5)
  \pscurve[linewidth=0.2,linestyle=solid]{-}(0,0)(2.5,2)(5,0)
  \pscurve[linewidth=0.4,linestyle=dotted]{-}(0,5)(2.5,3)(5,5)
\end{pspicture}} \;\; , \quad \nn \\
\smatrix_{\laa\lbb}^{\lcc\ldd} & = \Dsmatrix \,\delta_\laa^\lcc \delta_\lbb^\ldd + \Esmatrix \,\delta_\laa^\ldd \delta_\lbb^\lcc &&
\raisebox{-1mm}{\begin{pspicture}(5,5)
  \psline[linewidth=0.4,linestyle=dotted]{-}(0,0)(0,5)
  \psline[linewidth=0.4,linestyle=dotted]{-}(5,0)(5,5)
\end{pspicture}} \;\quad
\raisebox{-1mm}{\begin{pspicture}(5,5)
  \psline[linewidth=0.4,linestyle=dotted]{-}(0,0)(5,5)
  \psline[linewidth=0.4,linestyle=dotted]{-}(5,0)(0,5)
\end{pspicture}} \;\; , &
\smatrix_{\laa\lbb}^{\lCC\lDD} & = \Fsmatrix \,\levi_{\laa\lbb} \levi^{\lCC\lDD} &&
\raisebox{-1mm}{\begin{pspicture}(5,5)
  \pscurve[linewidth=0.4,linestyle=dotted]{-}(0,0)(2.5,2)(5,0)
  \pscurve[linewidth=0.2,linestyle=solid]{-}(0,5)(2.5,3)(5,5)
\end{pspicture}} \;\; , \label{smat0} \\
\smatrix_{\lAA\lbb}^{\lCC\ldd} & = \Gsmatrix \,\delta_\lAA^\lCC \delta_\lbb^\ldd &&
\raisebox{-1mm}{\begin{pspicture}(5,5)
  \psline[linewidth=0.2]{-}(0,0)(0,5)
  \psline[linewidth=0.4,linestyle=dotted]{-}(5,0)(5,5)
\end{pspicture}} \;\; , &
\smatrix_{\laa\lBB}^{\lcc\lDD} & = \Lsmatrix \,\delta_\laa^\lcc \delta_\lBB^\lDD &&
\raisebox{-1mm}{\begin{pspicture}(5,5)
  \psline[linewidth=0.4,linestyle=dotted]{-}(0,0)(0,5)
  \psline[linewidth=0.2]{-}(5,0)(5,5)
\end{pspicture}} \;\; , \nn \\
\smatrix_{\lAA\lbb}^{\lcc\lDD} & = \Hsmatrix \,\delta_\lAA^\lDD \delta_\lbb^\lcc &&
\raisebox{-1mm}{\begin{pspicture}(5,5)
  \psline[linewidth=0.2]{-}(0,0)(5,5)
  \psline[linewidth=0.4,linestyle=dotted]{-}(5,0)(0,5)
\end{pspicture}} \;\; , &
\smatrix_{\laa\lBB}^{\lCC\ldd} & = \Ksmatrix \,\delta_\laa^\ldd \delta_\lBB^\lCC &&
\raisebox{-1mm}{\begin{pspicture}(5,5)
  \psline[linewidth=0.4,linestyle=dotted]{-}(0,0)(5,5)
  \psline[linewidth=0.2]{-}(5,0)(0,5)
\end{pspicture}} \;\; . \nn
\end{align}

The first nontrivial order in the expansion of the S-matrix in the
sigma-model coupling constant $2\pi/\sqrt{\lambda}$ defines the
T-matrix
\[ \label{asmatrix_expansion}
 \Smatrix = \unit + \frac{2\pi i}{\sqrt{\lambda }} \, \Tmatrix +
\order\lrbrk{ \frac{1}{\lambda} } \; .
\]
The T-matrix should satisfy the classical limit of the YBE
(cYBE). Among the restrictions imposed by it is the requirement that
the T-matrix inherits the factorized form from the S-matrix:
\[ \label{factorize_at}
\Tmatrix = \unit\otimes\tmatrix + \tmatrix\otimes\unit \; .
\]
The components of $\tmatrix$ are parametrized similar to \eqref{smat0} by
\begin{align}
\tmatrix_{\lAA\lBB}^{\lCC\lDD} & = \Atmatrix \,\delta_\lAA^\lCC \delta_\lBB^\lDD + \Btmatrix \,\delta_\lAA^\lDD \delta_\lBB^\lCC \; , &
\tmatrix_{\lAA\lBB}^{\lcc\ldd} & = \Ctmatrix \,\levi_{\lAA\lBB} \levi^{\lcc\ldd} \; , \nn \\
\tmatrix_{\laa\lbb}^{\lcc\ldd} & = \Dtmatrix \,\delta_\laa^\lcc \delta_\lbb^\ldd + \Etmatrix \,\delta_\laa^\ldd \delta_\lbb^\lcc \; , &
\tmatrix_{\laa\lbb}^{\lCC\lDD} & = \Ftmatrix \,\levi_{\laa\lbb} \levi^{\lCC\lDD} \; , \label{eqn:tmatrix-coeff} \\
\tmatrix_{\lAA\lbb}^{\lCC\ldd} & = \Gtmatrix \,\delta_\lAA^\lCC \delta_\lbb^\ldd \; , &
\tmatrix_{\laa\lBB}^{\lcc\lDD} & = \Ltmatrix \,\delta_\laa^\lcc \delta_\lBB^\lDD \; , \nn \\
\tmatrix_{\lAA\lbb}^{\lcc\lDD} & = \Htmatrix \,\delta_\lAA^\lDD \delta_\lbb^\lcc \; , &
\tmatrix_{\laa\lBB}^{\lCC\ldd} & = \Ktmatrix \,\delta_\laa^\ldd \delta_\lBB^\lCC \; . \nn
\end{align}
The relation with the coefficients appearing in $\smatrix$ is given by an expansion similar to \eqref{asmatrix_expansion}.

\paragraph{A puzzle.} Before diffeomorphism and kappa gauge fixing
the worldsheet theory is classically integrable; since fixing a
unitary gauge may be interpreted as expanding around a classical
solution and solving some of the equations of motion, the gauge-fixed
theory is expected to be integrable at the classical level.
As such,
one is entitled to expect that it has a two-particle factorized
scattering matrix and that, despite the symmetry algebra being
centrally-extended
\cite{Arutyunov:2006ak}, the symmetry transformations act on
multi-excitation states via the Leibniz rule.\footnote{It is worth
mentioning that all these expectations are realized in theories with
centrally-extended algebras -- e.g. WZW models.}
It is moreover usually the case that symmetries fix the tensor
structure of the scattering matrix.

Quite surprisingly, the situation at hand is somewhat different: under the
assumption of a Leibniz rule action on multi-particle states, the
constraints imposed by the symmetry algebra -- while qualitatively
consistent with the structure of the world sheet Lagrangian -- are
not consistent with the explicit calculation of the S-matrix.
%
%
%
%
%

It appears therefore that the mere existence of a nontrivial (even
momentum dependent) center of the symmetry algebra is insufficient to
explain the results of worldsheet perturbation theory.
The resolution of this puzzle relies on the observation
that, even though their action on fields appears at first sight to be
local, the
$\algPSU(2|2)^2$ generators are in fact nonlocal objects.
Consequently, their action is subtle and may not follow the Leibniz
rule. We will also argue that the nonlocal structure of the symmetry
generators is special and it is not affected by perturbative quantum
corrections.

\paragraph{The tree-level S-matrix.}
The T-matrix can be explicitly calculated in perturbation theory. In
the gauge where $J_+=(1-a)J+aE$
is fixed%
\footnote{Here $E$ is the worldsheet energy and $J$ is the angular
momentum on $S^5$. The constant $a$ is a gauge parameter, which
allows one to interpolate between various gauges used in the
literature.}
and to leading order in $1/\sqrt{\lambda}$ we found
\begin{align}
\Atmatrix(p,p') & = \Quarter \biggsbrk{ (1-2a)\lrbrk{\cpp} + \frac{\lrbrk{p-p'}^2}{\cpp} } \; , \nn \\
\Btmatrix(p,p') & = -\Etmatrix(p,p')= \frac{pp'}{\cpp} \; , \nn \\
\Ctmatrix(p,p') & = \Ftmatrix(p,p') = \Half \frac{\sqrt{\left(\energy+1\right)\left(\energy'+1\right)}
\lrbrk{\cpp+p'-p}}{\cpp} \; , \label{ourfinalequation} \\
\Dtmatrix(p,p') & = \Quarter \biggsbrk{ (1-2a)\lrbrk{\cpp} - \frac{\lrbrk{p-p'}^2}{\cpp} } \; , \nn \\
\Gtmatrix(p,p') & = -\Ltmatrix(p',p) = \Quarter \biggsbrk{ (1-2a)\lrbrk{\cpp} - \frac{p^2-p'^2}{\cpp} } \; , \nn \\
\Htmatrix(p,p') & = \Ktmatrix(p,p') = \Half \, \frac{pp'}{\cpp} \, \frac{\left(\energy+1\right)\left(\energy'+1\right)-pp'}{\sqrt{\left(\energy +1\right)\left(\energy '+1\right)}} \; . \nn
\end{align}
Here $\energy = \sqrt{1+p^2}$ denotes the relativistic energy. The
S-matrix is gauge-dependent, since unlike the spectrum it is not a
physical object with clear target-space interpretation. However, the
S-matrix determines the spectrum via Bethe equations (at least
asymptotically for infinitely long strings) and its gauge-dependence
should be simple enough for the solutions of the Bethe equations to be
gauge invariant. Indeed, in the class of gauges discussed here, only
the diagonal matrix elements are gauge-dependent. The differences
between different gauges can be attributed to the gauge dependence of
the length of the string \cite{Staudacher:2004tk}. These two effects,
the difference in the length and the gauge-dependence of the S-matrix,
mutually cancel in the Bethe equations
\cite{Frolov:2006cc,Arutyunov:2006iu}.

\section{Hopf algebra}
\label{sec:Hopf}

The solution to the puzzle described in the previous section and an
explanation of the results outlined there turns out to be quite
interesting. At its foundation lies the fact that mutual nonlocality
of symmetry currents and fundamental fields leads to nontrivial
effects which introduce a natural ordering on the fixed-time spacial
slices of the worldsheet. This philosophy was applied extensively to
the analysis of the nonlocal integrals of motion of relativistic
two-dimensional integrable field theories (see e.g.
\cite{Bernard:1990tt,Fendley:1991ve}) where it was shown to be
equivalent to the YBE. Here we will identify a mild nonlocality of the
Noether currents of the $\algPSU(2|2)^2$ symmetry and analyze its
consequences.
We will argue that, in spite of being arrived at through a classical
treatment, the technique we use and the basic structure of the result hold
unmodified at the quantum level.

\begin{figure}[ht]
\centerline{\includegraphics[scale=0.5]{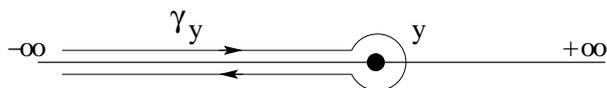}}\caption{The contour
$\gamma_y$ for the action of the global charges $Q_{(1)}$ on a field
inserted at the position $y$. \label{gammay}}\nn
\end{figure}

\begin{figure}[ht]
\centerline{\includegraphics[scale=0.45]{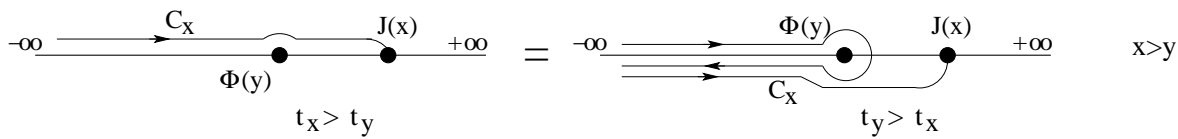}}
\caption{Contour manipulations leading to nontrivial braiding in the
product of mutually-nonlocal fields.  \label{c_manip}}\nn
\end{figure}

Quite generally, given a current $J$ and a field $\Phi$ on a
$(1+1)$-dimensional worldsheet, their left- and
right-multiplications are related by
\be
J{}^A{}_B(x)\,\Phi^C(y)=
\Theta{}^{ACF}_{BDE}\,
\Phi^D(y)J{}^E{}_F(x)~~~{\rm if}~~~x>y~,
\label{braiding} \ee where $\Theta$ is usually called the braiding
matrix. Obviously, if the current and the field are mutually local
the braiding matrix is trivial: \be
\Theta{}^{ACF}_{BDE}=\delta^A_E\delta^C_D\delta^F_B~~. \ee However,
if $J$ and $\Phi$ are mutually nonlocal, then the braiding can be
nontrivial. For example, in virtually all theories exhibiting
nonlocal conserved charges, the product between the current
$J_{(2)}$ whose conserved charge is the first nonlocal charge and
the fundamental field of the theory is \be J{}_{(2)}^a(x)\,\Phi(y)=
\Phi(y)J{}_{(2)}^a(x)-\half f^{abc}{\widehat Q}{}^b_{(1)}(\Phi(y))
J_{(1)}^c(x)~~~{\rm for}~~~x>y \; , \label{braiding_Q2} \ee where
$a,\,b$ and $c$ are adjoint indices, $J_{(1)}$ are the currents for
global symmetries, $f^{abc}$ are the structure constants of the
corresponding symmetry group and ${\widehat Q}_{(1)}(\Phi(y))$
denotes the usual action of the global symmetries on the fields
$\Phi(y)$: \be {\widehat Q}{}^b_{(1)}(\Phi(y))=\int_{\gamma_y}d z
J_{(1)}^b(z) \Phi(y)~~. \label{action} \ee Here the charges acting
on fields are defined by integrating the forms dual to the currents along
 a contour surrounding the point $y$
and not simply  along an equal-time
slice. The contour $\gamma_y$ starts and ends at $z=-\infty$ and
encircles the point $y$ (cf. \figref{gammay}). It is important to
note that as we are always considering conserved currents,
$\partial_\mu J^a_\mu=0$, the exact shape of the contour is
irrelevant, or in other words, given that we are integrating a closed form
the result only depends weakly on the shape of the contour.
To understand the origin of the nontrivial braiding matrix $\Theta$ in
\eqref{braiding} let us consider  a current $J$
whose definition involves a choice of contour $C_x$ starting at
$x=-\infty$ and ending at the location of the current. For any field
$\Phi$, the product $J(x)\,\Phi(y)$ comes equipped with the natural
time-ordering that a field located to the left of another is also at a
later time.%
\footnote{The reverse choice -- that a field located to the
left of another is also at an earlier time -- may also be made.}
Explicitly, $J(x)\,\Phi(y)\equiv J(x,t+\epsilon)\,\Phi(y,
t)\big|_{\epsilon\rightarrow 0}$ and similarly $\Phi(y) J(x)\equiv
\Phi(y, t+\epsilon)J(x,t)\big|_{\epsilon\rightarrow 0}$. In this
latter case one must make sure that the contour defining $J(x)$ also
sits in the past of $\Phi$. Let us then consider the left-hand side of
\eqref{braiding}, $J(x)\,\Phi(y)$ with $x>y$, and rearrange it such
that it is in the correct space-like and time-like order. The
necessary transformations are illustrated in \figref{c_manip}. In this
figure time runs upward.
The left-hand side of \figref{c_manip} accounts for the spatial
order $J{}_{(2)}^a(x,t+\epsilon)\,\Phi(y, t)\rightarrow
\Phi(y,t)J{}_{(2)}^a(x,t+\epsilon)$. The contour $C_x$ must then be
deformed to make sure that, as required by the fact that $\Phi$ is
located to the left of $J$, $\Phi(y)$ is always in the future of the
contour defining $J(x)$. The appearance of the contour starting and
ending at $x=-\infty$ and encircling $\Phi(y)$ is at the origin of
the braiding matrix $\Theta$; its precise expression depends on the
details of the current $J$.

In local quantum field theories it is typically the case that the
currents corresponding to the global symmetries are local with respect
to the fundamental fields of the theory and thus do not exhibit any
nontrivial braiding. This is equivalent to the fact
that the action of symmetry generators on fields is described by
commutators.
As we now describe, it turns out that a notable exception to this
rule is the worldsheet theory in light-cone gauge, where the
nonlocality is provided by the light-cone field $x^-$.

The gauge-invariant Hamiltonian of the worldsheet sigma-model
depends only on the derivatives of $x^-$; they are determined by the
solutions of the constraints and so -- order by order in the number
of fields -- are local operators.  As pointed out in
\cite{Arutyunov:2006ak} (see also \appref{sec:constJ}), the
$\algPSU(2,2|4)$ (super)currents whose supercharges generate
$\algPSU(2|2)^2$ depend on $x^-$ rather than its derivatives:
\be
J_{Q{}^A{}_B} =e^{i\sigma_{AB} x^-/2}{\widetilde J}_{Q{}^A{}_B}
~~~~~~~~ \sigma_{AB} =\grade{A}-\grade{B} ~~~~~~~~
x^-(x)=\int_{C_x}\,dw\, {{\acute x}^-}(w) \label{current_Q}
\ee
where ${\widetilde J}$ is a local combination of the transverse
fields and $\grade{A}$ denotes the grade of the index $A$:
$\grade{a}=0,\,\grade{\alpha}=1$. The contour $C_x$ starts at
negative infinity and is the one on the left-hand side of
\figref{c_manip}. Using the fact that the Virasoro constraints imply
that \be \{{{\acute
x}^-}(w),\,\Phi(y)\}=i\,\frac{2\pi}{\sqrt{\lambda}}
\delta(w-y)\,{\acute \Phi}(y) \ee it is trivial to find, using the
same contour manipulations as described in \figref{c_manip}, that
\be J_{Q{}^A{}_B}(x)
\Phi(y)=\left(e^{-\frac{\pi\sigma_{AB}}{\sqrt{\lambda}}
\partial_y}\Phi(y)\right)J_{Q{}^A{}_B}(x)
~~~{\rm for}~~~x>y \; .
\label{braidingxmin}
\ee
In this case the contour deformation is allowed
because the integrand of the contour integral is a total derivative.

\begin{figure}[ht]
\centerline{\includegraphics[scale=0.36]{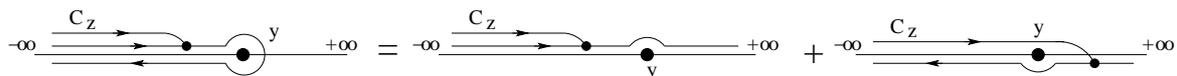}}
\caption{Contour manipulations for the action of a charge on single
field. \label{c_manip_charge}}\nn
\end{figure}

To find the action of the global symmetry generators on a generic
field $\Phi$ we use \eqref{action}. Integrating \eqref{braidingxmin}
the contour $\gamma_z$ described in \figref{gammay}, restoring the
indices \eqref{braiding} and using the fact that this contour may be
split as shown in \figref{c_manip_charge} one immediately arrives at
the conclusion that the worldsheet supercharges belonging to
$\algPSU(2|2)^2$ act as follows:
\be
{\widehat Q}{}_{(1)}^A{}_B(\Phi{}^C(y))=Q{}_{(1)}^A{}_B\Phi^C(y)-
\left(e^{-\frac{\pi\sigma_{AB}}{\sqrt{\lambda}}
\partial_y} \Phi{}^C(y)\right)Q{}_{(1)}^A{}_B
\label{single_field_action}
\ee
where $Q_{(1)}$ are the usual Noether charges associated to the
currents $J_{Q{}^A{}_B}$.
\be
Q{}_{(1)}^A{}_B=\int_{-\infty}^\infty dz
\,J_{Q{}^A{}_B}(z)~~.
\ee
Let us note that, had the $e^{i\sigma_{AB} x^-/2}$ factor been
absent from the Noether currents,  the equation
\eqref{single_field_action} reduced to the usual Poisson bracket
action of the Noether charge on fields.

These arguments can easily be repeated recursively for
multi-particle states. For our purpose only two-particle states are
of direct interest. Using the same logic as in \cite{Bernard:1990tt}
for the bilocal charges of various integrable field theories, the
action of supercharge on a product of fields
$\Phi^C(x_1)\Phi^D(x_2)$ requires picking a contour starting and
ending at negative infinity and encircling the points $x_1$ and
$x_2$. The contour is then deformed to separate the action on the
two fields; the same arguments as above lead to
\be {\widehat
Q}{}_{(1)}^A{}_B(\Phi{}^C(x)\Phi{}^D(x') )= {\widehat
Q}{}_{(1)}^A{}_B(\Phi{}^C(x))\,\Phi{}^D(x') +
\left(e^{-\frac{\pi\sigma_{AB}}{\sqrt{\lambda}}
\partial_{x}}\Phi{}^C(x)\right)
{\widehat Q}{}_{(1)}^A{}_B(\Phi{}^D(x'))
\ee

From a formal algebraic standpoint, this action defines a nontrivial
coproduct
\be
\Delta({\widehat Q}{}_{(1)}^A{}_B) ={\widehat
Q}{}_{(1)}^A{}_B\otimes \unit +
e^{-\frac{\pi\sigma_{AB}}{\sqrt{\lambda}}\partial_{x}}\unit\otimes {\widehat
Q}{}_{(1)}^A{}_B
\label{coprod}
\ee
thus promoting the $\algPSU(2|2)^2$ to a Hopf algebra (up to a
definition of antipode and counit). It is in fact easy to see that
this coproduct is precisely that constructed from gauge theory
algebraic considerations in \cite{GoHe}.\footnote{A coproduct
implementing the gauge theory symmetry algebra on
two-particle spin chain states was constructed in
\cite{Plefka:2006ze}. It is related to the one in \cite{GoHe} by a
nonlocal field redefinition.}

This result represents the resolution of the puzzle described in
the previous section. Most importantly, equation \eqref{coprod}
obviously implies that the $\algPSU(2|2)$ generators do not act on
products of fields following the Leibniz rule. It is with
this coproduct action that the result of the explicit calculation of
the $\tmatrix$ has to be consistent. More precisely, defining the
S-matrix as an operator \be S:W_p\otimes W_{p'}\rightarrow
W_p\otimes W_{p'}~~, \ee the requirement of invariance under global
symmetries translates into \footnote{ It is worth noting that,
taking expectation values of this equation between two-particle
states (located, respectively, $t=+\infty$ and $t=-\infty$), leads
to the same constraints on the S-matrix as in the gauge theory
analysis.}
\be \left( \unit\otimes {\widehat Q}{}_{(1)}{}^A{}_B + {\widehat
Q}{}_{(1)}{}^A{}_B \otimes
e^{-\frac{i\pi\sigma_{AB}}{\sqrt{\lambda}} p'}\unit \right)S =
S\left( {\widehat Q}{}_{(1)}^A{}_B\otimes \unit +
e^{-\frac{i\pi\sigma_{AB}}{\sqrt{\lambda}} p}\unit\otimes {\widehat
Q}{}_{(1)}^A{}_B \right) \label{S_invariance} \ee In
\secref{sec:symmetries} we will check, to leading order in the
$1/\sqrt{\lambda}$ expansion, that this is indeed so. As we will
discuss shortly, the conservation of the nonlocal charges should
also involve a modified action.
%
%

It is worth noting that the details of the coproduct \eqref{coprod}
depend on the choice of gauge, in particular on the expression of $x^-$
in terms of the transverse fields. This is one source of
gauge-dependence of the worldsheet S-matrix and it
is the reflection at the algebraic level of the gauge dependence
observed in its explicit calculation.

The arguments used above work just as well at the quantum level
provided that no nonlocal contributions to the global symmetry
currents are generated by quantum corrections. Assuming that
perturbation theory in the light-cone gauge-fixed worldsheet theory
is well-defined, it is not hard to construct a two-step argument that
this is indeed the case. First, we notice that at any finite order in
perturbation theory the relevant part of the Lagrangian is local and
it does not depend on $x^-$ but only on its derivatives. Consequently,
the (finite or infinite) renormalization of the currents cannot
involve $x^-$ and thus must be local (in the sense that they do not
require a choice of contour). The second observation is that $x^-$ is
the only field exhibiting nontrivial boundary conditions \be
x^-(-\infty)-x^-(+\infty)=p_{\rm ws}~~. \ee From this standpoint it
behaves similarly to a soliton whose corresponding topological charge
is the worldsheet momentum. Since perturbative effects in a massive
theory are local,
one may safely expect that they will not affect the action of
$x^-$ on local fields.

Putting together these two pieces of argument we reach the
conclusion that the structure \eqref{current_Q} of the Noether
supercurrents survives quantum corrections and consequently so does
the structure of the coproduct \eqref{coprod}. Quantum corrections
affect only the action of the global charges on single fields ${\widehat
Q}{}_{(1)}^A{}_B(\Phi{}^C(x_i))$, which is braiding-independent when
evaluated on the vacuum.

While formally implying an agreement between gauge and string
theory to all orders in perturbation theory (up to gauge artifacts),
the discussion above
does not directly address the consistency of the resulting S-matrix
with integrability.
One way to settle this issue
is to check whether the S-matrix commutes with the
bilocal (and consequently with the higher nonlocal) charge(s).

This is a cumbersome and tedious calculation and we will only
briefly outline the necessary steps for defining the action of
bilocal charges on the asymptotic states leaving the details of the
calculation and the constraints following from them to the
interested reader.
%
Evidence for the consistency of the Hopf algebra with integrability in
specific examples was previously discussed in e.g. \cite{BeFe}.  There
it was argued that, while the YBE for the monodromy matrix was not
modified, its logarithmic derivative (describing charge conservation)
is modified by the inclusion of braiding matrices similar to those in
\eqref{cop_bil}. The origin of this modification was traced to the
rules of differential calculus over the quantum group.
%
It is therefore reasonable to expect that if the asymptotic states
form a representation of the Hopf algebra, the YBE is satisfied. It
would be interesting to see if it is possible to choose states for the
S-matrix described in \cite{Beisert:su22-S-matrix}.

As in the case of the conservation of global charges, the
conservation of the first nonlocal charge can be expressed as
\be
\kern-10pt  \lim_{T\to\infty} \bra{A,p;B,p'} e^{iHT}\Delta({\widehat Q}_{(2)})
\ket{C,p;D,p'}
= \lim_{T\to\infty} \bra{A,p;B,p'} \Delta({\widehat Q}_{(2)}) e^{iHT}
\ket{C,p;D,p'} \, .
\label{cons}
\ee
where $\Delta({\widehat Q}_{(2)})$ denotes the action
of $Q_{(2)}$ on a product of two fields. For relativistic field
theories with local Noether currents it is known that this analysis
leads to the same result as the Yang-Baxter equation \cite{AbLi}.

To understand how the action of $Q_{(2)}$ is modified
by the presence of the coproduct \eqref{coprod}
let us first recall
that if $J_{(1)}$ denotes the generic global symmetry currents, the
first nonlocal charge is
\be
(Q_{(2)}){}^A{}_B&=&\int_{-\infty}^\infty d x \int_{-\infty}^x dy\,
J_{(1)}{}_0^A{}_C(x) J_{(1)}{}_0^C{}_B(y)+\int dx
\Sigma{}^A{}_B(x)\cr &\equiv& (Q_{(2)}^{\rm
bil}){}^A{}_B+(Q_{(2)}^{\rm loc}){}^A{}_B~~,  \vphantom{{}^{\big|}}
\label{bilocal_charges} \ee where $\Sigma$ is a functional of fields
which may be determined by the requirement that $Q_{(2)}$ is
conserved. In the absence of kappa-gauge fixing and in conformal
gauge $\Sigma$ has a simple expression in terms of the coset
vielbein; classically, in a gauge-fixed framework it is a
combination of the space-like components of Noether currents. In a
covariant quantization framework it was argued that $\Sigma$ exists
at the quantum level \cite{Berk2}.

The expressions of the global symmetry currents depend on the
details of both the kappa and diffeomorphism gauges and the
expression of $\Sigma$ inherits this dependence. As for the case of
massive relativistic field theories, it is natural to expect that the
expression of the bilocal part of $Q_{(2)}$ receives quantum
corrections only through the quantum corrections to the expressions
of the global symmetry currents while the corrections to $\Sigma$
however are sensitive to the OPE of the global symmetry currents.

An interesting question is whether it is possible to truncate
$Q_{(2)}$ such that it involves only a subalgebra of the full symmetry
algebra. Explicit calculation shows that
their conservation requires that the currents $J_{(1)}$ represent
the complete symmetry algebra; it does not seem possible to truncate
$J$ to a subalgebra of $\algPSU(2,2|4)$ while maintaining the
conservation of $Q_{(2)}$. This appears to complicate the conservation
equation \eqref{cons}, since some of the generators of $\algPSU(2,2|4)$
change the number of excitations of the state they act on.
Substantial simplification occurs however if we notice that such
effects are irrelevant if we pick two-excitation in- and out-states
with both excitations belonging to $\algPSU(2|2)^2\subset \algPSU(2,2|4)$.
Indeed, states with more than two excitations -- potentially created
by the action of the  components of the currents outside
$\algPSU(2|2)^2$ -- are orthogonal on our chosen out-state. Thus, for
the purpose of evaluating the two sides of the equation \eqref{cons}
it suffices to consider in \eqref{bilocal_charges} only the currents
$J_{(1)}$ generating $\algPSU(2|2)^2$.

The bilocal charge exhibits two kinds on nonlocality and
they must be properly taken into account. First, since the currents
appearing in $Q_{(2)}^{\rm bil}$ are the global symmetry currents,
their $x^-$ dependence introduces a contour $C_z$ similar to that on
the left-hand side of \figref{c_manip}. Secondly, we have the
inherent nonlocality of \eqref{bilocal_charges}  which in the
absence of the previous contours leads to equation \eqref{braiding_Q2}. The
contour associated to the left-most current $J_0$ in
\eqref{bilocal_charges}  ends on this last contour.

Similarly to $Q_{(1)}$, one first finds the action of $Q_{(2)}$ on a
single field. The result may then be used to express as a coproduct the
action  of $Q_{(2)}$ on a product of two fields. The contour
manipulations lead to the following structure:
\be \Delta({\widehat Q}_{(2)}{}^A{}_B)={\widehat Q}_{(2)}{}^A{}_B\otimes
\unit+\Psi_0{}^{AD}_{BC}\otimes {\widehat Q}_{(2)}{}^C{}_D
+\Psi_1{}^{AD}_{BC}\otimes {\widehat Q}_{(1)}{}^C{}_D \label{cop_bil}
\ee
 where the formal braiding matrices $\Psi_0{}^{AD}_{BC}$ and
$\Psi_1{}^{AD}_{BC}$ include further actions of the global charges
as well as of spatial derivatives.

At the classical level, finding the action on states is in principle
straightforward. This action however receives quantum corrections and
they are currently unknown. Nevertheless, following the example of
bosonic sigma-models \cite{Zamolodchikov:1978fd}, one may leave them
arbitrary and determine them consistently together with the S-matrix.
We will however not pursue here this direction and leave it for
future work.

The algebraic structure uncovered in the beginning of this section,
while of a rather different origin, is similar to that of the gauge
theory spin chain. The main difference related to the fact that, even
though in \cite{Arutyunov:2006ak} the contribution of the zero-mode of
$x^-$ to $\,{\rm e}\,^{\pm ix^-/2}$ was identified as a
length-changing operator, the factors $\,{\rm e}\,^{\pm ix^-/2}$ in
the supersymmetry generators act directly on the oscillators building
the state rather than by changing the (already infinite) length of the
string. In other words, they directly produce momentum-dependent phase
factors rather than insert length-changing markers denoted by ${\cal
Z}^\pm$ in \cite{Beisert:su22-S-matrix}. Thus, the phases found on the world
sheet are somewhat analogous to those appearing in the nonlocal or
cumulative notation of \cite{Beisert:2006qh}. Indeed, the action of
the supercharge on the $k$-th factor in a product of fields will be
multiplied by a phase depending on all momenta of the excitations to
the left of this excitation. In the other (twisted, in terminology of
\cite{Beisert:2006qh}) realization of braiding, the ${\cal Z}^\pm$
markers are crucial for the verification of the YBE as well as for the
derivation of the spin chain Bethe equations
\cite{Beisert:su22-S-matrix}. It is therefore interesting to see how
the Bethe equations arise on the worldsheet.

There is in fact a fairly straightforward procedure to reconstruct the
nested Bethe equations given the information already available.  To
this end it is useful to recall that the usual procedure of
constructing the Bethe equations starts with an arbitrary state and
imposes that the state is mapped into itself by the scattering of one
excitation past all the other ones. Enforcing this condition requires
the diagonalization of a product of scattering matrices which is, in
fact, the multi-particle S-matrix.
For this purpose one chooses an arbitrary type of excitation and
treats the states containing only this type of excitation as a new
vacuum state; the other excitations are interpreted as excitations
above this level-2 vacuum. One then imposes the periodicity condition
on these new states. These steps are further repeated until all types
of excitations are accounted for.
In other words, at each step in the construction of the nested Bethe
equations one finds the multi-particle S-matrix with respect to a new
vacuum and sets its eigenvalues to unity.

In the presence of the coproduct \eqref{coprod}, the knowledge of
(almost) factorization of a scattering matrix (such that the spin
chain S-matrix) following from the YBE allows in principle the
construction of all the required multi-particle scattering
matrices. The main departure from the usual relation between
two-particle and multi-particle S-matrices is the need for additional
phases depending on all the excitations building the state. Their
appearance can be justified by the fact that, while the 2-particle
S-matrix depends only of the two excitations being scattered, the
coproduct introduces a phase depending on all the excitation to their
left. Thus, these additional phases must be explicitely included in
the relation between the multi-particle and two-particle S-matrices.

\section{Scattering of bosons}
\label{sec:scalars}

We start with the bosonic part of the sigma-model action:
\[
 S_\sigma =\frac{\sqrt{\lambda }}{4\pi }
 \int_{}^{}d\tau \int_{-\pi }^{\pi }d\sigma \,
 \sqrt{-h}\,h^{\mathbf{a}\mathbf{b}}G_{MN}\partial _\mathbf{a}
 \mathbb{X}^M\partial
 _\mathbf{b}\mathbb{X}^N \; ,
\]
where $\mathbb{X}^M=(T,\Phi,Y^m,Z^\mu)$. The metric is taken from
the equation~\eqref{eqn:adsxs}. The worldsheet metric
$h^{\mathbf{ab}}$ has the signature $(+-)$ and the Levi-Civita
symbol $\varepsilon^{\mathbf{ab}}$, used later, is defined such that
$\varepsilon^{01}=\varepsilon_{10}=1$.

\subsection{The a-gauge}

We consider the gauge in which the angular momentum is uniformly
distributed along the string, which is best suited for studying the
near-BMN limit \cite{Callan:2003xr}. For various purposes it is
interesting to look at a one-parameter family of interpolating
gauges introduced in \cite{Arutyunov:2006gs}, which includes the
uniform gauge from \cite{Callan:2003xr} and its light-cone
modification \cite{Frolov:2006cc} as particular cases. The uniform
momentum density in that gauge is associated with
\[\label{jplus}
 J_+=(1-a)J+aE
\]
The pure uniform gauge corresponds to $a=0$, whereas $a=1/2$ gives
the light-cone gauge.

To find the gauge-fixed Lagrangian, we follow the procedure outlined
in \cite{Kruczenski:2004cn}: T-dualize in the direction canonically
conjugate to \eqref{jplus}, integrate out the worldsheet metric, and
fix the gauge in the resulting Nambu-Goto action. The T-duality
transformation is facilitated by integrating in a field whose on-shell
value is
\[\label{piclass}
 \Pi_\mathbf{a}^{{\rm (cl)}}=\frac{\sqrt{\lambda }}{2\pi }\
 \sqrt{-h}\,\varepsilon _{\mathbf{a}\mathbf{b}}
 h^{\mathbf{b}\mathbf{c}}
 \left[(1-a)G_{\varphi \varphi }\,\partial _\mathbf{c}\Phi
 +aG_{tt}\,\partial _\mathbf{c}T\right] \; ,
\]
so that
\begin{equation}\label{j+++}
J_+=\int_{-\pi }^{+\pi }d\sigma \,\Pi_1^{{\rm (cl)}} \; .
\end{equation}
Adding
\[
 S_\Pi=\frac{\pi }{\sqrt{\lambda }}\int_{}^{}d^2\sigma \,
 \frac{\sqrt{-h}\,h^{\mathbf{a}\mathbf{b}}
 \left(\Pi_\mathbf{a}-\Pi_\mathbf{a}^{{\rm (cl)}}\right)
 \left(\Pi_\mathbf{b}-\Pi_\mathbf{b}^{{\rm (cl)}}\right)}
 {(1-a)^2G_{\varphi \varphi }-a^2G_{tt}}
\]
to the sigma-model action changes nothing since the additional term
is quadratic in $\Pi_\mathbf{a}$. On the other hand, addition of
this term in conjunction with the linear field redefinition
\[
 T=X^+ - \frac{a}{1-a} \, \Phi
\]
eliminates the quadratic dependence on $\Phi$, leaving only the
linear term:
\[
 S_\Phi =-\frac{1}{1-a}\int_{}^{}d^2\sigma \,
 \varepsilon ^{\mathbf{a}\mathbf{b}}\Pi_\mathbf{a}
 \partial _\mathbf{b}\Phi \; .
\]
Integrating out $\Phi $ imposes the constraint $\partial
_\mathbf{a}\Pi_\mathbf{b}-\partial_\mathbf{b}\Pi_\mathbf{a}=0$,
which is solved by
\begin{equation}\label{fifi}
 \Pi_\mathbf{a}=\partial _\mathbf{a}\tilde{\Phi } \; .
\end{equation}
Substituting this back into the action gives the sigma-model with
the T-dual metric \cite{Buscher:1987sk}:
\begin{align}
 G_{++} & = \frac{(1-a)^2G_{\varphi \varphi }G_{tt}}
 {(1-a)^2G_{\varphi \varphi }-a^2G_{tt}} \; ,
 \nn \\
 G_{\tilde{\varphi }\tilde{\varphi }} & =
 \frac{4\pi ^2}{\lambda }\,\,
 \frac{1}{(1-a)^2G_{\varphi \varphi }-a^2G_{tt}} \; ,
\end{align}
and the B-field
\[
 B_{\tilde{\varphi}+}=\frac{2\pi }{\sqrt{\lambda }}\,\,
 \frac{aG_{tt}}{(1-a)^2G_{\varphi \varphi }-a^2G_{tt}} \; .
\]
According to \eqref{fifi} and \eqref{j+++} the dual field satisfies
the boundary condition
\[\label{bctilde}
 \tilde{\Phi}(\tau,\sigma+2\pi)=\tilde{\Phi}(\tau,\sigma) + J_+ \; .
\]

We can now start with the Nambu-Goto action in the T-dual
coordinates:
\[
 S_{NG}=\frac{\sqrt{\lambda }}{2\pi }\int_{}^{}d^2\sigma \,L_{NG},
\]
\[
 L_{NG}=-\sqrt{-\det_{\mathbf{a}\mathbf{b}}G_{MN}\,\partial _\mathbf{a}
 \tilde{\mathbb{X}}^M\partial _\mathbf{b}\tilde{\mathbb{X}}^N}-\frac{1}{2}\,
 \varepsilon ^{\mathbf{a}\mathbf{b}}B_{MN}\partial _\mathbf{a}
 \tilde{\mathbb{X}}^M\partial _\mathbf{b}\tilde{\mathbb{X}}^N,
\]
where $\tilde{\mathbb{X}}^M=(X^+,\tilde{\Phi},Y^m,Z^\mu)$. The
natural gauge condition, consistent with \eqref{bctilde}, is
\[
 X^+ = \frac{\tau}{1-a} \comma \tilde{\Phi} = \frac{J_+\sigma}{2\pi} \; .
\]
After imposing this gauge condition it is convenient to rescale
$\sigma $ by $J_+/\sqrt{\lambda }$, so that the worldsheet
coordinate changes in the interval $-\pi J_+/\sqrt{\lambda }<\sigma
\leq\pi J_+/\sqrt{\lambda }$. Then $J_+/\sqrt{\lambda }$ appears
only in the length of the string and $\sqrt{\lambda }/2\pi $ enters
only as an overall factor in front of the action. We shall further
consider the limit $J_+/\sqrt{\lambda }\to \infty $, in which the
worldsheet becomes an infinite plane and the dependence on $J_+$
completely disappears. $2\pi /\sqrt{\lambda }$ remains, as a loop
counting parameter.

After all the rescalings, the gauge-fixed Lagrangian does not depend
on any parameters at all:
\[ \label{fulla}
\begin{split}
 L_{{\rm g.f.}} = \ & -\frac{\sqrt{G_{\varphi \varphi }G_{tt}}}
 {(1-a)^2G_{\varphi \varphi }-a^2G_{tt}}\,
 \left\{1-\frac{{(1-a)^2G_{\varphi \varphi }-a^2G_{tt}}}{2}
 \vphantom{\frac{\left[{(1-a)^2G_{\varphi \varphi }-a^2G_{tt}}\right]^2}
 {2G_{\varphi \varphi }G_{tt}}}
 \right. \\
 & \times\left.
 \left[\left(1+\frac{1}{G_{\varphi \varphi }G_{tt}}\right)
 \partial _\mathbf{a}X\cdot\partial ^\mathbf{a}X
 -\left(1-\frac{1}{G_{\varphi \varphi }G_{tt}}\right)
 \left(\tim{X}\cdot\tim{X}+\spa{X}\cdot\spa{X}\right)\right]
 \right. \\
 & + \left.
 \frac{\left[{(1-a)^2G_{\varphi \varphi }-a^2G_{tt}}\right]^2}
 {2G_{\varphi \varphi }G_{tt}}\left[
 \left(\partial _\mathbf{a}X\cdot\partial ^\mathbf{a}X\right)^2
 -\left(\partial _\mathbf{a}X\cdot
 \partial _\mathbf{b}X\right)^2
 \right]\right\}^{1/2} \\
 & + \frac{a}{1-a}\,\,\frac{G_{tt}}{{(1-a)^2G_{\varphi \varphi
 }-a^2G_{tt}}} \; .
\end{split}
\]
Here, the index contractions on
$X=(Y^m,Z^\mu )$
are done with the metric \eqref{eqn:adsxs}. Finally, to the quartic
order in the fields we get:
\[\label{lagmain}
\begin{split}
 L = \ &
  \frac{1}{2}\,\left(\partial _\mathbf{a}X\right)^2
 -\frac{1}{2}\,X^2+\frac{1}{4}\,Z^2\left(\partial _\mathbf{a}Z\right)^2
 -\frac{1}{4}\,Y^2\left(\partial _\mathbf{a}Y\right)^2
 +\frac{1}{4}\,\left(Y^2-Z^2\right)
 \left(\tim{X}^2+\spa{X}^2\right) \\
 & -\frac{1-2a}{8}\,\left(X^2\right)^2
+\frac{1-2a}{4}\,\left(\partial _\mathbf{a}X\cdot\partial
 _\mathbf{b}X\right)^2
 -\frac{1-2a}{8}\,\left[\left(\partial _\mathbf{a}X\right)^2\right]^2 \; .
\end{split}
\]
Here, unlike in \eqref{fulla}, target-space indices are contracted
with the flat Euclidian metric. The quadratic part of the Lagrangian
is 2d Lorentz invariant and $\grSO(8)$ symmetric. These symmetries
are broken by the interaction terms, many of which however preserve
$\grSO(8)$ and/or Lorentz invariance. In particular the
gauge-dependent part of the Lagrangian is Lorentz and $\grSO(8)$
invariant. This part disappears at $a=1/2$, which reflects
relative simplification
of the string action in the light-cone gauge
\cite{Arutyunov:2006gs}. The full action in any $a$-gauge is only
invariant under $\algSO(4)^2=\algSU(2)^4$.

\subsection{S-matrix}

Computing the tree-level S-matrix for the Lagrangian \eqref{lagmain}
is a fairly straightforward exercise.
The calculation can be done by applying LSZ reduction to the quartic
vertices in \eqref{lagmain}, which produces various tensor
structures with the $\grSO(4)^2$ indices. At the end we want to
transform the $\grSO(4)^2$ vector indices into the $\grSU(2)^4$
spinor notations according to \eqref{indexiana}, which in effect
trades combinations of $\delta _\mu ^\nu $ and $\delta _m^n$ for
combinations of $\delta _\laa^\lbb $, $\delta _\raa^\rbb$,
$\delta_\lAA^\lBB$ and $\delta_\rAA^\rBB$. The basic
$\grSU(2)$-invariants are the the identity and the permutation
operators:
\begin{equation}\label{}
\unit_{ab}^{cd}=\delta _b^d\delta _a^c
\comma
P_{ab}^{cd}=\delta _b^c\delta _a^d \; ,
\end{equation}
and analogous operators acting on the dotted indices. The T-matrix
acts in the tensor product and we will use the notations like
$\unit\otimes P$, $P\otimes \unit$ or $P\otimes P$ to denote
permutations that act on dotted, undotted or both types of indices.
Written in the $\grSO(4)^2$ notations, these operators parameterize all
possible combinations of the $\grSO(4)$ indices that arise in the
scattering amplitudes:
\begin{align}
 (\unit\otimes P+P\otimes \unit)^{mn}_{kl} & = \delta^m_k\delta^n_l + \delta^m_l\delta^n_k - \delta^{mn}\delta_{kl} \; , \nn \\
 (P\otimes P)_{kl}^{mn} & = \delta_l^m\delta_k^n \; , \\
 (\unit\otimes\unit)^{mn}_{kl} & = \delta^m_k\delta ^n_l \; . \nn
\end{align}
With the use of these formulas we find:
\begin{align} \label{SMart}
 \Tmatrix_{YY\rightarrow YY} & = \Half\lrsbrk{(1-2a) (\cpp) + \frac{(p-p')^2}{\cpp}} \unit\otimes\unit
 + \frac{pp'}{\cpp} \left(\unit\otimes P+P\otimes \unit\right) \; , \nn \\
 \Tmatrix_{ZZ\rightarrow ZZ} & = \Half\lrsbrk{(1-2a) (\cpp) - \frac{(p-p')^2}{\cpp}} \unit\otimes\unit
 - \frac{pp'}{\cpp} \left(\unit\otimes P+P\otimes \unit\right) \; , \nn \\
 \Tmatrix_{ZY\rightarrow ZY} & = \Half\lrsbrk{(1-2a) (\cpp) + \frac{p^2-p'^2}{\cpp}} \; , \nn \\
 \Tmatrix_{ZY\rightarrow YZ} & = 0 \; , \nn \\
 \Tmatrix_{ZZ\rightarrow YY} & = 0 \; .
\end{align}
The (bosonic) T-matrix appears to have a factorized form
\eqref{factorize_at}. We should emphasize that this is a result of
very delicate cancelations among different diagrams. From
\eqref{SMart} we can extract coefficients $\Atmatrix$, $\Btmatrix$,
$\Dtmatrix$, $\Etmatrix$, $\Gtmatrix$ and $\Ltmatrix$ in
\eqref{eqn:tmatrix-coeff}, see \eqref{ourfinalequation}.
$\Ctmatrix$, $\Ftmatrix$, $\Htmatrix$, and $\Ktmatrix$ only appear
in the scattering of
fermions.

\subsection{Absence of particle production}

In this section we offer some arguments for the factorization of the
bosonic S-matrix beyond leading order, in particular the absence of
$2\rightarrow 4$ particle production at tree level and the
corresponding factorization of the $3\rightarrow3$ tree level
amplitude. It is well known that the factorization of the S-matrix
follows from the selection rules that the number of particles of a
given mass is unchanged and that the final momenta are the same as
the initial ones, see for example \cite{Zamolodchikov:1978xm}. It is
straightforward to keep higher terms in the expansion of the
light-cone Lagrangian provided we restrict our attention to the
bosonic part. Using the uniform light-cone gauge $a=\tfrac{1}{2}$
for convenience we find the Lagrangian density describing only
fields on the $\Sphere^5$ \be
\Lagr_{\rm lc}&=&P_y {\dot Y}-{\cal H}_{\rm lc} \\
&=& -\frac{1}{2}\left(-{\dot Y}^2+{\acute
Y}^2+Y^2\right)+\frac{1}{2\sqrt{\lambda}}Y^2 {\acute Y}^2\nn\\
& & +\frac{1}{32 \lambda }\left(-Y^2 {\acute Y}^4 + Y^4 {\dot Y}^2
- Y^2 {\dot Y}^4 -{\acute Y}^2(9 Y^4+2 Y^2 {\dot Y}^2)+4 Y^2 ({\dot
Y} \cdot {\acute Y})^2 \right)  
+ \ldots \nn
\ee
and the
analogous Lagrangian density for fields on the $\AdS_5$ is
\be
\Lagr_{\rm lc} &=& -\frac{1}{2}\left(-{\dot Z}^2+{\acute
Z}^2+Z^2\right)-\frac{1}{2\sqrt{\lambda}}Z^2 {\acute Z}^2 \\
& & +\frac{1}{32 \lambda }\left(-Z^2 {\acute Z}^4 + Z^4 {\dot Z}^2
- Z^2 {\dot Z}^4 -{\acute Z}^2(9 Z^4 + 2 Z^2 {\dot Z}^2)+4 Z^2
({\dot Z} \cdot {\acute Z})^2 \right) 
+ \ldots \nn
\ee
The dots refer to terms of higher order in $1/\sqrt{\lambda}$. The mixed terms
can also be easily found but we do not record them here. We
further restrict our attention to two directions on the sphere,
$Y^5,\ Y^6$, and consider the scattering of the complex
coordinate $Y=\tfrac{1}{2}(Y^5+iY^6)$. The vertices for the above
interactions are given by 
%
\be \pic{19}{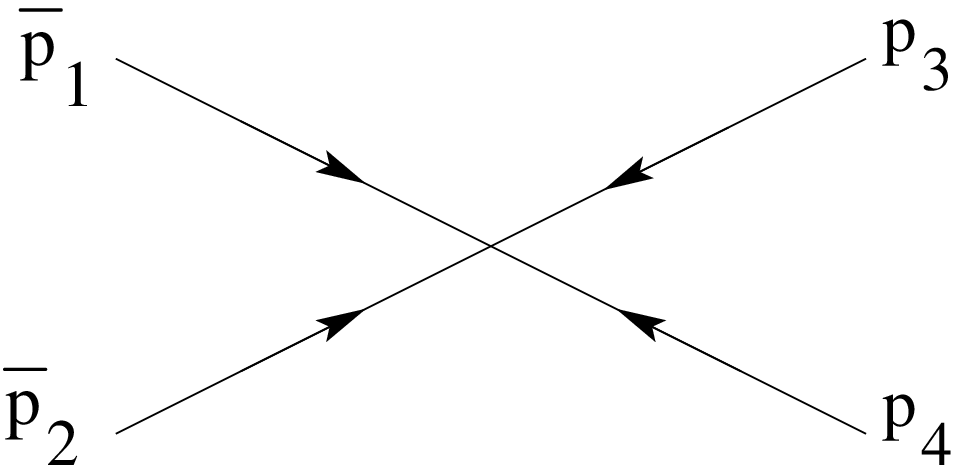}{0.35}\; &=&\frac{-i }{ \sqrt{\lambda}}
(p_1+p_2)(p_3+p_4)\\
\pic{22}{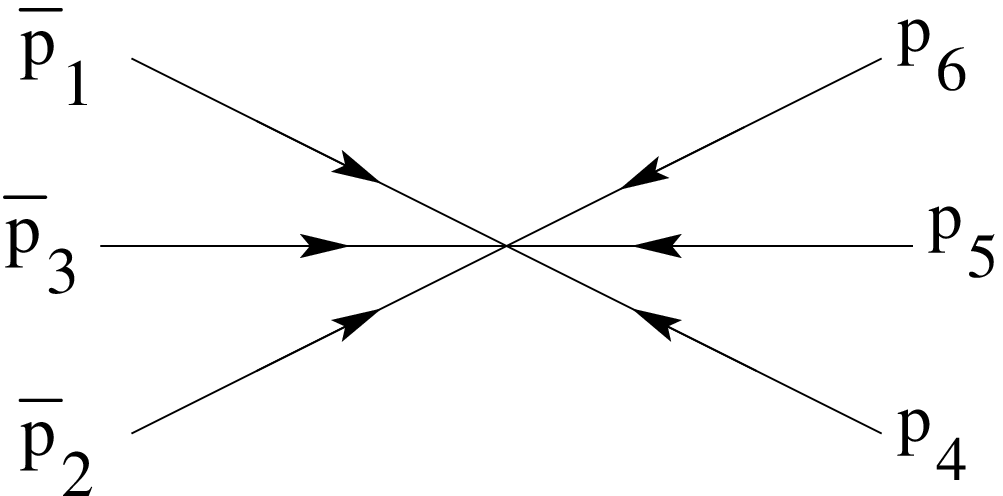}{0.35}&=&
\frac{i}{32 \lambda} \Big[ 8(\energy_1
\energy_2+\energy_1\energy_3+\energy_2\energy_3-(p_1p_2+p_1p_3+p_2
p_3))\times\nn
\ee
\vspace{-13truemm}
\be
& &  \kern28pt
(\energy_4 \energy_5+\energy_4\energy_6+\energy_5\energy_6-(p_4
p_5+p_4 p_6+p_5 p_6)) \\
~~~~~~~~~~~~~~~~~~~~~~~~~~~~~~~~~
& & \kern-55pt +8 (\energy_1
+\energy_2+\energy_3)(\energy_4+
\energy_5+\energy_6)-64(p_1+p_2+p_3)(p_4+p_5+p_6)\Big] \nn
\ee
%
%
where the two-momenta are the pairs $(\energy_i,p_i)$. The contributions to
the $2\rightarrow4$ scattering involving two four-vertices are
given by
\begin{figure}[htbp]
  \begin{center}
    \mbox{
{\includegraphics[width=4.0in,height=2.1in,angle=0]{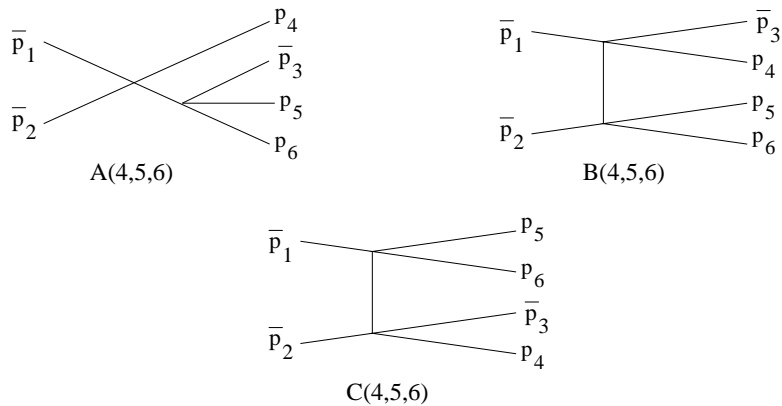} }}
    \caption{Three of the diagrams contributing to $2\rightarrow4$ scattering.}
    \label{diag}
  \end{center}
\end{figure}
\be {\mathbf A(4,5,6):}\kern10pt \frac{1}{ \sqrt{\lambda}
}\left(\frac{(p_1+p_2)^2(p_5+p_6)^2}{(\energy_1+\energy_2-\energy_4)^2-(p_1+p_2-p_4)^2-1}\right)+(4\leftrightarrow
5)+(4 \leftrightarrow 6) \ee \be {\mathbf B(4,5,6):}\kern10pt
\frac{1}{\sqrt{\lambda}
}\left(\frac{(p_1-p_3)^2(p_5+p_6)^2}{(\energy_1-\energy_3-\energy_4)^2-(p_1-p_3-p_4)^2-1}\right)+(4\leftrightarrow
5)+(4 \leftrightarrow 6) \ee \be {\mathbf C(4,5,6):}\kern10pt
\frac{1}{\sqrt{\lambda}
}\left(\frac{(p_2-p_3)^2(p_5+p_6)^2}{(\energy_1-\energy_5-\energy_6)^2-(p_1-p_5-p_6)^2-1}\right)+(4\leftrightarrow
5)+(4 \leftrightarrow 6) \ee and from the six-vertex we get the
contribution \be {\mathbf D:}& &
\frac{i }{32 \lambda} \left( 8(\energy_1 \energy_2-\energy_1\energy_3-\energy_2\energy_3-(p_1p_2-p_1p_3-p_2 p_3))\right.\times\nn\\
&&\left.\kern15pt (\energy_4 \energy_5+\energy_4\energy_6+\energy_5\energy_6-(p_4 p_5+p_4 p_6+p_5 p_6))\right.\nn\\
& &\left.-8 (\energy_1 +\energy_2-\energy_3)(\energy_4+
\energy_5+\energy_6)+64 (p_1+p_2-p_3)(p_4+p_5+p_6)\right)\ . \ee
We can now (analytically and numerically) check that
$A(4,5,6)+B(4,5,6)+C(4,5,6)+(4\leftrightarrow 5)+(4\leftrightarrow
6)+D=0$ for generic values of the external momenta. We can see this explicitly in some simple cases; for example set
$p_1=-p_2, p_5=-p_6$. In this case $p_3=-p_4$ and
$\energy_1=\energy_3+\energy_5$ (on-shell
$\energy_i=\sqrt{1+p_i^2}$) and we can see that all diagrams of
type A vanish as do $B(4,5,6)$ and $C(4,5,6)$. The remaining
contribution from $B(5,6,4)$, $B(6,5,4)$, $C(5,6,4)$ and  $C(6,5,4)$
can be simplified using
\be
p_5=\left(\left(\sqrt{1+p_1^2}-\sqrt{1+p_3^2}\right)^2-1\right)^{\frac{1}{2}}
\ee
to
\be -\frac{i}{\sqrt{\lambda}}\left(p_1^4+(p_1^2+3 p_3^2)+
p_1^2(p_3^3-2\energy_1\energy_3)\right) \; . \ee
Using energy and
momentum conservation it is straightforward to show that the
six-vertex gives the negative of this result. We can also examine
the case when all the momenta are much larger than the mass and
again $p_1=-p_2$ in this case $\energy_i\simeq |p_i|$ and by
examining specific cases it is straightforward to see that the
$2\rightarrow4$ amplitude vanishes. Thus we have shown complete
cancellation between the diagrams for
$2\rightarrow4$ scattering.

In fact we have shown more than the
absence of particle production, if we consider $3\rightarrow3$
scattering we find the exact same cancellations as above except for
the special kinematical region where the outgoing momenta are equal
to the incoming. In this case the internal propagator in the two
vertex diagrams become singular giving an amplitude which splits
into a finite part canceled by the six-vertex term and  momentum
$\delta$-function. Hence we also see the factorization of the
$3\rightarrow3$ tree level amplitude into $2\rightarrow2$ events which
in this case is
equivalent to the absence of particle production.

We should mention
that the authors of
\cite{Frolov:2006cc} were able to construct a unitary transformation
which, quite generically, removed
particle producing terms from the light-cone Hamiltonian. The
existence of such a transformation does
not  require any particular symmetries of the Hamiltonian but only
relies upon the existence of the quantum theory and the non-zero mass
of the free particles. Indeed the remaining terms in the Hamiltonian
can have quite generic coefficients and so in this case the absence of
particle production does not seem to necessarily imply the
factorization of multi-particle scattering processes. This is an
important distinction as it is this factorization which is equivalent
to the existence of higher conserved charges and so integrability.

\section{Scattering of fermions}
\label{sec:fermions}

\subsection{Physical degrees of freedom}

We now turn to the scattering of fermions. For the sake of
simplicity we shall only consider the uniform light-cone gauge that
corresponds to $a=1/2$ in \eqref{jplus}. The results for the
constant-$J$ gauge ($a=0$) are displayed in appendix B. The degrees
of freedom that are left after fixing the uniform light-cone gauge
are given by the fields $Y_{\lAA\rAA}$, $Z_{\laa\raa}$,
$\Psi_{\lAA\raa}$ and $\Bsi_{\laa\rAA}$. See \secref{sec:results}
and \tabref{fig:fields-indices} for more details.

We use northeast-southwest conventions to raise and lower
$\algSU(2)$ indices
\[
  x^a = \levi^{ab} x_b
  \comma
  x_a = x^b \levi_{ba}
  \; ,
\]
where $\levi^{12} = \levi_{12} = 1$, and likewise for all other
indices. Also complex conjugation changes the position of the index,
e.g. $(Y_{\lAA\rAA})^* \equiv Y^{*\lAA\rAA}$. It is important not to
confuse this with $Y_{\lAA\rAA}^* \equiv Y^{*\lBB\rBB} \levi_{\lBB\lAA}
\levi_{\rBB\rAA}$. Moreover, the bosonic fields satisfy the reality
condition
\[ \label{eqn:reality-condition}
Y_{\lAA\rAA}^* = Y_{\lAA\rAA} \qquad\mbox{and}\qquad Z_{\laa\raa}^*
= Z_{\laa\raa} \; .
\]

\subsection{Action and quantization}

For the superstring calculation in uniform light-cone gauge, we use
the action derived in \cite{Frolov:2006cc}. In
\appref{app:rewriting-light-cone-action} we rewrite this action in a
second order formalism and obtain
\[ \label{eqn:action-quadratic-matrix}
 \Action = \sqrt{\lambda}
           \int\limits_{-\infty}^{\infty} \!d\tau
           \int\limits_{-\pi J_+/\sqrt{\lambda}}^{\pi J_+/\sqrt{\lambda}} \frac{d\sigma}{2\pi}\: \Lagr
\]
with
\[ \label{eqn:lagr-quadratic-matrix}
\begin{split}
\Lagr_0 = \ & \str \biggsbrk{
   \frac{1}{4} \tim{X} \tim{X}
 - \frac{1}{4} \spa{X} \spa{X}
 - \frac{1}{4}   {X}     {X}
 - \frac{i}{2} \Sigma_+ \chi \tim{\chi}
 - \frac{1}{2} \Sigma_+ \chi \spa{\chi}^\conj
 - \frac{1}{2} \chi \chi
 } \; ,
\\[3mm]
\Lagr_\mathrm{int} = \ &
 - \frac{1}{8} \str \Sigma_8 X X \str \spa{X} \spa{X}
\\ &
 + \frac{1}{8}  \str \chi \spa{\chi} \chi \spa{\chi}
 + \frac{1}{8}  \str \chi \chi \spa{\chi} \spa{\chi}
 + \frac{1}{16} \str \comm{\chi}{\spa{\chi}} \comm{\chi^\conj}{\spa{\chi}^\conj}
 + \frac{1}{4}  \str \chi \spa{\chi}^\conj \chi \spa{\chi}^\conj
\\ &
 - \frac{1}{8} \str \Sigma_8 X X \str \spa{\chi}\spa{\chi}
 + \frac{1}{4} \str \comm{X}{\spa{X}} \comm{\chi}{\spa{\chi}}
 + \str X \spa{\chi} X \spa{\chi}
\\ &
 + \frac{i}{8} \str \comm{X}{\tim{X}} \comm{\chi^\conj}{\spa{\chi}}
 - \frac{i}{8} \str \comm{X}{\tim{X}} \comm{\chi}{\spa{\chi}^\conj}
 \; .
\end{split}
\]
The conjugation $\chi^\conj$ as well as the constant matrices
$\Sigma_+$ and $\Sigma_8$ are defined in the
\appref{app:rewriting-light-cone-action}. This action contains only
the physical fields introduced above. They are written as elements
of two $\grSU(2,2|4)$ supermatrices. The bosons are contained in
\[
X = \matr{cccc|cccc}{
0         & 0         & +  Z^{\lz\rw} & +i Z^{\lz\rz} & 0 & 0 & 0 & 0 \\
0         & 0         & +i Z^{\lw\rw} & -  Z^{\lw\rz} & 0 & 0 & 0 & 0 \\
-  Z^{\lw\rz} & -i Z^{\lz\rz} & 0         & 0         & 0 & 0 & 0 & 0 \\
-i Z^{\lw\rw} & +  Z^{\lz\rw} & 0         & 0         & 0 & 0 & 0 &
0 \\ \hline
0 & 0 & 0 & 0 & 0         & 0         & +i Y^{\lx\ry} & -  Y^{\lx\rx} \rule{0mm}{5mm} \\
0 & 0 & 0 & 0 & 0         & 0         & -  Y^{\ly\ry} & -i Y^{\ly\rx} \\
0 & 0 & 0 & 0 & -i Y^{\ly\rx} & +  Y^{\lx\rx} & 0         & 0         \\
0 & 0 & 0 & 0 & +  Y^{\ly\ry} & +i Y^{\lx\ry} & 0         & 0 } \; ,
\]
and the fermions in
\[
\chi = e^{\frac{i\pi}{4}} \matr{cccc|cccc}{
0 & 0 & 0 & 0 & 0 & 0 & +  \Bsi^{\lz\ry} & +i \Bsi^{\lz\rx} \\
0 & 0 & 0 & 0 & 0 & 0 & +i \Bsi^{\lw\ry} & -  \Bsi^{\lw\rx} \\
0 & 0 & 0 & 0 & +i\Psi^{*\ly\rz} & - \Psi^{*\lx\rz} & 0 & 0 \\
0 & 0 & 0 & 0 & - \Psi^{*\ly\rw} & -i\Psi^{*\lx\rw} & 0 & 0 \\
\hline
0 & 0 & +  \Psi^{\lx\rw} & +i \Psi^{\lx\rz} & 0 & 0 & 0 & 0 \rule{0mm}{5mm} \\
0 & 0 & +i \Psi^{\ly\rw} & -  \Psi^{\ly\rz} & 0 & 0 & 0 & 0 \\
-i\Bsi^{*\lw\rx} & + \Bsi^{*\lz\rx} & 0 & 0 & 0 & 0 & 0 & 0 \\
+ \Bsi^{*\lw\ry} & +i\Bsi^{*\lz\ry} & 0 & 0 & 0 & 0 & 0 & 0 } \; .
\]
Plugging in these expression, the free part of the Lagrangian
becomes
\[\label{free-Lag-uniform}
\begin{split}
\Lagr_0 = \ & + \tfrac{1}{2} \tim{Y}^*_{\lAA\rAA} \tim{Y}^{\lAA\rAA}
- \tfrac{1}{2} \spa{Y}^*_{\lAA\rAA} \spa{Y}^{\lAA\rAA} -
\tfrac{1}{2}     {Y}^*_{\lAA\rAA}     {Y}^{\lAA\rAA}
\\ &
+ \tfrac{1}{2} \tim{Z}^*_{\laa\raa} \tim{Z}^{\laa\raa} -
\tfrac{1}{2} \spa{Z}^*_{\laa\raa} \spa{Z}^{\laa\raa} - \tfrac{1}{2}
{Z}^*_{\laa\raa}     {Z}^{\laa\raa}
\\ &
+ i \Psi^*_{\lAA\raa} \tim{\Psi}^{\lAA\raa} - \tfrac{i}{2} \brk{
\Psi^*_{\lAA\raa} \spa{\Psi}^{*\lAA\raa}
                   + \Psi  _{\lAA\raa} \spa{\Psi}^{ \lAA\raa} }
- \Psi^*_{\lAA\raa} \Psi^{\lAA\raa}
\\ &
+ i \Bsi^*_{\laa\rAA} \tim{\Bsi}^{\laa\rAA} - \tfrac{i}{2} \brk{
\Bsi^*_{\laa\rAA} \spa{\Bsi}^{*\laa\rAA}
                   + \Bsi  _{\laa\rAA} \spa{\Bsi}^{ \laa\rAA} }
- \Bsi^*_{\laa\rAA} \Bsi^{\laa\rAA} \; .
\end{split}
\]
The corresponding equations of motion are solved by the following
mode expansion:
\begin{align}
Y_{\lAA\rAA}(\vx)      & = \int\frac{dp}{2\pi}
\frac{1}{\sqrt{2\energy}} \:
                           \Bigbrk{ a_{\lAA\rAA}(p)      \, e^{-i\vp\cdot\vx}
                                  + a^\dag_{\lAA\rAA}(p) \, e^{+i\vp\cdot\vx} } \; , \\
Z_{\laa\raa}(\vx)      & = \int\frac{dp}{2\pi}
\frac{1}{\sqrt{2\energy}} \:
                           \Bigbrk{ a_{\laa\raa}(p)      \, e^{-i\vp\cdot\vx}
                                  + a^\dag_{\laa\raa}(p) \, e^{+i\vp\cdot\vx} } \; , \\
\Psi_{\lAA\raa}(\vx)   & = \int\frac{dp}{2\pi}
\frac{1}{\sqrt{\energy}} \:
                           \Bigbrk{ b_{\lAA\raa}(p)      \, u(p) \, e^{-i\vp\cdot\vx}
                                  + b^\dag_{\lAA\raa}(p) \, v(p) \, e^{+i\vp\cdot\vx} } \; , \\
\Bsi_{\laa\rAA}(\vx)   & = \int\frac{dp}{2\pi}
\frac{1}{\sqrt{\energy}} \:
                           \Bigbrk{ b_{\laa\rAA}(p)      \, u(p) \, e^{-i\vp\cdot\vx}
                                  + b^\dag_{\laa\rAA}(p) \, v(p) \, e^{+i\vp\cdot\vx} } \; ,
\end{align}
where the energy is $\energy = \sqrt{1 + p^2}$, the wave functions
are
\[
  u(p) = \cosh{\tfrac{\theta}{2}}
  \comma
  v(p) = \sinh{\tfrac{\theta}{2}}
\]
and the rapidity $\theta$ is defined through $p = \sinh\theta$. The
scalar product in the exponentials is $\vp\cdot\vx = \energy \tau +
p \sigma$. The canonical commutation relations are given by
\begin{align}
 & \comm{ a^{\lAA\rAA}(p)}{a^\dag_{\lBB\rBB}(p')} = 2\pi \, \delta^{\lAA}_{\lBB} \delta^{\rAA}_{\rBB} \, \delta(p-p') \; , &
 & \acomm{b^{\lAA\raa}(p)}{b^\dag_{\lBB\rbb}(p')} = 2\pi \, \delta^{\lAA}_{\lBB} \delta^{\raa}_{\rbb} \, \delta(p-p') \; , \nn \\
 & \comm{ a^{\laa\raa}(p)}{a^\dag_{\lbb\rbb}(p')} = 2\pi \, \delta^{\laa}_{\lbb} \delta^{\raa}_{\rbb} \, \delta(p-p') \; , &
 & \acomm{b^{\laa\rAA}(p)}{b^\dag_{\lbb\rBB}(p')} = 2\pi \, \delta^{\laa}_{\lbb} \delta^{\rAA}_{\rBB} \, \delta(p-p') \; .
\end{align}

The above choice of labeling the modes has some very nice features.
Firstly, bosons and fermions are treated identically. All indices
are carried by the mode operators. The wave functions are scalar
functions and no Dirac matrices are required. Secondly, particles
and anti-particles can be considered at once without notational
differences. The particle/anti-particle relationship is determined
by which pair of oscillators occurs in the expansion of one field.
Looking, for example, at the field $Y_{\lAA\rAA}$ (for fixed $\lAA$ and
$\rAA$), we see that the oscillator $a^\dag_{\lAA\rAA}$ creates the
``anti-excitation'' of the ``excitation'' that is destroyed by the
oscillator $a_{\lAA\rAA}$. Note, however, that these two oscillators
do not form a canonical pair. Rather $a^\dag_{\lAA\rAA}$ and
$a^{\lAA\rAA} = \levi^{\lAA\lBB} \levi^{\rAA\rBB} a_{\lBB\rBB}$ are
conjugate to each other as it can be seen from the commutation
relations. This is after all a consequence of $(a^{\lAA\rAA})^* =
a^\dag_{\lAA\rAA}$. It is interesting to observe the different
origin of the latter relation for bosons and fermions. For the
bosons is originates from the reality condition
\eqref{eqn:reality-condition}. The fermions $\Psi_{\lAA\raa}$ and
$\Psi^*_{\lAA\raa}$, however, are independent. In this case it is
the equations of motion which require $(b^{\lAA\raa})^* =
b^\dag_{\lAA\raa}$.

\subsection{Tree-level S-matrix}
\label{sec:full-T}

We compute the (65536) components of the T-matrix as defined in \eqref{smatrix_components} and \eqref{asmatrix_expansion}, relying only on the manifest $\grSU(2)^4$ symmetry. There are four kinds of particles that we can scatter: $Y_{\lAA\rAA}$, $Z_{\laa\raa}$, $\Psi_{\lAA\raa}$,
$\Bsi_{\laa\rAA}$. Let us consider the scattering two $Y$'s. There are four different channels, which can be found by taking the tensor product of the corresponding representations, cf. \tabref{fig:fields-indices}:
\[
(\rep{2,2,1,1}) \otimes (\rep{2,2,1,1}) =  (\rep{3,3,1,1}) \oplus
(\rep{3,1,1,1}) \oplus (\rep{1,3,1,1}) \oplus (\rep{1,1,1,1}) \; .
\]
These four representations can be realized by the following states%
\footnote{The brackets $\{\;\}$ and $[\;]$ denote symmetrization and
anti-symmetrization of two undotted or two dotted indices. The prime
is written to distinguish different particle momenta.}
\begin{center}
\begin{tabular}{l|l|l|l}
  $(\rep{3,3,1,1})$
& $(\rep{3,1,1,1})$ & $(\rep{1,3,1,1})$ & $(\rep{1,1,1,1})$ \\
\hline
  $\ket{Y^{\vphantom{|}}_{\{\lAA\{\rAA} Y'_{\lBB\}\rBB\}}}$
& $\ket{Y^{\vphantom{|}}_{\{\lAA [\rAA} Y'_{\lBB\}\rBB ]}}$ & $\ket{Y^{\vphantom{|}}_{[\lAA\{\rAA} Y'_{\lBB ]\rBB\}}}$
& $\ket{Y^{\vphantom{|}}_{ [\lAA [\rAA} Y'_{\lBB ]\rBB ]}}$ \\
& $\ket{\Psi^{\vphantom{|}}_{\{\lAA [\raa} \Psi'_{\lBB\}\rbb ]}}$ & $\ket{\Bsi^{\vphantom{|}}_{[\laa\{\rAA} \Bsi'_{\lbb ]\rBB\}}}$
& $\ket{\Psi^{\vphantom{|}}_{ [\lAA [\raa} \Psi'_{\lBB ]\rbb ]}}$ \\
& &
& $\ket{\Bsi^{\vphantom{|}}_{ [\laa [\rAA} \Bsi'_{\lbb ]\rBB ]}}$ \\
& & & $\ket{Z^{\vphantom{|}}_{ [\laa [\raa} Z'_{\lbb ]\rbb ]}}$
\end{tabular}
\end{center}
Hence the action of the T-matrix is of the form
\[
\begin{split}
\Tmatrix \ket{Y^{\vphantom{|}}_{\lAA\rAA} Y'_{\lBB\rBB}} = & + \#\,
\ket{Y^{\vphantom{|}}_{\{\lAA\{\rAA} Y'_{\lBB\}\rBB\}}} + \#\,
\ket{Y^{\vphantom{|}}_{\{\lAA[\rAA} Y'_{\lBB\}\rBB ]}} + \#\,
\ket{Y^{\vphantom{|}}_{ [\lAA\{\rAA} Y'_{\lBB]\rBB\}}} + \#\,
\ket{Y^{\vphantom{|}}_{ [\lAA [\rAA} Y'_{\lBB ]\rBB ]}} \\ & +
\#\, \half \levi_{\rAA\rBB} \levi^{\raa\rbb}
\ket{\Psi^{\vphantom{|}}_{\{\lAA\,\raa} \Psi'_{\lBB\}\rbb\,}} + \#\, \half
\levi_{\lAA\lBB} \levi^{\laa\lbb} \ket{\Bsi^{\vphantom{|}}_{\,\laa\{\rAA}
\Bsi'_{\lbb\,\rBB\}}} \\ & + \#\, \half \levi_{\rAA\rBB}
\levi^{\raa\rbb}
\ket{\Psi^{\vphantom{|}}_{ [\lAA\,\raa} \Psi'_{\lBB ]\rbb\,}} +
\#\, \half \levi_{\lAA\lBB} \levi^{\laa\lbb}
\ket{\Bsi^{\vphantom{|}}_{\,\laa[\rAA} \Bsi'_{\lbb\,\rBB ]}} \\ & +
\#\,
\quarter \levi_{\rAA\rBB}
\levi_{\lAA\lBB} \levi^{\raa\rbb} \levi^{\laa\lbb}
\ket{Z^{\vphantom{|}}_{\laa\raa}
Z'_{\lbb\rbb}} \; .
\end{split}
\]
The explicit computation yields
\[
\begin{split}
\Tmatrix \ket{Y^{\vphantom{|}}_{\lAA\rAA} Y'_{\lBB\rBB}} =
\ \frac{1}{\cpp} \Bigsbrk{ &
  \half \bigbrk{ (p-p')^2 + 4pp' }
  \ket{Y^{\vphantom{|}}_{\{\lAA\{\rAA} Y'_{\lBB\}\rBB\}}} \\[-1mm] &
  + \half (p-p')^2 \bigbrk{
  \ket{Y^{\vphantom{|}}_{\{\lAA [\rAA} Y'_{\lBB\}\rBB ]}} +
  \ket{Y^{\vphantom{|}}_{ [\lAA\{\rAA} Y'_{\lBB ]\rBB\}}} } \\ &
  + \half \bigbrk{ (p-p')^2 - 4pp' }
  \ket{Y^{\vphantom{|}}_{ [\lAA [\rAA} Y'_{\lBB ]\rBB ]}} \\ &
  - 2pp' \sinh\tfrac{\theta-\theta'}{2} \bigbrk{
  \half \levi_{\rAA\rBB} \levi^{\raa\rbb} \ket{\Psi^{\vphantom{|}}_{\{\lAA\,\raa} \Psi'_{\lBB\}\rbb\,}} +
  \half \levi_{\lAA\lBB} \levi^{\laa\lbb} \ket{\Bsi^{\vphantom{|}}_{\,\laa\{\rAA} \Bsi'_{\lbb\,\rBB\}}} \\[-1mm] &
  \qquad\qquad\qquad\quad +
  \half \levi_{\rAA\rBB} \levi^{\raa\rbb} \ket{\Psi^{\vphantom{|}}_{ [\lAA\,\raa} \Psi'_{\lBB ]\rbb\,}} +
  \half \levi_{\lAA\lBB} \levi^{\laa\lbb} \ket{\Bsi^{\vphantom{|}}_{\,\laa [\rAA} \Bsi'_{\lbb\,\rBB ]}} }
  } \; , \nn
\end{split}
\]
which simplifies to
\[
\begin{split}
\Tmatrix \ket{Y^{\vphantom{|}}_{\lAA\rAA} Y'_{\lBB\rBB}} = \ &
  \Half \, \frac{(p-p')^2}{\cpp} \,
\ket{Y^{\vphantom{|}}_{\lAA\rAA} Y'_{\lBB\rBB}}
  + \frac{pp'}{\cpp} \, \bigbrk{
  \ket{Y^{\vphantom{|}}_{\lAA\rBB} Y'_{\lBB\rAA}} +
  \ket{Y^{\vphantom{|}}_{\lBB\rAA} Y'_{\lAA\rBB}} } \\ &
  - \frac{pp'}{\cpp} \, \sinh\frac{\theta-\theta'}{2} \, \bigbrk{
  \levi_{\rAA\rBB} \levi^{\raa\rbb}
\ket{\Psi^{\vphantom{|}}_{\lAA\raa} \Psi'_{\lBB\rbb}} +
  \levi_{\lAA\lBB} \levi^{\laa\lbb}
\ket{\Bsi^{\vphantom{|}}_{\laa\rAA} \Bsi'_{\lbb\rBB}} }
  \; .
\end{split}
\]
The bosonic part reproduces, of course, the computation in
\secref{sec:scalars} for $a=\half$. Notice that the coefficients
are such that the states on the right hand side do not differ in both
undotted and dotted indices from the in-state on the left hand
side. Such terms would prevent group-factorization
\eqref{factorize_at} of the T-matrix.

The other components of $\Tmatrix$ are listed in
\appref{app:full-T}. The entire result can be written concisely by
giving the coefficient functions as defined in
\eqref{eqn:tmatrix-coeff} for one $\algPSU(2|2)$ factor. We find%
\footnote{Notice that there arise signs when $\tmatrix$ acts across a
fermionic index:\\
$\Tmatrix \ket{\Phi_{A\dot{A}}\Phi'_{B\dot{B}}} =
(-)^{\grade{\dot{A}}(\grade{B}+\grade{D})}       \,
\ket{\Phi_{C\dot{A}}\Phi'_{D\dot{B}}} \, \tmatrix_{AB}^{CD} +
(-)^{\grade{B}(\grade{\dot{A}}+\grade{\dot{C}})} \,
\ket{\Phi_{A\dot{C}}\Phi'_{B\dot{D}}} \,
\tmatrix_{\dot{A}\dot{B}}^{\dot{C}\dot{D}} $.}
\begin{align}
    \Atmatrix(p,p') & = -\Dtmatrix(p,p') =  \Quarter \, \frac{(p-p')^2}{\cpp} \; , \nn \\
    \Btmatrix(p,p') & = -\Etmatrix(p,p') =  \frac{pp'}{\cpp} \; , \nn \\
    \Ctmatrix(p,p') & = +\Ftmatrix(p,p') = - \frac{pp'}{\cpp} \, \sinh\frac{\theta-\theta'}{2} \; ,  \label{eqn:factorized} \\
    \Gtmatrix(p,p') & = -\Ltmatrix(p,p') = -\Quarter \, \frac{p^2-p'^2}{\cpp} \; , \nn \\
    \Htmatrix(p,p') & = +\Ktmatrix(p,p') =  \frac{pp'}{\cpp} \, \cosh\frac{\theta-\theta'}{2} \; . \nn
\end{align}
In order to compare with the form in \secref{sec:results}, note the following kinematical identities
\begin{align}
\frac{1}{2}
\sqrt{\left(\energy+1\right)\left(\energy'+1\right)}\lrbrk{\cpp+p'-p}
& = - p p' \sinh\tfrac{\theta-\theta'}{2} \; , \\
\frac{1}{2}
\frac{\left(\energy+1\right)\left(\energy'+1\right)-pp'}{\sqrt{\left(\energy+1\right)\left(\energy
'+1\right)}} & = \cosh\tfrac{\theta-\theta'}{2} \; .
\end{align}

\subsection{Symmetries}
\label{sec:symmetries}

The string states are constructed from oscillators that individually
do not satisfy the level-matching condition (i.e. they carry
nonvanishing worldsheet momentum). In this sense they can be
considered off-shell. The symmetry algebra in the absence of the
level-matching constraint in the uniform light-cone gauge-fixed theory
is $\algPSU(2|2)_L\times\algPSU(2|2)_R\ltimes\Reals^2$
\cite{Arutyunov:2006ak}.
The total momentum is measured by an operator $\genP$ which
appears as one of the central generators of the symmetry algebra.
The other central generator is the total energy $\genH$.

In \cite{Arutyunov:2006ak} the symmetry generators were found in term
of the worldsheet fields. We would like to act with the symmetry
generators on the scattering states; consequently, we need to know the
oscillator representation of the symmetry generators.  Since the
nonlocal nature of the symmetry generators has already been taken into
account, it suffices to focus on their local part. In the notation of
\secref{sec:Hopf} this corresponds to analyzing the currents
generically denoted by ${\tilde J}$ in equation \eqref{current_Q}.
Computing the integrals along fixed-time slices, the oscillator representation
of the generators of $\algPSU(2|2)_L$ is (to quadratic order)
\begin{align}
  \genL{\lAA}{\lBB} & = \int\frac{dp}{2\pi}\: \frac{1}{2}\! \lrsbrk{ c^\dag_{\lAA\dot{C}} \, c^{\lBB\dot{C}}
           - c^{\dag\lBB\dot{C}} \, c_{\lAA\dot{C}} } \; , &
  \genQ{\laa}{\lBB} & = \int\frac{dp}{2\pi}\: (-)^{\grade{\dot{C}}} \lrsbrk{ u \, c^\dag_{\laa\dot{C}} \, c^{\lBB\dot{C}}
                                                                           - v \, c^{\dag\lBB\dot{C}} \, c_{\laa\dot{C}} } \; , \nn \\
  \genR{\laa}{\lbb} & = \int\frac{dp}{2\pi}\: \frac{1}{2}\! \lrsbrk{ c^\dag_{\laa\dot{C}} \, c^{\lbb\dot{C}}
                                                                   - c^{\dag\lbb\dot{C}} \, c_{\laa\dot{C}} } \; , &
  \genS{\lAA}{\lbb} & = \int\frac{dp}{2\pi}\: (-)^{\grade{\dot{C}}} \lrsbrk{ u \, c^\dag_{\lAA\dot{C}} \, c^{\lbb\dot{C}}
                                                                           - v \, c^{\dag\lbb\dot{C}} \, c_{\lAA\dot{C}} } \; , \nn
\end{align}
the generators of $\algPSU(2|2)_R$ are
\begin{align}
  \genLd{\rAA}{\rBB} & = \int\frac{dp}{2\pi}\: \frac{1}{2}\! \lrsbrk{ c^\dag_{C\rAA} \, c^{C\rBB}
                                                                    - c^{\dag C\rBB} \, c_{C\rAA} } \; , &
  \genQd{\raa}{\rBB} & = \int\frac{dp}{2\pi}\: \lrsbrk{ u \, c^\dag_{C\raa} \, c^{C\rBB}
                                                      - v \, c^{\dag C\rBB} \, c_{C\raa} } \; , \nn \\
  \genRd{\raa}{\rbb} & = \int\frac{dp}{2\pi}\: \frac{1}{2}\! \lrsbrk{ c^\dag_{C\raa} \, c^{C\rbb}
                                                                    - c^{\dag C\rbb} \, c_{C\raa} } \; , &
  \genSd{\rAA}{\rbb} & = \int\frac{dp}{2\pi}\: \lrsbrk{ u \, c^\dag_{C\rAA} \, c^{C\rbb}
                                                      - v \, c^{\dag C\rbb} \, c_{C\rAA} } \; , \nn
\end{align}
and the two dentral charge generators generators read
\begin{align}
  \genP             & = \int\frac{dp}{2\pi}\: p \, c^\dag_{A\dot{A}} \, c^{A\dot{A}} \; , &
  \genH             & = \int\frac{dp}{2\pi}\: \energy \, c^\dag_{A\dot{A}} \, c^{A\dot{A}} \; . \nn
\end{align}
As before we have
\[
 u = \cosh{\tfrac{\theta}{2}} \comma
 v = \sinh{\tfrac{\theta}{2}} \comma
 p = \sinh\theta \comma
 \energy = \cosh\theta \; .
\]
These formulas give the oscillator representation of the centrally
extended $\algPSU(2|2)^2$ algebra. Since $\genQ{}{}$ and $\genS{}{}$
transform as the components of a Lorentz spinor
\cite{Hofman:2006xt}, one can see from this form that this
representation is related to a representation of the non-centrally
extended algebra ($\genP=0$) by a Lorentz boost. Recall that the
free worldsheet Lagrangian indeed possesses Lorentz invariance.

\paragraph{The algebra.} The generators $\genP$ and $\genH$ are the
two central charges corresponding to total momentum and total energy
of a representation. The rotation generators $\genR{\laa}{\lbb}$ and
$\genL{\lAA}{\lBB}$ act onto a generic generator $\genJ$ canonically
as
\begin{align}
  \comm{\genL{\lAA}{\lBB}}{\genJ_\lCC} & =  \delta_\lCC^\lBB \, \genJ_\lAA - \half \delta_\lAA^\lBB \, \genJ_\lCC \; ,&
  \comm{\genL{\lAA}{\lBB}}{\genJ^\lCC} & = -\delta_\lAA^\lCC \, \genJ^\lBB + \half \delta_\lAA^\lBB \, \genJ^\lCC \; , \\
  \comm{\genR{\laa}{\lbb}}{\genJ_\lcc} & =  \delta_\lcc^\lbb \, \genJ_\laa - \half \delta_\laa^\lbb \, \genJ_\lcc \; , &
  \comm{\genR{\laa}{\lbb}}{\genJ^\lcc} & = -\delta_\laa^\lcc \, \genJ^\lbb + \half \delta_\laa^\lbb \, \genJ^\lcc \; .
\end{align}
The supercharges satisfy
\begin{align}
  \acomm{\genQ{\laa}{\lAA}}{\genQ{\lbb}{\lBB}} & = - \half \levi_{\laa\lbb} \levi^{\lAA\lBB} \, \genP \; , \\
  \acomm{\genS{\lAA}{\laa}}{\genS{\lBB}{\lbb}} & = - \half \levi_{\lAA\lBB} \levi^{\laa\lbb} \, \genP \; , \\
  \acomm{\genQ{\laa}{\lAA}}{\genS{\lBB}{\lbb}} & = \delta_\laa^\lbb \, \genL{\lBB}{\lAA}
                                                 + \delta_\lBB^\lAA \, \genR{\laa}{\lbb}
                                                 + \half \delta_\laa^\lbb \delta_\lBB^\lAA \, \genH \; .
\end{align}
Here we have the same central charge appearing in the anticommutator
of $\genQ{}{}$ and $\genS{}{}$ with itself. This is due to the
quadratic approximation made here. Including higher orders,
one would find that the two central charges (both denoted here by
$\genP$) are the complex conjugate of each other. The generators of
$\algPSU(2|2)_R$ obey identical algebra relations.

\paragraph{Single excitation representation.}
If we act with the supercharges defined above on a single oscillator, we find
\begin{align}
  \genLd{\rAA}{\rBB} \ket{c^\dag_\rCC} & = \delta_\rCC^\rBB \ket{c^\dag_\rAA} -  \half \delta_\rAA^\rBB \ket{c^\dag_\rCC} \; , &
  \genRd{\raa}{\rbb} \ket{c^\dag_\rcc} & = \delta_\rcc^\rbb \ket{c^\dag_\raa} - \half \delta_\raa^\rbb \ket{c^\dag_\rcc} \; , \\
  \genQd{\raa}{\rBB} \ket{c^\dag_\rCC} & = u \, \delta_\rCC^\rBB \ket{c^\dag_\raa} \; , &
  \genQd{\raa}{\rBB} \ket{c^\dag_\rcc} & = -v \, \levi_{\raa\rbb} \levi^{\rBB\rCC} \ket{c^\dag_\rCC} \; , \\
  \genSd{\rAA}{\rbb} \ket{c^\dag_\rCC} & = -v \, \levi_{\rAA\rBB} \levi^{\rbb\rcc} \ket{c^\dag_\rcc} \; , &
  \genSd{\rAA}{\rbb} \ket{c^\dag_\rcc} & = u \, \delta_\rcc^\rbb \ket{c^\dag_\rAA} \; ,
\end{align}
where the undotted indices remain unaffected and have been suppressed. This is the same representation as for a single excitation of the spin chain \cite{Beisert:su22-S-matrix}. Comparing the two cases, we see that the coefficients $a,b,c,d$ of \cite{Beisert:su22-S-matrix} take the values $a=d=u$ and $b=c=-v$. Note that indeed $ad-bc=1$ (required by the closure of the algebra), $\half(ad+bc)=\half \energy$ (first central charge) and $ab=cd=-\half p$ (two more central charges).

\paragraph{Invariance of the T-matrix.} With the ingredients described
above we can now verify that the tree-level worldsheet S-matrix
derived in \secref{sec:full-T} is invariant under $\algPSU(2|2)^2$
transformations with the the coproduct action \eqref{coprod}:
\begin{align}
\comm{\genQ{\laa}{\lBB}\otimes \genF + \genF \otimes\genQ{\laa}{\lBB}}{\tmatrix}
 & = + (\half\genP \genF) \otimes \genQ{\laa}{\lBB} -
\genQ{\laa}{\lBB} \otimes (\half\genP \genF) \\
\comm{\genS{\lAA}{\lbb}\otimes \genF + \genF \otimes\genS{\lAA}{\lbb}}{\tmatrix}
 & = - (\half\genP \genF) \otimes \genS{\lAA}{\lbb} +
\genS{\lAA}{\lbb} \otimes (\half\genP \genF)
\end{align}
where $\genF$ acts as $\genF \ket{c^\dag_A} = (-)^{\grade{A}}
\ket{c^\dag_A}$. These equations arise from the expansion of
\eqref{S_invariance} at large 't~Hooft coupling; the terms appearing
on the right hand side of these equations are a direct consequence of
the non-trivial co-product for the action of the supercharges on
multi-excitation states , cf. \secref{sec:Hopf}.

\section{Comparison with SYM}
\label{sec:SpinChainSmatrices}

As we have mentioned previously the S-matrix of planar ${\cal N}=4$
SYM can be constructed by using the fact that a choice of
ferromagnetic spin chain vacuum state breaks the full $\algPSU(2,2|4)$
to its $\algPSU(2|2)^2$ subgroup. However this novel spin chain is
"dynamic" in the sense that the number of lattice sites can
change. This induces two additional central charges shared
by both factors of the
symmetry group.

For the spin chain, the scattering processes $\phi_a
\phi_b\rightarrow \psi_\alpha\psi_\beta$ and
$\psi_\alpha\psi_\beta\rightarrow  \phi_a \phi_b$ involve the
creation or annihilation of a vacuum lattice site, denoted by ${\cal
Z}^\pm$, and it is these insertions which give rise to the
non-trivial coproduct for the global charges in the spin chain (see
\cite{Gomez:2006va},\cite{Plefka:2006ze}) while at the same time
being responsible for the appearence of the central charges. The
S-matrix of \cite{Beisert:su22-S-matrix} for a single $\grSU(2|2)$
sector involving the scalar fields, $\phi_a$, and the fermions,
$\psi_{\alpha}$ with $a,\alpha=1,2$, is uniquely
determined up to an overall phase by demanding
that the $S$-matrix is invariant under the action of the $\algSU(2|2)$
algebra where the generators act on the fields as:
\begin{align}
  \genL{\lAA}{\lBB} \ket{\phi_\lCC} & = \delta_\lCC^\lBB
\ket{\phi_\lAA} - \half \delta_\lAA^\lBB \ket{\phi_\lCC} \; , &
  \genR{\laa}{\lbb} \ket{\psi_\lcc} & = \delta_\lcc^\lbb
\ket{\psi_\laa} - \half \delta_\laa^\lbb \ket{\psi_\lcc} \; , \\
  \genQ{\laa}{\lBB} \ket{\phi_\lCC} & = a \, \delta_\lCC^\lBB \ket{\psi_\laa} \; , &
  \genQ{\laa}{\lBB} \ket{\psi_\lcc} & = b \, \levi_{\laa \lbb}
\levi^{\lBB\lCC} \ket{\phi_\lCC{\cal Z}^+} \; , \\
  \genS{\lAA}{\lbb} \ket{\phi_\lCC} & = c \, \levi_{\lAA\lBB}
\levi^{\lbb\lcc} \ket{\psi_\lcc{\cal Z}^-} \; , &
  \genS{\lAA}{\lbb} \ket{\psi_\lcc} & = d \, \delta_\lcc^\lbb \ket{\phi_\lAA} \; ;
\end{align}
and the extra central charges $\genP$ and $\genK$ required by the presence
of the length-changing operators ${\cal Z}^\pm$ act as
\[
  \genP\ket{\chi}= ab\, \ket{\chi{\cal Z}^+} \comma
\genK\ket{\chi}=cd\, \ket{\chi{\cal Z}^-} \; .
\]
When these generators act on multi-particle states the ${\cal Z}^\pm$ operators introduce
additional momentum-dependent phases which promote this algebra to a Hopf subalgebra.

The resulting $S$-matrix is given by
\begin{align}
\smatrix^B\ket{\phi_a\phi'_b} &=
 \Asmatrix^B\ket{\phi'_{\{a}\phi_{b\}}}
+\Bsmatrix^B\ket{\phi'_{[a}\phi_{b]}}
+\half
\Csmatrix^B\varepsilon_{ab}\varepsilon^{\alpha\beta}\ket{\psi'_\alpha\psi_\beta{\cal
Z}^{-}} \; ,
\\
\smatrix^B\ket{\psi_\alpha\psi'_\beta} &=
 \Dsmatrix^B\ket{\psi'_{\{\alpha}\psi_{\beta\}}}
+\Esmatrix^B\ket{\psi'_{[\alpha}\psi_{\beta]}}
+\half
\Fsmatrix^B\varepsilon_{\alpha\beta}\varepsilon^{ab}\ket{\phi'_a\phi_b{\cal
Z}^{+}} \; ,
\\
\smatrix^B\ket{\phi_a\psi'_\beta} &=
 \Gsmatrix^B\ket{\psi'_\beta\phi_{a}}
+\Hsmatrix^B\ket{\phi'_{a}\psi_\beta} \; ,
\\
\smatrix^B\ket{\psi_\alpha\phi'_b} &=
 \Ksmatrix^B\ket{\psi'_\alpha\phi_{b}}
+\Lsmatrix^B\ket{\phi'_{b}\psi_\alpha} \; ,
\label{SU22_S_matrix}
\end{align}
with the coefficients
\begin{align}\label{allfromsu22}
 \Asmatrix^B&= S^0_{pp'}\,\frac{x_{p'}^+-x_{p}^-}{x_{p'}^--x_{p}^+}\; , \nn \\
 \Bsmatrix^B&= S^0_{pp'}\,\frac{x_{p'}^+-x_{p}^-}{x_{p'}^--x_{p}^+}\, \left(1
             -2 \,\frac{1-\frac{1}{x^+_{p}x^-_{p'}}}{1-\frac{1}{x^+_{p}x^+_{p'}}}\,
             \frac{x_{p'}^--x_{p}^-}{x_{p'}^+-x_{p}^-}\
             \right) \; , \nn \\
 \Csmatrix^B&= S^0_{pp'}\,\frac{ 2\gamma_p\gamma_{p'}}{x^+_{p}x^+_{p'}}
                \,\frac{1}{1-\frac{1}{x^+_{p}x^+_{p'}}}
                \,\frac{x^-_{p'}-x^-_{p}}{x_{p'}^--x_{p}^+}\; , \nn \\
 \Dsmatrix^B&= -S^0_{pp'},\nn \\
 \Esmatrix^B&= -S^0_{pp'}\, \left(1
-2\,\frac{1-\frac{1}{x^-_{p}x^+_{p'}}}{1-\frac{1}{x^-_{p}x^-_{p'}}}
                              \, \frac{x_{p'}^+-x_{p}^+}{x_{p'}^--x_{p}^+}\ \right) \; , \nn \\
 \Fsmatrix^B&= -S^0_{pp'}\,\frac{2}{\gamma_{p}\gamma_{p'}x^-_{p}x^-_{p'}}
                 \,\frac{(x_{p}^+-x_{p}^-)(x_{p'}^+-x_{p'}^-)}{1-\frac{1}{x^-_{p}x^-_{p'}}}
                 \,\frac{x^+_{p'}-x^+_{p}} {x_{p'}^--x_{p}^+} \; , \nn \\
 \Gsmatrix^B&= S^0_{pp'} \,\frac{x_{p'}^+-x_{p}^+}{x_{p'}^--x_{p}^+} \; ,
 ~~~~~~~~~~~~~~
 \Hsmatrix^B = S^0_{pp'}
\,\frac{\gamma_p}{\gamma_{p'}}\,\frac{x_{p'}^+-x_{p'}^-}{x_{p'}^--x_{p}^+}
\; ,
 \nn \\
 \Ksmatrix^B&= S^0_{pp'}
\,\frac{\gamma_{p'}}{\gamma_p}\,\frac{x_{p}^+-x_{p}^-}{x_{p'}^--x_{p}^+}
\; ,
 ~~~~~~~~~~
 \Lsmatrix^B = S^0_{pp'} \,\frac{x_{p'}^--x_{p}^-}{x_{p'}^--x_{p}^+} \; ,
\end{align}
where
\[
\gamma_p=\abs{x_p^- - x_p^+}^{1/2}
\]
and
\[\label{x+-}
x_p^\pm=\frac{\pi e^{\pm \ihalf p}}{\sqrt{\lambda }\sin\frac{p}{2}}
\left(1+\sqrt{1+\frac{\lambda }{\pi ^2}\,\sin^2\frac{p}{2}}\right) \; .
\]
As mentioned before, the phase factor $S^0$ is undetermined by the
algebraic construction. The one that correctly reproduces the
semiclassical string spectrum via Bethe equations
\cite{Kazakov:2004qf} is
\[\label{snol}
S^0_{pp'}=\frac{1-\frac{1}{x^+_{p'}x^-_{p}}}{1-\frac{1}{x^-_{p'}x^+_{p}}}
\,e^{i\theta\left(p,p'\right)}
\]
with the dressing phase given by \cite{Arutyunov:2004vx} (to the
leading order in $1/\sqrt{\lambda }$)
\begin{eqnarray}\label{afsphase}
 \theta(p,p')&=&\frac{\sqrt{\lambda }}{2\pi
 }\sum_{r,s=\pm}^{}rs\,\chi (x_{p}^r,x_{p'}^s),
 \nn \\
 \chi
 (x,y)&=&(x-y)\left[\frac{1}{xy}+\left(1-\frac{1}{xy}\right)
 \ln\left(1-\frac{1}{xy}\right)\right] \; .
\end{eqnarray}

In the comparison with the worldsheet calculation we are actually
interested in the coefficients of ${\cal P}_{{\rm g}}P^u_{pp'}S^B$,
where ${\cal P}_{{\rm g}}$ is the graded permutation operator and
$P^u_{pp'}$ exchanges the excitation momenta. Furthermore, in order to
find the S-matrix for the full $\grPSU(2,2|4)$ theory we use the
relation%
\footnote{The full S-matrix has to be divided by $A^B$, because the
$\alg{psu}(2|2)$ S-matrix was
defined in \cite{Beisert:su22-S-matrix} as the physical
scattering matrix of the fields $\Phi _{A\dot{1}}$. In addition to
$\smatrix$ for the left $\alg{psu}(2|2)$ indices, the scattering of
this field receives contribution from
$\smatrix_{\dot{1}\dot{1}}^{\dot{1}\dot{1}}=A^B$.}%
,
\be
 \Smatrix=\frac{1}{\Asmatrix^B}\smatrix^B\otimes \smatrix^B \comma
 \Smatrix_{A\dot{A}B\dot{B}}^{C\dot{C}D\dot{D}}(p,p')=\frac{1}{\Asmatrix^B}
(\smatrix^B)_{AB}^{CD}(p,p') (\smatrix^B)_{\dot{A}\dot{B}}^{\dot{C}\dot{D}}(p,p') \; .
\ee
Consequently we can relate the above coefficients to those of
$\smatrix$ used in \secref{sec:results}
\begin{align}
\Asmatrix&= \tfrac{1}{2\sqrt{\Asmatrix^B}}\brk{\Asmatrix^B-\Bsmatrix^B} \; , &
\Bsmatrix&= \tfrac{1}{2\sqrt{\Asmatrix^B}}\brk{\Asmatrix^B+\Bsmatrix^B} \; , &
\Csmatrix&= \tfrac{1}{ \sqrt{\Asmatrix^B}}\Csmatrix^B\; , \nn \\
\Dsmatrix&= \tfrac{1}{2\sqrt{\Asmatrix^B}}\brk{-\Dsmatrix^B+\Esmatrix^B} \; , &
\Esmatrix&= \tfrac{1}{2\sqrt{\Asmatrix^B}}\brk{-\Dsmatrix^B-\Esmatrix^B} \; , &
\Fsmatrix&=-\tfrac{1}{ \sqrt{\Asmatrix^B}}\Fsmatrix^B\; , \nn\\
\Gsmatrix&= \tfrac{1}{ \sqrt{\Asmatrix^B}}\Gsmatrix^B\; , &
\Hsmatrix&= \tfrac{1}{ \sqrt{\Asmatrix^B}}\Hsmatrix^B\; , \nn \\
\Lsmatrix&= \tfrac{1}{ \sqrt{\Asmatrix^B}}\Lsmatrix^B\; , &
\Ksmatrix&= \tfrac{1}{ \sqrt{\Asmatrix^B}}\Ksmatrix^B\; .
\label{relation}
\end{align}

In order to compare our worldsheet results with those of the spin
chain S-matrix of \cite{Beisert:su22-S-matrix} we must expand the
latter in $1/\sqrt{\lambda }$. But first we should understand how the
spin chain momenta are related to the worldsheet momenta. As part of
the gauge fixing of the worldsheet theory we chose the density of the
light-cone momentum to be a constant which in turn
fixed the string length to be $\mathcal{J}=2\pi J_+/\sqrt{\lambda}$
where $J_+$ is the momentum. Then, we took
$\mathcal{J}$ to be infinite, which
allowed for a sensible definition of the $S$-matrix, and
expanded in powers of $\tfrac{2\pi}{\sqrt{\lambda}}$ which acts as a
loop counting parameter. This should be contrasted with the spin chain
whose length $L$ is identified with the momentum $J$ plus an
additional term that depends on the number of
excitations%
\footnote{Various excitations contribute differently to the
length, see \cite{longrange} for the precise definition. For the sake
of our argument, it is enough to known that $J\rightarrow \infty $ and
$M=O(1)$ in the decompactification limit. The difference between $L$
and $J$ then becomes negligible.}%
: $L=J+M$. Going from
the spin chain to the string worldsheet involves the rescaling by a
factor of $\sqrt{\lambda }/2\pi $, which affects all dimensional
quantities and in particular all momenta in \eqref{allfromsu22}, which
should be rescaled as
\[
 p\,\,\longrightarrow\,\, \frac{2\pi p}{\sqrt{\lambda }}
~~~~~~~~
p_{\rm chain}=\frac{2\pi}{\sqrt{\lambda }}p_{\rm string}\;.
\]
Indeed, once we impose the periodic boundary conditions, the
spin chain momentum is quantized in the units of $2\pi /J$, while in
the sigma-model the momentum quantization unit is $\sqrt{\lambda
}/J$. We should therefore first rescale as above all momenta in the
spin chain
S-matrix and then expand it in $1/\sqrt{\lambda }$. The matrix
elements in \eqref{allfromsu22} depend on $1/\sqrt{\lambda }$ only
through $x_p^\pm$. Therefore, the strong-coupling expansion is
equivalent to the low-momentum expansion of the spin chain
S-matrix. For the kinematical variables
\eqref{x+-} the rescaling of momenta yields:
\begin{equation}\label{}
 x_p^\pm=\frac{1+\varepsilon }{p}\left(1\pm\frac{i\pi p}
 {\sqrt{\lambda }}+O\left(\frac{1}{\lambda }\right)\right)~~.
\end{equation}
Note that in the limit we are considering here all information about
bound states appears at higher orders in the $1/\sqrt{\lambda}$
expansion.
%
%

The expansion of \eqref{allfromsu22}--\eqref{afsphase} in
$1/\sqrt{\lambda}$ is a tedious but straightforward exercise. The
small-momentum expansion of the dressing phase \eqref{afsphase} was
computed in \cite{Roiban:2006yc}:
\begin{eqnarray}\label{}
 \theta (p,p')&=&-\frac{2\pi }{\sqrt{\lambda }}\,
 \left(1+\varepsilon \right)\left(1+\varepsilon '\right)
 \,\left.\frac{\partial ^2\chi }{\partial x\,\partial y}
 \right|_{x=\frac{1+\varepsilon }{p}\,,\,\,y=\frac{1+\varepsilon '}{p'}}
 \nonumber \\ &=&\frac{2\pi }{\sqrt{\lambda }}\,\,
 \frac{\frac{1}{2}\left(p-p'\right)^2-(p-p')(\varepsilon 'p-\varepsilon p')
 +\frac{1}{2}(\varepsilon 'p-\varepsilon p')^2}{\varepsilon 'p-\varepsilon
 p'}\,.
\end{eqnarray}
Expanding the matrix elements we find for the components
\eqref{eqn:tmatrix-coeff} of the T-matrix:
\begin{align}
\Atmatrix(p,p') & = \Quarter \biggsbrk{\lrbrk{\cpp} -2(p-p') + \frac{(p-p')^2}{\cpp} } \; , \nn \\
\Btmatrix(p,p') & = -\Etmatrix(p,p')= \frac{pp'}{\cpp} \; , \nn \\
\Ctmatrix(p,p') & = \Ftmatrix(p,p') = \Half \frac{\sqrt{\left(\energy+1\right)\left(\energy'+1\right)} \lrbrk{\cpp+p'-p}}{\cpp} \; , \label{ourfinalequation1} \\
\Dtmatrix(p,p') & = \Quarter \biggsbrk{\lrbrk{\cpp} - \frac{(p-p')^2}{\cpp} } \; , \nn \\
\Gtmatrix(p,p') & = -\Ltmatrix(p',p) = \Quarter \biggsbrk{\lrbrk{\cpp} +2p' - \frac{p^2-p'^2}{\cpp} } \; , \nn \\
\Htmatrix(p,p') & = \Ktmatrix(p,p') = \Half \, \frac{pp'}{\cpp} \, \frac{\left(\energy+1\right)\left(\energy'+1\right)-pp'}
                                          {\sqrt{\left(\energy +1\right)\left(\energy '+1\right)}} \; . \nn
\end{align}
This should be compared with the string calculation in the
constant-$J$ gauge given in \eqref{ourfinalequation} for $a=0$. We
note that the results almost agree, the only difference being terms
which are linear in the momenta.
These terms should not affect the physical spectrum when the
S-matrix is plugged into the asymptotic Bethe equations. We suspect
that when the linear terms are promoted to the linear phases in the
exact S-matrix, they account for the
%
difference between the length of the spin chain and the internal
length of the string.
This
can be seen most clearly in the rank-one sectors; for example, the
bosonic $\algSU(2)$ sector of the spin chain is described by the
Bethe equation
\[
e^{i L p_i}=\prod_{j\not= i}^M \smatrix^B_{\algSU(2)}(p_i,p_j).
\]
The string length is proportional to the R-charge $J$, but for the
spin chain the length is $L=J+M$, where $M$ is the number of
impurities. In order to compare with the string theory we must
rewrite the equation as
\[
e^{i J p_i}=\prod_{j\not= i}^M
\smatrix^B_{\algSU(2)}(p_i,p_j)e^{i(p_j-p_i)}.
\]
Thus there are new terms in the $S$-matrix, which after the
rescaling described above, contribute terms linear in momentum at
order $1/\sqrt{\lambda}$. The appropriate change in length is
different for the various impurities and one should carefully trace
through the effects in the Bethe equations to see exactly how the
string and spin chain $S$-matrices are related, which is beyond the
scope of the present paper.

If, instead of \eqref{snol}, we take
\[
 S^0_{pp'}=e^{i\frac{p-p'}{2}},
\]
we find the resulting $\tmatrix$-matrix is given by
\begin{align}
\Atmatrix(p,p') & = \Quarter \, \frac{(p-p')^2}{\cpp}  \; , \nn \\
\Btmatrix(p,p') & = -\Etmatrix(p,p')= \frac{pp'}{\cpp} \; , \nn \\
\Ctmatrix(p,p') & = \Ftmatrix(p,p') = \Half \frac{\sqrt{\left(\energy+1\right)\left(\energy'+1\right)}\lrbrk{\cpp+p'-p}}{\cpp} \; , \label{ourfinalequation2} \\
\Dtmatrix(p,p') & = \Quarter \biggsbrk{2(p-p') - \frac{(p-p')^2}{\cpp} } \; , \nn \\
\Gtmatrix(p,p') & = -\Ltmatrix(p',p) = \Quarter \biggsbrk{\lrbrk{\cpp} +2p' - \frac{p^2-p'^2}{\cpp} } \; , \nn \\
\Htmatrix(p,p') & = \Ktmatrix(p,p') = \Half \, \frac{pp'}{\cpp} \, \frac{\left(\energy+1\right)\left(\energy'+1\right)-pp'}
          {\sqrt{\left(\energy +1\right)\left(\energy '+1\right)}} \; . \nn
\end{align}
This agrees with the string calculation in the light-cone gauge
($a=1/2$), again up to terms linear in momenta.

\section{Conclusions and discussion}
\label{sec:conclusions}

The gauge-fixed sigma-model in $\AdS_5\times \Sphere^5$ is a rather
complicated 2d quantum field theory. Even at tree level, the
calculations of the finite-size spectrum in \cite{Callan:2003xr}
and of the S-matrix here involve complicated combinatorics. We have
analyzed the latter calculation in detail and explicitly shown that
the scattering matrix has all the required factorization properties
consistent with integrability. A crucial ingredient was the fact that the
action of the symmetry algebra on multi-particle states is given by
a nontrivial coproduct.

A similar (albeit not identical) nonstandard realization of the
symmetry algebra occurs on the gauge theory side of the duality
\cite{Beisert:su22-S-matrix}. Though different, these nonstandard
realizations are crucial for the positive comparison of the worldsheet
and the spin chain picture of the dual SYM theory. The
difference we noted between the realization of symmetries can
therefore be identified as a gauge degree of freedom. Indeed, there
appears to exist a nonlocal field redefinition \cite{GoHe} that
relates the two coproducts.

It would be very interesting to extend our calculations to include loop
effects. Such a calculation would provide further nontrivial checks on
standing conjectures regarding the S-matrices appearing in the context
of the AdS/CFT correspondence. As mentioned above, at least one
nonlocal field redefinition is necessary to directly identify at the
classical level the fields and the algebraic structures on the two
sides of the duality. Such redefinitions have the potential of
altering the quantum theories. It is thus not immediately obvious
which worldsheet one should use for computing quantum corrections to
the scattering matrix. A possible guide is provided by the symmetry
algebra described in \secref{sec:Hopf}. There we argued that
perturbative quantum corrections cannot alter the coproduct derived
classically. Imposing this as a constraint on the definition of the
quantum theory may uniquely identify it.
Higher loop corrections in this theory should reproduce the higher
order terms in the $1/\sqrt{\lambda}$ expansion of the spin chain
S-matrix described in section \secref{sec:SpinChainSmatrices}.

Two important issues that we have not discussed here are crossing
symmetry and analyticity of the S-matrix. While the former is a
kinematical restriction, the analytic properties of the S-matrix
contain information about the spectrum of bound states of the
theory. In the bootstrap approach to integrable relativistic quantum
field theories these properties are very important, along with the
quantum Yang-Baxter equation,  in determining the S-matrix (up to a
smooth phase) and the spectrum
\cite{Zamolodchikov:1978xm}.

The AdS/CFT sigma-model in the light-cone (or any other unitary) gauge
lacks relativistic invariance. This is reflected in the structure of
the S-matrix, which depends
on the individual momenta of the incoming particles rather than on
their particular combination such as the difference of rapidities in
relativistic theories.
Nevertheless, based on algebraic considerations, the S-matrix of
AdS/CFT was conjectured to satisfy a crossing relation
\cite{Janik:2006dc}. Recalling that in relativistic quantum field
theory the crossing symmetry is a simple consequence of the Feynman
rules \cite{Landafshits} and noting that two-dimensional Lorentz
invariance is only broken spontaneously on the worldsheet, it is not
unnatural to hope that diagrammatic perturbation theory of the type
used in this paper may be helpful in understanding the behavior of
the S-matrix
under particle-antiparticle transformation.
A different perspective on the connection with the crossing symmetry
of relativistic field theories may be provided by the construction
of \cite{MannPolchinski}.

Comparatively little is known
about the analytic properties of the S-matrix. According to the
standard bootstrap philosophy, bound states of the theory correspond
to (complex) simple poles of the scattering matrix%
\footnote{The reverse does not always hold and simple poles do not
always correspond to bound states.}%
. While the physical meaning of some higher order poles is known, an
interpretation of more serious singularities of a two-dimensional
S-matrix is currently lacking. Our tree-level calculation of the
S-matrix does not yield direct information on its analytic
structure. In particular, the two-magnon bound state present in the
gauge theory spin chain exhibits, in our expansion, a difference of
order $i/\sqrt{\lambda}$ between the corresponding world sheet
momenta of its constituents; thus, it must be a quantum effect and
should be accessible only after the perturbative series is (partly)
resumed. An efficient way to gain access to the analytic structure
of the S-matrix at the classical level is the analysis of the
scattering of worldsheet solitons.
In the limit of small charges and small momenta, the
$1/\sqrt{\lambda}$ expansion of the soliton S-matrix reduces -- in the
appropriate gauge -- to the results of perturbative calculations of
the type described here.

\bigskip
\subsection*{Acknowledgments}
\bigskip
We would like to thank S.~Frolov, V.~Kazakov, U.~Lindstr\"om and
A.~Tseytlin for interesting discussions. The work of K.Z. was
supported in part by the Swedish Research Council under contracts
621-2004-3178 and 621-2003-2742, and by grant NSh-8065.2006.2 for
the support of scientific schools, and by RFBR grant 06-02-17383.
The work of T.K. and K.Z. was supported by the G\"oran Gustafsson
Foundation. The work of R.R. is supported in part by the National
Science Foundation under grant PHY-0608114.

\appendix

\section{Symmetry considerations }
\label{sec:puzzles}

As mentioned in \secref{sec:results} the gauge-invariant worldsheet theory as well as
the worldsheet
theory in conformal gauge are classically  integrable.
Formally, one may think of gauge-fixing the diffeomorphism invariance
as equivalent with expanding around some classical solution; for the
light-cone gauge this is the BMN point-like string \cite{Frolov:2002av}
 combined with solving a subset of the classical equations of
motion.
It is therefore reasonable to expect that
diffeomorphism-gauge-fixed  worldsheet theory remains integrable.
$\kappa$-symmetry gauge-fixing cannot be understood in a similar
language; however, in explicit examples, it is possible to see that
integrability is preserved.

Furthermore, there exist known examples in which
despite the symmetry algebra being centrally-extended
\cite{Arutyunov:2006ak},
the symmetry transformations act on multi-excitation states via the
Leibniz rule. It is worth stressing that virtually in all known
continuum integrable models this expectation is in fact realized and
it relies only on the fact that in a quantum theory operators act via
commutators.

Additionally, one may expect \cite{AbLi,Zamolodchikov:1978fd}
that the S-matrix is determined up to a phase by the
symmetries of the theory, in particular the nonlocal symmetries and
it again turns out that this expectation is  realized in most
existing integrable field theories.

\subsection{Leibniz rule and symmetry constraints}

Under the assumption that the global symmetries act following the
Leibniz rule it is quite trivial to impose their conservation in the
scattering process. For this purpose we need some sufficiently
general action on single excitations.
Denoting by $B^C(p)$ the creation operators, the symmetry
generators $Q_{(1)}^A{}_B$\footnote{Here $Q_{(1)}$ uniformly covers
both the bosonic and the fermionic $\algPSU(2|2)$ generators.} act
on these excitations as
\[
\begin{split}
\acomm{Q_{(1)}{}^A{}_B}{B^C(p)} =\ & \acomm{Q_{(1)}{}^A{}_B}{B^C(p)}_0 \\
& + E_-^{AC}E_+{}_{BD} \acomm{c}{B^D(p)} + E_+^{AC}E_-{}_{BD} \acomm{c^*}{B^D(p)} \; .
\end{split}
\label{aaa}
\]
where $E_\pm$ are defined as
\[
E_+=\matr{cc}{\levi_{ab} & 0 \\ 0 & 0 } \;,\;\;
E_-=\matr{cc}{0 & 0 \\ 0 & \levi_{\alpha\beta}}\;,\;\;
I_+=\matr{cc}{ \unit_2 & 0 \\ 0 & 0 }   \;,\;\;
I_-=\matr{cc}{0 & 0 \\ 0 & \unit_2 }    \;
\]
and have we introduced $I_\pm$ for later convenience.

In the equation (\ref{aaa})
\[
\acomm{Q_{(1)}{}^A{}_B}{B^C(p)}_0 = f^{AC}_{BD}(p)B^D(p)
\comma
\acomm{c}{B^C(p)}_0 = c(p)B^C(p)
\label{struct_fct}
\]
represent the action of symmetry generators in the absence
of the central extension and of the central charge, respectively.
The coefficients $f^{AC}_{BD}(p)$ have manifest $\algSU(2)\oplus \algSU(2)$
symmetry and may be written as
\[
\begin{split}
f^{AC}_{BD}(p) =\ & \Bigbrk{I_+{}^C_BI_+{}^A_D-\half I_+{}^A_BI_+{}^C_D}+
                    \Bigbrk{I_-{}^C_BI_-{}^A_D-\half I_-{}^A_BI_-{}^C_D}\\
                  & +a(p)I_-{}^A_DI_+{}^C_B+d(p)I_+{}^A_DI_-{}^C_B \; .
\end{split}
\]
The functions $c(p)$, $a(p)$ and $d(p)$ are determined by the
worldsheet sigma-model together with its Poisson structure. The
momentum displayed as their argument is that of the excitation the
generators act on.

Among the many consequences of the vanishing of the commutation of the
S-matrix and the $\algPSU(2|2)$ generators are
\begin{align}
\Csmatrix(p,p') & = \hphantom{-}\frac{1}{2a(p)} \Bigsbrk{c(p')\Lsmatrix(p,p')-c(p)\Hsmatrix(p,p')-\bigbrk{\Dsmatrix(p,p')-\Esmatrix(p,p')}c^*(p')} \nn \\
           & =-\frac{1}{2a(p')} \Bigsbrk{c(p')\Ksmatrix(p,p')-c(p)\Gsmatrix(p,p')+\bigbrk{\Dsmatrix(p,p')-\Esmatrix(p,p')}c^*(p)} \\
\Fsmatrix(p,p') & = \hphantom{-}\frac{1}{2d(p)} \Bigsbrk{c^*(p')\Gsmatrix(p,p')-c^*(p)\Ksmatrix(p,p') -\bigbrk{\Asmatrix(p,p')-\Bsmatrix(p,p')}c(p')} \nn \\
           & = -\frac{1}{2d(p')} \Bigsbrk{c^*(p')\Hsmatrix(p,p')-c^*(p)\Lsmatrix(p,p')+\bigbrk{\Asmatrix(p,p') -\Bsmatrix(p,p')}c(p)} \; . \label{constraints_global}
\end{align}

The conservation of the first nonlocal charge provides further
constraints on the scattering matrix which  may be derived
by considering the conservation of the
first nonlocal charge in the scattering process. As we discuss in
more detail in \secref{sec:Hopf}, under the assumption that
the $\algPSU(2|2)$ generators act following the Leibniz rule, general
considerations \cite{Luscher} imply that the action of the bilocal
charge on 2-particle states is
\[
\begin{split}
Q_{(2)}{}^S{}_T \ket{\Phi_A(p)} \otimes \ket{\Phi_B(p')} =\
 & \bigbrk{Q_{(2)}{}^S{}_T \ket{\Phi_A(p)}} \otimes \ket{\Phi_B(p')} \\
 & + \ket{\Phi_A(p)} \otimes \bigbrk{Q_{(2)}{}^S{}_T \ket{\Phi_B(p')}} \\
 & + \bigbrk{Q_{(1)}{}^S{}_M \ket{\Phi_A(p)}} \otimes \bigbrk{Q_{(1)}{}^M{}_T \ket{\Phi_B(p')}}
\end{split}
\]
It is somewhat less trivial to extract information
that is not already included in the conservation of the
global symmetry generators\footnote{This is so
because it necessarily requires knowledge of
the action of the bilocal charge on single-particle states.}.
It is however easy to argue on general grounds that, in the presence
of the central charges, the conservation
of $Q_{(2)}$
\[ \label{conservation_general}
  \smatrix Q_{(2)} \ket{\Phi_A(p)\Phi_B(p')}
= Q_{(2)} \smatrix \ket{\Phi_A(p)\Phi_B(p')}
\]
that the $\Csmatrix(p,p')$ and/or $\Fsmatrix(p,p')$ be nonvanishing.

Indeed, one may break the action of $Q_{(2)}$ into two parts, with
even and odd parity in flavor space and similarly for the S-matrix:
\[
Q_{(2)} = Q_{(2)}^{\rm even} + Q_{(2)}^{\rm odd}
\comma
\smatrix = \smatrix^{\rm even} + \smatrix^{\rm odd} \; .
\]
The odd-parity component of \eqref{conservation_general}
\[
\comm{Q_{(2)}^{\rm odd}}{\smatrix^{\rm even}} +
\comm{Q_{(2)}^{\rm even}}{\smatrix^{\rm odd}} = 0
\]
is then an inhomogeneous linear equation for the unknown functions
$\Csmatrix(p,p')$ and $\Fsmatrix(p,p')$ with the inhomogeneous term
provided by the central charges of the algebra. It is worth pointing
out that the nontriviality of this equation arises from the fact that
the structure functions \eqref{struct_fct} are momentum-dependent.
This departs from previous analyses of the relation between the
Yang-Baxter equation and nonlocal integrals of motion.

While this discussion was rather qualitative, it points to the
possibility that the Lagrangian, the centrally-extended
$\algPSU(2|2)^2$ symmetry and
existence of nonlocal charges have a chance of being consistent with
each other in the context of the assumptions listed here. A more
detailed analysis shows that to find exact agreement one must
also include the effects of the non-trivial coproduct as was described
in  \secref{sec:Hopf}.

\section{Scattering of fermions in constant-$J$ light-cone gauge}
\label{sec:constJ}

In this appendix we will consider the superstring in the constant-$J$
light-cone gauge and show that up to terms linear in momenta the
S-matrix is that of ref. \cite{Beisert:su22-S-matrix} when we choose the
overall phase factor to be that conjectured by AFS \cite{Arutyunov:2004vx}. We
 start with the light-cone Hamiltonian described in \cite{Callan:2003xr}, restrict to
a $\grSU(2|2)$ sector and calculate the S-matrix for this subsector.  For the
constant-$J$ gauge we introduce the light-cone coordinates
\[
x^+ = t \comma x^- = \phi-t
\]
 and fix the gauge,
\[
x^+ = \tau \comma p_- = 1 \comma \Gamma^+\theta = 0
\]
where $p_-$ is the light-cone momentum density, $\theta$ is a complex
positive chirality spinor and $\Gamma^A$ are the ten dimensional Dirac
matrices.
The light-cone Lagrangian is written in terms of the physical fields
which are the eight bosons $z^i,\ i=1,\ldots,4$, $y^{i'},\
i'=5,\ldots,8$ and the eight component spinors $\psi$ and
$\psi^\dagger$.  The fermions further break  into $ {\hat \psi}$ and
$\tilde{\psi}$ which are even or odd under the action of
$\Pi=\gamma^1\gamma^2\gamma^3\gamma^4$ where the $\gamma^i$ are the
$8\times 8$ $\gamma$-matrices: \be \Pi \hat{\psi}=\hat{\psi},\qquad
\Pi \tilde{\psi}=-\tilde{\psi}. \ee The spinors $\hat{\psi}$
transform in the $(1,2;1,2)$ of the $\grSU(2)^4$, while $\tilde{\psi}$
transform as $(2,1;2,1)$. In what follows we will restrict our
attention to the $y^{i'}$ bosons and the $\tilde{\psi}$ fermions.
The relevant part of the Lagrangian (dropping the tilde on the
$\psi$) is $\Lagr= \Lagr_0+\Lagr_{int}$, where,
\begin{eqnarray}\label{eqn:lag-quad}
\Lagr_0=\frac{1}{2}\left({\dot y}^2-{\acute y}^2-y^2\right) +i
\psi^\dagger\dot\psi+\frac{i}{2}(\psi\acute\psi+\psi^\dagger\acute
\psi^\dagger)+\psi^\dagger\psi \ee and $\Lagr_{int}={\cal
L}_{BB}+\Lagr_{FF}+\Lagr_{BF}$ with
\begin{eqnarray}\label{eqn:lag-bb}
\Lagr_{BB} = \frac{1}{\sqrt{\lambda}} \left[ -\frac{1}{2}
y^2\acute{y}^2 + \frac{1}{8}(y^2)^2 - \frac{1}{8} \left(
(\dot{y}^2)^2 + 2\dot{y}^2\acute{y}^2 + (\acute{y}^2)^2\right) +
\frac{1}{2} (\acute{y}\cdot\dot{y})^2 \right]
\end{eqnarray}

\begin{eqnarray}\label{eqn:lag-ff}
\Lagr_{FF} &=& -\frac{1}{4 \sqrt{\lambda}} \Biggl[ -i \left[
(\acute{\psi}\psi) + (\acute{\psi}^\dagger\psi^\dagger) \right]
(\psi^\dagger\psi) - \frac{1}{2} (\acute{\psi}\psi +
\acute{\psi}^\dagger\psi^\dagger)^2 \Biggr. \cr & &+ \frac{1}{2}
\left( (\psi^\dagger\acute{\psi}) -
(\acute{\psi}^\dagger\psi)\right)^2 + \frac{i}{12}
(\psi\gamma^{jk}\psi^\dagger)
(\psi^\dagger\gamma^{jk}\acute{\psi}^\dagger) \cr \Biggl. & &+
\frac{i}{48} (\psi\gamma^{jk}\psi +
\psi^\dagger\gamma^{jk}\psi^\dagger)
(\acute{\psi}^\dagger\gamma^{jk}\psi -
\psi^\dagger\gamma^{jk}\acute{\psi}) - (j,k \Leftrightarrow j'k')
\Biggr]
\end{eqnarray}

\begin{eqnarray}\label{eqn:lag-bf}
\Lagr_{BF} &=& \frac{1}{\sqrt{\lambda}} \Biggl[ -\frac{i}{4}
[\dot{y}^2 + \acute{y}^2 + y^2] (\psi\acute{\psi} +
\psi^\dagger\acute{\psi}^\dagger) -\frac{i}{2}
(\dot{y}\cdot\acute{y}) (\psi^\dagger\acute{\psi} +
\psi\acute{\psi}^\dagger) \Biggr. \cr & &-\frac{1}{2} \acute{y}^2
(\psi^\dagger\psi) - \frac{i}{4} (\acute{y}_{j'}y_{k'})
(\psi\gamma^{j'k'}\psi + \psi^\dagger\gamma^{j'k'}\psi^\dagger) \cr
\Biggl. & &+\frac{1}{4} (\dot{y}^{j'}\acute{y}^{k'})
(\psi\gamma^{j'k'}\psi - \psi^\dagger\gamma^{j'k'}\psi^\dagger)
\Biggr].
\end{eqnarray}

To properly identify the $\grSU(2|2)$ sector it is necessary to
identify how the fields transform under the $\grSU(2)^2$ symmetries.
We will use the representation for the $8\times 8$ $\gamma$-matrices
\begin{align}
\gamma^1 & = \epsilon \times \epsilon\times \epsilon &
\gamma^5 & = \tau_3   \times \epsilon\times \unit    \nn\\
\gamma^2 & = \unit    \times \tau_1  \times \epsilon &
\gamma^6 & = \epsilon \times \unit   \times \tau_1   \label{cliffmat} \\
\gamma^3 & = \unit    \times \tau_3  \times \epsilon &
\gamma^7 & = \epsilon \times \unit   \times \tau_3   \nn\\
\gamma^4 & = \tau_1   \times \epsilon\times \unit    &
\gamma^8 & = \unit    \times \unit   \times \unit    \nn
\end{align}
with
\be
\epsilon = \matr{cc}{ 0 & 1 \\ -1 & 0 } \comma
\tau_1   = \matr{cc}{ 0 & 1 \\  1 & 0 } \comma
\tau_3   = \matr{cc}{ 1 & 0 \\ 0 & -1 } \; .
\ee
The generators of the four $\grSU(2)$ factors symmetry can be
expressed as $8\times 8$ $\grSO(8)$ matrices:
\begin{align}
\Sigma_1^\pm & = -\frac{1}{4i}(\gamma^2\gamma^3\pm\gamma^1\gamma^4) &
\Omega_1^\pm & = \frac{1}{4i}(-\gamma^6\gamma^7\pm\gamma^8\gamma^5) \nn\\
\Sigma_2^\pm & = -\frac{1}{4i}(\gamma^3\gamma^1\pm\gamma^2\gamma^4) &
\Omega_2^\pm & = \frac{1}{4i}(-\gamma^7\gamma^5\pm\gamma^8\gamma^6) \label{xplctgen} \\
\Sigma_3^\pm & = -\frac{1}{4i}(\gamma^1\gamma^2\pm\gamma^3\gamma^4) &
\Omega_3^\pm & = \frac{1}{4i}(-\gamma^5\gamma^6\pm\gamma^8\gamma^7) \; . \nn
\end{align}
and we can rewrite the fermions $\tilde{\psi}$ in the
$(2,1;2,1)$ representation in a notation closer to that used previously, e.g. \secref{sec:fermions},
\be
\tilde{\psi} = \frac{1}{2}
\left( \begin{array}{c}
-\Bsi_{\lw\ry}+\Bsi_{\lz\rx} \\
-i( \Bsi_{\lz\ry}-\Bsi_{\lw\rx})\\
i ( \Bsi_{\lz\ry}-\Bsi_{\lw\rx}) \\
-\Bsi_{\lw\ry}+\Bsi_{\lz\rx}  \\
-i(\Bsi_{\lw\ry}+\Bsi_{\lz\rx} ) \\
-\Bsi_{\lz\ry}-\Bsi_{\lw\rx} \\
-\Bsi_{\lz\ry}-\Bsi_{\lw\rx} \\
i (\Bsi_{\lw\ry}+\Bsi_{\lz\rx} )
\end{array} \right)
\comma
\tilde{\psi}^{\dagger} = \frac{1}{2}
\left( \begin{array}{c}
-\Bsi^*_{\lw\ry}+\Bsi^*_{\lz\rx} \\
-i( \Bsi^*_{\lz\ry}-\Bsi^*_{\lw\rx})\\
i ( \Bsi^*_{\lz\ry}-\Bsi^*_{\lw\rx}) \\
-\Bsi^*_{\lw\ry}+\Bsi^*_{\lz\rx}  \\
-i(\Bsi^*_{\lw\ry}+\Bsi^*_{\lz\rx} ) \\
-\Bsi^*_{\lz\ry}-\Bsi^*_{\lw\rx} \\
-\Bsi^*_{\lz\ry}-\Bsi^*_{\lw\rx} \\
i (\Bsi^*_{\lw\ry}+\Bsi^*_{\lz\rx} )
\end{array} \right) \; .
\ee
The $\Bsi_{\laa\rAA} $ transform non-trivially under the
$\grSU(2)$'s generated by $\Sigma^+$, which acts on the undotted index,
and $\Omega^+$, which acts on the dotted index. We will be
interested in the $\Bsi_{\laa\rx}$ which have $\Omega^+_3$ charge
$-\tfrac{1}{2}$. Using the corresponding action of the $\grSU(2)$
generators on the $y$ bosons transforming as $(1,1;2,2)$:
\begin{align}
\Omega_1^+ & = \frac{1}{2i}\matr{cccc}{ 0 & 0 & 0 & -1 \\ 0 & 0 & -1 & 0 \\ 0 & 1 & 0 & 0 \\ 1 & 0 & 0 & 0 } &
\Omega_1^- & = \frac{1}{2i}\matr{cccc}{ 0 & 0 & 0 & 1 \\ 0 & 0 & -1 & 0 \\ 0 & 1 & 0 & 0 \\ -1 & 0 & 0 &  0 } \nn \\
\Omega_2^+ & = \frac{1}{2i}\matr{cccc}{ 0 & 0 & 1 & 0 \\ 0 & 0 & 0  & -1 \\ -1 & 0 & 0 & 0 \\ 0 & 1 & 0 &  0 } &
\Omega_2^- & = \frac{1}{2i}\matr{cccc}{ 0 & 0 & 1 & 0 \\ 0 & 0 & 0 & 1 \\ -1 & 0 & 0 & 0 \\ 0 & -1 & 0 &  0 } \nn\\
\Omega_3^+ & = \frac{1}{2i}\matr{cccc}{ 0 & -1 & 0 & 0 \\ 1 & 0 & 0 & 0 \\ 0 & 0 & 0 & -1 \\ 0 & 0 & 1 &  0 } &
\Omega_3^- & = \frac{1}{2i}\matr{cccc}{ 0 & -1 & 0 & 0 \\ 1 & 0 & 0 & 0 \\0 & 0 & 0 & 1 \\ 0 & 0 & -1 &  0 }
\end{align}
and we can introduce the following complex bosons:
\begin{align}
Y^{\lz\rx} & =\frac{ 1}{\sqrt{2}}\left(y^5-i y^6\right) \; , & Y^{\lw\ry} & =\frac{1}{\sqrt{2}}\left(y^5+i y^6\right) \; , \nn \\
Y^{\lz\ry} & =\frac{-1}{\sqrt{2}}\left(y^7+i y^8\right) \; , & Y^{\lw\rx} & =\frac{-1}{\sqrt{2}}\left(y^7-i y^8\right) \; .
\end{align}
The action of the $\grSU(2)$ generators is a little complicated but we will be
only interested in the bosons with $\Omega^+_3$ charge
$-\tfrac{1}{2}$, that is $Y_{\lz\rx}$ and $Y_{\lw\rx}$ and these
satisfy
\be
\Omega_+^-\ Y_{\lw\rx}=Y_{\lz\rx} \comma
\Omega_-^-\ Y_{\lz\rx}=Y_{\lw\rx} \; .
\ee
In this notation the free part of the
Lagrangian takes a similar form to that in (\ref{free-Lag-uniform})
\[
\begin{split}
\Lagr_0 = &\ + \tfrac{1}{2} \tim{Y}^*_{\lAA\rAA} \tim{Y}^{\lAA\rAA}
- \tfrac{1}{2} \spa{Y}^*_{\lAA\rAA} \spa{Y}^{\lAA\rAA} -
\tfrac{1}{2}     {Y}^*_{\lAA\rAA}     {Y}^{\lAA\rAA}
\\ &
+ i \Bsi^*_{\laa\rAA} \tim{\Bsi}^{\laa\rAA} + \tfrac{i}{2} \brk{
\Bsi^*_{\laa\rAA} \spa{\Bsi}^{*\laa\rAA}
                  + \Bsi  _{\laa\rAA} \spa{\Bsi}^{ \laa\rAA} }
+\Bsi^*_{\laa\rAA} \Bsi^{\laa\rAA} \; .
\end{split}
\]
and so the equations of motion can be solved by a similar mode
expansion:
\begin{align}
Y_{\lAA\rAA}(\vx)      & = \int\frac{dp}{2\pi}
\frac{1}{\sqrt{2\energy}} \:
                           \Bigbrk{ a_{\lAA\rAA}(p)      \, e^{-i\vp\cdot\vx}
                                  + a^\dag_{\lAA\rAA}(p) \, e^{+i\vp\cdot\vx} } \; , \\
\Bsi_{\laa\rAA}(\vx)   & = \int\frac{dp}{2\pi}
\frac{1}{\sqrt{2\energy}} \:
                           \Bigbrk{ b_{\laa\rAA}(p)      \, u(p) \, e^{-i\vp\cdot\vx}
                                  + b^\dag_{\laa\rAA}(p) \, v(p) \, e^{+i\vp\cdot\vx} } \; , \\
\Bsi^*_{\laa\rAA}(\vx) & = \int\frac{dp}{2\pi}
\frac{1}{\sqrt{2\energy}} \:
                         \Bigbrk{ b_{\laa\rAA}(p)      \, v(p) \, e^{-i\vp\cdot\vx}
                                + b^\dag_{\laa\rAA}(p) \, u(p) \, e^{+i\vp\cdot\vx} } \; .
\end{align}
The energy is still $\energy = \sqrt{1+p^2}$ but the wave functions
are slightly different than previously
\[
  v(p) = \sqrt{2} \, \cosh{\tfrac{\theta}{2}}
  \comma
  u(p) = -\sqrt{2} \,  \sinh{\tfrac{\theta}{2}}~.
\]
The rapidity $\theta$ is still defined through $p = \sinh\theta$
and the scalar product in the exponentials is $\vp\cdot\vx = \energy
\tau + p \sigma$. The canonical commutation relations are given, as
before,  by
\begin{align}
 & \comm{ a^{\lAA\rAA}(p)}{a^\dag_{\lBB\rBB}(p')} = 2\pi \, \delta^{\lAA}_{\lBB} \delta^{\rAA}_{\rBB} \, \delta(p-p') \; , &
 & \acomm{b^{\lAA\raa}(p)}{b^\dag_{\lBB\rbb}(p')} = 2\pi \, \delta^{\lAA}_{\lBB} \delta^{\raa}_{\rbb} \, \delta(p-p') \; , \nn \\
 & \comm{ a^{\laa\raa}(p)}{a^\dag_{\lbb\rbb}(p')} = 2\pi \, \delta^{\laa}_{\lbb} \delta^{\raa}_{\rbb} \, \delta(p-p') \; , &
 & \acomm{b^{\laa\rAA}(p)}{b^\dag_{\lbb\rBB}(p')} = 2\pi \, \delta^{\laa}_{\lbb} \delta^{\rAA}_{\rBB} \, \delta(p-p') \; .
\end{align}
We  focus on the fields $Y_{\lAA\rx}$ and $\Bsi_{\laa\rx}$
which comprise a closed $\grSU(2|2)$ subsector of the full theory and
which makes comparison with Beisert's S-matrix most transparent. Parameterizing the T-matrix as
\begin{align}
\tmatrix \ket{Y_{\lAA\rx} Y'_{\lBB\rx}} = \ &
   \Atmatrix(p,p') \ket{Y_{\lAA\rx} Y'_{\lBB\rx}}
 + \Btmatrix(p,p') \ket{Y_{\lBB\rx} Y'_{\lAA\rx}}
 + \Ctmatrix(p,p') \levi_{\lAA\lBB} \levi^{\laa\lbb} \ket{\Bsi_{\laa\rx} \Bsi'_{\lbb\rx}}
 \\[1mm]
\tmatrix \ket{Y_{\lAA\rx} \Bsi'_{\lbb\rx}} = \ &
  \Gtmatrix(p,p') \ket{Y_{\lAA\rx} \Bsi'_{\lbb\rx}}
 +\Htmatrix(p,p') \ket{\Bsi_{\lbb\rx} Y'_{\lAA\rx}}
 \\[1mm]
\tmatrix \ket{\Bsi_{\laa\rx} Y'_{\lBB\rx}} = \ &
  \Ktmatrix(p,p') \ket{Y_{\lBB\rx}\Bsi'_{\laa\rx} }
 +\Ltmatrix(p,p') \ket{\Bsi_{\laa\rx}Y'_{\lBB\rx} }
 \\[1mm]
\tmatrix \ket{\Bsi_{\laa\rx} \Bsi'_{\lbb\rx}} = \ &
  \Dtmatrix(p,p') \ket{\Bsi_{\laa\rx} \Bsi'_{\lbb\rx}}
 +\Etmatrix(p,p') \ket{\Bsi_{\lbb\rx} \Bsi'_{\laa\rx}}
 +\Ftmatrix(p,p') \levi_{\laa\lbb} \levi^{\lAA\lBB} \ket{Y_{\lAA\rx} Y'_{\lBB\rx}} \; ,
\end{align}
we find
\begin{align}
\Atmatrix(p,p') & = \Half \biggsbrk{\cpp + \frac{p'^2+p^2}{\cpp}} \; , \\
\Btmatrix(p,p') & = \Etmatrix(p,p') = \frac{pp'}{\cpp} \; , \\
\Ctmatrix(p,p') & = \Ftmatrix(p,p') = - \Half \, \frac{\sqrt{(\energy+1)(\energy'+1)} \lrbrk{\cpp+p'-p}}{\cpp} \; , \\
\Dtmatrix(p,p') & = \Half \biggsbrk{\cpp - \frac{2pp'}{\cpp}} \; , \\
\Gtmatrix(p,p') & = \Ltmatrix(p',p) = \Half \biggsbrk{\cpp+ \frac{(p+p')p'}{\cpp}} \; , \\
\Htmatrix(p,p') & = \Ktmatrix(p,p') = \Half \, \frac{pp'}{\cpp} \, \frac{(\energy+1)(\energy'+1)-p p'}{\sqrt{(\energy'+1)(\energy+1)}} \; .
\end{align}
We note that in this case,
as we are explicitly restricting our fields to lie in a single $\algSU(2|2)$ rather
than calculating the factorized T-matrix, there is no additional $\tfrac{1}{2}(\Atmatrix-\Btmatrix)\ \left(\unit\otimes\unit\right)$
which must be included to get the expressions used in \secref{sec:results}. There is an ambiguity in the sign of $\Ftmatrix$ which is due to the choice of the fermion ordering; here we have used the convention
\be
\ket{\Bsi_{\laa\rx}\Bsi'_{\lbb\rx}}=b^{\dagger}_{\laa\rx}b'^{\dagger}_{\lbb\rx}\ket{J_+}
\ee
and so it follows from the hermiticity of the Hamiltonian that
$\Ftmatrix=\Ctmatrix$.  These expressions are in good agreement with the
gauge theory, there are of course the terms linear in momenta which are
presumably related to the difference between the definition of the string
length and that of the gauge theory spin chain.

Now, we construct the supersymmetry generators in this $\algSU(2|2)$ sector. The
analogous calculation for the uniform gauge was explicitly carried
out in \cite{Arutyunov:2006ak} and we will here repeat their
calculation for the constant-$J$ gauge, at least to lowest order. We will
start with the Noether currents corresponding to left
multiplication in the gauge unfixed theory and give  expressions
in terms of all ten bosonic coordinates, $x^\mu$, and the sixteen component complex spinor $\theta$.
We can then gauge fix these currents to find their action on the physical fields which are scattered by the
S-matrix.  The Noether currents are given by
$ j=p+*q+*{\bar q}$ where \be
j&=&g(x,\theta)\ J\ g(x,\theta)^{-1}\nn\\
 &=&g(x,\theta)\left( L^A P_A +*L^{\alpha}Q_{\alpha}+*{ \bar L}^{\alpha}{\bar Q}_{\alpha}\right)g(x,\theta)^{-1}
\ee and $g(x, \theta)={\rm{exp}}(\tfrac{1}{2} (x^+
P^-+x^-P^+)){\rm{exp}}(x^I P^I){\rm{exp}}(\theta{
\bar Q}+{\bar \theta} Q)$. For compactness it is useful to define
$\theta^\alpha F_{\alpha}=\theta^\alpha
{\bar Q}_{\alpha}+{\bar \theta}^\alpha Q_{\alpha}$ and introduce
the quantities
 \be
\pi_A(\theta)&=&{\rm{ e}}^{\theta F} P_A{\rm{ e}}^{-\theta F}\nn\\
            &=&\pi^B_A P_A+\pi^{\alpha}_{A} F_{\alpha}\ {\rm{ and}}\nn\\
\pi_{\alpha}(\theta) &=&{\rm{ e}}^{\theta F} F_{\alpha}{\rm{e}}^{-\theta F}\nn \\
            &=&\pi^B_{\alpha} P_A+\pi^{\beta}_{\alpha} F_{\beta}
\ee
so that
\be
j&=&g(x)\left(\left(L^A\pi^B_A+*L^{\alpha}\pi^B_{\alpha}\right)P_B+\left(L^A\pi^{\beta}_A+*L^{\alpha}\pi_{\alpha}^{\beta}\right)F_\beta\right)g(x)^{-1}.
\ee
Now, using the usual trick of scaling the fermions,
$\theta\rightarrow t\theta$, taking  the derivative and
integrating using the boundary conditions
 \be
\pi^A_B(t=0)=\delta^A_B\qquad \pi^{\alpha}_A(t=0)=0\nn\\
\pi^A_\alpha(t=0)=0\qquad
\pi^{\alpha}_\beta(t=0)=\delta^\alpha_\beta \ee
 we can find the closed expressions
\be
\pi^A_B=\cos\left(\sqrt{\alpha \beta}\right),\qquad { \pi}^{\alpha}_A=\frac{\sin\sqrt{\beta \alpha}}{\sqrt{\beta \alpha}}\beta \\
\pi^A_{\alpha}=-\frac{\sin \sqrt{\alpha
\beta}}{\sqrt{\alpha\beta}}\alpha\, \qquad \pi^{\beta}_{\alpha}=\cos
\sqrt{\beta\alpha} \ee
 with the short hand
 \be
\beta^{\alpha}_A=f_{\gamma A}^{\alpha}\theta^{\gamma}\ ,\qquad
\alpha^A_{\beta}=\theta^{\gamma}f_{\gamma\beta}^A \ee
and the $f_{(A,\alpha)(B,\beta)}^{(C,\gamma)}$ are the $\algPSU(2,2|4)$
structure constants.
We are particularly interested in the current corresponding to the
conserved charges $Q^-=\tfrac{1}{2}{\bar \gamma}^+{\gamma}^-Q$
so we consider the truncation
\be
{\bar {\cal Q^-}}&=&j|_{Q^-}\nn\\
&=&\frac{1}{2} \exp\left(\frac{-i x^-\Pi}{2}\right){\rm
exp}\left(\frac{i x^I }{2}{\bar \gamma}^0 \Pi {\bar \gamma}^I\right)( \pi^\alpha_A L^A+\pi^\alpha_{\beta} *L)~.
\ee
We have used the  $\algPSU(2,2|4)$ algebra, in particular the
relation
\be
[Q,P^\mu]=\frac{i}{2}Q{\bar \gamma}^0\Pi{\bar \gamma}^\mu,
\qquad \Pi=\gamma^1{\bar \gamma^2}\gamma^3{\bar \gamma}^4~.
\ee
which implies for our choice of coset representative that
\be
 g(x) Q^-g(x)=
\tfrac{1}{2}Q^-\exp\left(\frac{-i x^-\Pi}{2}\right){\rm
exp}\left(\frac{i x^I }{2}{\bar \gamma}^0 \Pi {\bar \gamma}^I\right)~.
\ee
The most important result  is the occurrence of the $e^{i
\frac{x^-\Pi}{2}}$ factor in the definition of the Noether current.
As discussed in \secref{sec:Hopf} it is this factor which is responsible for
the non-trivial coproduct and hence the non-trivial realization of
integrability. It is worth noting that this factor does not occur in
the pp-wave background as there $\comm{P^+}{Q^-}=0$.  In order to get
manageable expressions and to check that we have sensible results we
expand the time component of the current
in powers of the physical fields and keep only the lowest, quadratic,
part
 \be
 \pi^{\beta}_{\alpha}=\delta^\beta_{\alpha}\ ,\qquad {\bar \pi}^{\alpha}_A= -\frac{i}{2}{\bar \gamma}^0 \Pi {\bar \gamma}_A{\bar \theta}\nn\\
 L^+=2 dx^+\ ,\  L^I=dx^I, \ L=d\theta+i dx^+ \Pi \theta
\ee \be
{\bar {\cal Q}}^-_0&=&  e^{\tfrac{-i x^-\Pi}{2}}e^{\tfrac{i x^I {\bar \gamma}^0\Pi{\bar \gamma}^I}{2}}\left( {\bar \pi}^\alpha_A L_0^A+\pi^\alpha_{\beta} L_1\right)\nn\\
&=&-\frac{i}{2} e^{\tfrac{-i x^-\Pi}{2}}\Pi( p^I
{\bar \gamma}^I{\bar \theta}-i x^I{\bar \gamma}^I \Pi {\bar \theta}+x^I{\bar \gamma}^I
\acute\theta ) \ee which we can compare with the results of Metsaev for the total charge
\cite{Metsaev:2001bj}  (up to an overall normalization) \be
 {\bar Q}^-_{p.p.}=\int d\sigma\  \left(p^I {\bar \gamma}^I{\bar \theta}-i x^I{\bar \gamma}^I \Pi{\bar \theta}-{\acute x}^I{\bar \gamma}^I \theta\right)
\ee and which agrees with our result if we drop the $e^{i
\frac{x^-\Pi}{2}}$ and integrate the last term by parts; there is of
course a similar expression for the conjugate supercharge. It is interesting
to further note that even in the plane-wave geometry there is a central extension of the
$\algPSU(2,2|4)$ algebra as can be easily seen if we
calculate the Poisson bracket of two holomorphic or anti-holomorphic
supercharges
\be
\tr \{ Q^-,Q^-\}&\propto& \int d\sigma\ \left(p^I \acute{x}^I+i { \bar \theta}\acute{\theta}\right) \nn\\
            &=&-\int d\sigma \ \acute{x}^-
\ee using the constraint equation in the last line. However in the plane wave limit there is no
non-trivial coproduct as there is no non-local $e^{i x^-}$ term. We can further restrict our charges so
that they lie in a single $\algSU(2|2)$ by imposing $\Pi Q^-=-Q^-$ so that they now only depend on
the fermionic  fields $\Bsi_{\laa\rx}$. In the full geometry the charges are
 $Q^-=\int d\sigma \: e^{\pm \tfrac{i}{2} x^-} \Omega\left(Y, Y^*, \Bsi,\Bsi^*\right)$ where $\Omega$ is a local function of the physical fields and including the effect of the exponential factor gives rise to the non-trivial phase factor, cf. \secref{sec:Hopf}. We note that even at higher orders in fields there are no additional non-local terms depending on $x^-$ and so the effects of the non-trivial coproduct are entirely captured by including the $e^{i x^-}$ terms.

\section{Rewriting the uniform light-cone gauge action}
\label{app:rewriting-light-cone-action}

For the superstring computation in uniform light-cone gauge, we make
use of the result of \cite{Frolov:2006cc}. The authors of that paper
wrote the Green-Schwarz superstring in a first order formalism and
fixed the uniform light-cone gauge and the kappa-symmetry. In order to
quantize the theory, they considered the near-plane wave limit.  The
Lagrangian was expanded in the transverse fields and the fermions were
shifted $\chi \mapsto \chi + \Phi(p,x,\chi)$ to obtain a canonical
kinetic term. Furthermore the fields were rescaled approriately and a
canonical transformation was applied to the bosonic sector to remove
all non-derivative quartic terms. The results we are interested in are
given in (5.4) with rescaling (5.6), in (5.13), and in (5.16) of
\cite{Frolov:2006cc}. In the notation of \cite{Frolov:2006cc}, the
Green-Schwarz superstring in the uniform light-cone gauge up to
quartic%
\footnote{A discussion of the Dirac brackets in light-cone gauge
to all orders in fields has appeared in \cite{Itoyama:2006cg}.}
order in the fields reads
\[
\Action = \int\limits_{-\infty}^{\infty} \!d\tau
\int\limits_{-\pi}^{\pi} \frac{d\sigma}{2\pi}\: \Lagr \comma \Lagr =
\Lagr_\mathrm{kin} - \Ham \comma \Ham = \Ham_2 + \Ham_4
\]
with%
\footnote{In formula (5.16) of \cite{Frolov:2006cc} there is
actually a factor of $\half$ missing in front of the second term in
the second line.}
\[ \label{eqn:hepth0603008-1}
\begin{split}
\Lagr_\mathrm{kin} = &\ p_M \tim{x}_M - \frac{i}{2} \str \bigbrk{
\Sigma_+ \chi\tim{\chi} }
\\[3mm]
\Ham_2 = &\
   \frac{1}{2} p_M^2
 + \frac{\tl}{2} \spa{x}_M^2
 + \frac{1}{2} x_M^2
 + \frac{\kappa\sqrt{\tl}}{2} \str \bigbrk{ \Sigma_+ \chi \widetilde{K}_8 \spa{\chi}^t K_8 }
 + \frac{1}{2} \str \bigbrk{ \chi^2 }
\\[3mm]
\Ham_4 = &\ \frac{1}{2P_+} \Biggsbrk{ 2 \tl \lrbrk{ \spa{y}^2 z^2 -
\spa{z}^2 y^2 + \spa{z}^2 z^2 - \spa{y}^2 y^2 }
\\ &
 - \tl \str \lrbrk{ \frac{1}{2} \chi \spa{\chi} \chi \spa{\chi}
                  + \frac{1}{2} \chi^2 \spa{\chi}^2
                  + \frac{1}{4} \comm{\chi}{\spa{\chi}} K_8 \comm{\chi}{\spa{\chi}}^t K_8
                  + \chi \widetilde{K}_8 \spa{\chi}^t K_8 \chi \widetilde{K}_8 \spa{\chi}^t K_8 }
\\ &
 + \tl \str \lrbrk{ (z^2-y^2) \spa{\chi}\spa{\chi}
                  + \frac{1}{2} \spa{x}_M x_N \comm{\Sigma_M}{\Sigma_N} \comm{\chi}{\spa{\chi}}
                  - 2 x_M x_N \Sigma_M \spa{\chi} \Sigma_N \spa{\chi} }
\\ &
 + \frac{i\kappa\sqrt{\tl}}{4} (x_M p_N)' \str \bigbrk{ \comm{\Sigma_M}{\Sigma_N} \comm{\widetilde{K}_8 \chi^t K_8}{\chi} }
 } \; .
\end{split}
\]
Here
\[
  \tl = \frac{4\lambda}{P_+^2}
\]
is the effective coupling constant which is kept finite in the
plane-wave limit $P_+ \to \infty$. The parameter $P_+ := J + E$
itself defines the light-cone gauge, and corresponds to $P_+ = 2
J_+$ in our conventions \eqref{jplus}.

All gauge symmetries are fixed in \eqref{eqn:hepth0603008-1} and we
are left with 16 real bosonic and 16 real fermionic degrees of
freedom given by the following fields. The bosonic coordinates and
their canonical conjugate momenta are denoted by
\[
  x_M, p_M \comma M = 1,\ldots,8 \; .
\]
These are the coordinates transverse to the light-cone. They are
divided into coordinates $z_a$ with $a=1,\ldots,4$ on $\AdS$ and
coordinates $y_s$ with $s=1,\ldots,4$ on $\Sphere^5$. The (complex)
fermionic variables are contained in the matrix
\[ \label{eqn:def_chi}
  \chi = \matr{cc}{0 & \Theta \\ - \Theta^\dag \Sigma & 0}
  \comma
  \Theta = \matr{cccc}{0 & 0 & \theta_{13} & \theta_{14} \\
                       0 & 0 & \theta_{23} & \theta_{24} \\
                       \theta_{31} & \theta_{32} & 0 & 0 \\
                       \theta_{41} & \theta_{42} & 0 & 0 } \; .
\]
The various constant matrices $\Sigma$ and $K$ used in these
formulas are defined as follows:
\begin{align}
\gamma_1 & = \matr{cccc}{ 0 & 0 & 0 &-1 \\
                          0 & 0 & 1 & 0 \\
                          0 & 1 & 0 & 0 \\
                         -1 & 0 & 0 & 0 }
\comma
\gamma_2   = \matr{cccc}{ 0 & 0 & 0 & i \\
                          0 & 0 & i & 0 \\
                          0 &-i & 0 & 0 \\
                         -i & 0 & 0 & 0 }
\comma
\gamma_3   = \matr{cccc}{ 0 & 0 & 1 & 0 \\
                          0 & 0 & 0 & 1 \\
                          1 & 0 & 0 & 0 \\
                          0 & 1 & 0 & 0 }
\nn \\
\gamma_4 & = \matr{cccc}{ 0 & 0 &-i & 0 \\
                          0 & 0 & 0 & i \\
                          i & 0 & 0 & 0 \\
                          0 &-i & 0 & 0 }
\comma
\gamma_5   = \Sigma = \matr{cccc}{1 & 0 & 0 & 0 \\
                                  0 & 1 & 0 & 0 \\
                                  0 & 0 &-1 & 0 \\
                                  0 & 0 & 0 &-1 }
\end{align}
\begin{align}
& \Sigma_M = \lrbrk{ \matr{cc}{\gamma_a & 0 \\ 0 & 0} \, , \, \matr{cc}{0 & 0 \\ 0 & i\gamma_s} } \\
& \Sigma_+ = \matr{cc}{\Sigma & 0 \\ 0 & \Sigma} \comma \Sigma_- =
\matr{cc}{-\Sigma & 0 \\ 0 & \Sigma} \comma \Sigma_8 =
-\Sigma_+\Sigma_- = \matr{cc}{\unit_4 & 0 \\ 0 & -\unit_4}
\end{align}
\[
K = \matr{cccc}{ 0 & 1 & 0 & 0 \\
                -1 & 0 & 0 & 0 \\
                 0 & 0 & 0 & 1 \\
                 0 & 0 &-1 & 0 }
\comma K_8 = \matr{cc}{ K & 0 \\ 0 & K } \comma \widetilde K_8 =
\matr{cc}{ K & 0 \\ 0 & -K } \; .
\]

We will now change back to a second order formalism. Using
$\tim{x}_M = \partial \Ham / \partial p_M$ we find the momentum to
cubic order in the fields
\[
  p_M = \tim{x}_M
      + \frac{i\kappa\sqrt{\tl}}{8J_+} x_N \partial_\sigma \str \bigbrk{ \comm{\Sigma_N}{\Sigma_M} \comm{\widetilde{K}_8 \chi^t K_8}{\chi} } \; .
\]
Plugging this into the Lagrangian yields
\[
  \Lagr = \Lagr_0 + \Lagr_\mathrm{int}
\]
with
\[ \label{eqn:lagr-quadratic-chi}
\begin{split}
\Lagr_0 = &\
   \frac{1}{2} \tim{x}_M^2
 - \frac{\lambda}{2J_+^2} \spa{x}_M^2
 - \frac{1}{2} x_M^2
 - \frac{i}{2} \str \bigbrk{ \Sigma_+ \chi\tim{\chi} }
 - \frac{\kappa\sqrt{\lambda}}{2J_+} \str \bigbrk{ \Sigma_+ \chi \spa{\chi}^\conj }
 - \frac{1}{2} \str \bigbrk{ \chi^2 }
\\[3mm]
\Lagr_\mathrm{int} = &\
 - \frac{\lambda}{2J_+^3} \lrbrk{ \spa{y}^2 z^2 - \spa{z}^2 y^2 + \spa{z}^2 z^2 - \spa{y}^2 y^2 }
\\ &
 + \frac{\lambda}{4J_+^3} \str \lrbrk{ \frac{1}{2} \chi \spa{\chi} \chi \spa{\chi}
                                     + \frac{1}{2} \chi^2 \spa{\chi}^2
                                     + \frac{1}{4} \comm{\chi}{\spa{\chi}} \comm{\chi^\conj}{\spa{\chi}^\conj}
                                     + \chi \spa{\chi}^\conj \chi \spa{\chi}^\conj }
\\ &
 - \frac{\lambda}{4J_+^3} \str \lrbrk{ (z^2-y^2) \spa{\chi}\spa{\chi}
                                     + \frac{1}{2} \spa{x}_M x_N \comm{\Sigma_M}{\Sigma_N} \comm{\chi}{\spa{\chi}}
                                     - 2 x_M x_N \Sigma_M \spa{\chi} \Sigma_N \spa{\chi} }
\\ &
 + \frac{i\kappa\sqrt{\lambda}}{16J_+^2} x_M \tim{x}_N \partial_\sigma \str \bigbrk{ \comm{\Sigma_M}{\Sigma_N} \comm{\chi^\conj}{\chi} }
\end{split}
\]
where we used $\tl = \lambda/J_+^2$ and introduced the conjugation
$()^\conj$. For bosonic ($X$) and fermionic ($\chi$) supermatrices
\[
  X    = \matr{cc}{A & 0 \\ 0 & D}
  \comma
  \chi = \matr{cc}{0 & B \\ C & 0}
\]
this is defined as
\[
  X^\conj := K_8 X^t K_8 = \matr{cc}{K A^t K & 0 \\ 0 & K D^t K}
  \comma
  \chi^\conj := \widetilde{K}_8 \chi^t K_8 = \matr{cc}{0 & K C^t K \\ - K B^t K & 0}
  \; .
\]
If the bosonic matrix is a product of fermionic ones, we can use
$(\chi_1 \chi_2)^\conj = -\chi_2^\conj \chi_1^\conj$.

To clean up the notation, we finally put the bosonic degrees of
freedom into a supermatrix

\[ \label{eqn:def_X}
  X := x_M \Sigma_M \; ,
\]
rescale $X\to\sqrt{2J_+}\,X$, $\chi\to\sqrt{J_+}\,\chi$,
$\sigma\to\sqrt{\lambda}/J_+ \, \sigma$ and fix $\kappa=1$. Then the
action takes the form \eqref{eqn:action-quadratic-matrix} given in
the main text.

\section{$\grSU(2)^4$ T-matrix in uniform light-cone gauge}
\label{app:full-T}

Here we list our results for the full T-matrix in uniform light-cone
gauge. There are some identities, which are useful in this context:
\begin{align}
  & \energy' p - \energy p' = \sinh(\theta-\theta') \nn \\
  & (p-p') \cosh\tfrac{\theta-\theta'}{2} = (\energy+\energy') \sinh\tfrac{\theta-\theta'}{2} \nn \\
  & \sinh\tfrac{\theta}{2} = \half \sqrt{\energy + p} - \half \sqrt{\energy - p} \\
  & \cosh\tfrac{\theta}{2} = \half \sqrt{\energy + p} + \half \sqrt{\energy - p} \nn \\
  & \sinh\tfrac{\theta-\theta'}{2} = \half \sqrt{(\energy + p)(\energy' - p')} - \half \sqrt{(\energy - p)(\energy' + p')} \nn \\
  & \cosh\tfrac{\theta-\theta'}{2} = \half \sqrt{(\energy + p)(\energy' - p')} + \half \sqrt{(\energy - p)(\energy' + p')} \nn
\end{align}

\newcommand{\ppcpp}{\tfrac{pp'}{\cpp}}

\subsubsection*{Boson-Boson}
\[
\begin{split}
\Tmatrix \ket{Y_{\lAA\rAA} Y'_{\lBB\rBB}} = \ &
  + \half \tfrac{(p-p')^2}{\cpp}
  \ket{Y_{\lAA\rAA} Y'_{\lBB\rBB}}
  + \ppcpp \Bigbrk{
  \ket{Y_{\lAA\rBB} Y'_{\lBB\rAA}} +
  \ket{Y_{\lBB\rAA} Y'_{\lAA\rBB}} } \\ &
  - \ppcpp \sinh\tfrac{\theta-\theta'}{2} \Bigbrk{
  \levi_{\rAA\rBB} \levi^{\raa\rbb} \ket{\Psi_{\lAA\raa} \Psi'_{\lBB\rbb}} +
  \levi_{\lAA\lBB} \levi^{\laa\lbb} \ket{\Bsi_{\laa\rAA} \Bsi'_{\lbb\rBB}} } \\[1mm]
\Tmatrix \ket{Z_{\laa\raa} Z'_{\lbb\rbb}} = \ &
  - \half \tfrac{(p-p')^2}{\cpp}
  \ket{Z_{\laa\raa} Z'_{\lbb\rbb}}
  - \ppcpp \Bigbrk{
  \ket{Z_{\laa\rbb} Z'_{\lbb\raa}} +
  \ket{Z_{\lbb\raa} Z'_{\laa\rbb}} } \\ &
  + \ppcpp \sinh\tfrac{\theta-\theta'}{2} \Bigbrk{
  \levi_{\raa\rbb} \levi^{\rAA\rBB} \ket{\Bsi_{\laa\rAA} \Bsi'_{\lbb\rBB}} +
  \levi_{\laa\lbb} \levi^{\lAA\lBB} \ket{\Psi_{\lAA\raa} \Psi'_{\lBB\rbb}} } \\[1mm]
\Tmatrix \ket{Y_{\lAA\rAA} Z'_{\laa\raa}} = \ &
  - \half \tfrac{p^2-p'^2}{\cpp} \ket{Y_{\lAA\rAA} Z'_{\laa\raa}}
  + \ppcpp \cosh\tfrac{\theta-\theta'}{2}
  \lrbrk{ \ket{\Bsi_{\laa\rAA} \Psi'_{\lAA\raa}}
        - \ket{\Psi_{\lAA\raa} \Bsi'_{\laa\rAA}} } \\[1mm]
\Tmatrix \ket{Z_{\laa\raa} Y'_{\lAA\rAA}} = \ &
  + \half \tfrac{p^2-p'^2}{\cpp} \ket{Z_{\laa\raa} Y'_{\lAA\rAA}}
  - \ppcpp \cosh\tfrac{\theta-\theta'}{2}
  \lrbrk{ \ket{\Psi_{\lAA\raa} \Bsi'_{\laa\rAA}}
        - \ket{\Bsi_{\laa\rAA} \Psi'_{\lAA\raa}} }
\end{split}
\]

\subsubsection*{Fermion-Fermion}
\[
\begin{split}
\Tmatrix \ket{\Psi_{\lAA\raa} \Psi'_{\lBB\rbb}} = \ &
  + \ppcpp \Bigbrk{
  \ket{\Psi_{\lBB\raa} \Psi'_{\lAA\rbb}} -
  \ket{\Psi_{\lAA\rbb} \Psi'_{\lBB\raa}} } \\ &
  - \ppcpp \sinh\tfrac{\theta-\theta'}{2} \Bigbrk{
  \levi_{\raa\rbb} \levi^{\rAA\rBB} \ket{Y_{\lAA\rAA} Y'_{\lBB\rBB}} -
  \levi_{\lAA\lBB} \levi^{\laa\lbb} \ket{Z_{\laa\raa} Z'_{\lbb\rbb}} } \\[1mm]
\Tmatrix \ket{\Bsi_{\laa\rAA} \Bsi'_{\lbb\rBB}} = \ &
  - \ppcpp \Bigbrk{
  \ket{\Bsi_{\lbb\rAA} \Bsi'_{\laa\rBB}} -
  \ket{\Bsi_{\laa\rBB} \Bsi'_{\lbb\rAA}} } \\ &
  + \ppcpp \sinh\tfrac{\theta-\theta'}{2} \Bigbrk{
  \levi_{\rAA\rBB} \levi^{\raa\rbb} \ket{Z_{\laa\raa} Z'_{\lbb\rbb}} -
  \levi_{\laa\lbb} \levi^{\lAA\lBB} \ket{Y_{\lAA\rAA} Y'_{\lBB\rBB}} } \\[1mm]
\Tmatrix \ket{\Psi_{\lAA\raa} \Bsi'_{\lbb\rBB}} = \ &
  - \ppcpp \cosh\tfrac{\theta-\theta'}{2} \Bigbrk{
  \ket{Y_{\lAA\rBB} Z'_{\lbb\raa}} +
  \ket{Z_{\lbb\raa} Y'_{\lAA\rBB}} } \\[1mm]
\Tmatrix \ket{\Bsi_{\laa\rAA} \Psi'_{\lBB\rbb}} = \ &
  + \ppcpp \cosh\tfrac{\theta-\theta'}{2} \Bigbrk{
  \ket{Z_{\laa\rbb} Y'_{\lBB\rAA}} +
  \ket{Y_{\lBB\rAA} Z'_{\laa\rbb}} }
\end{split}
\]

\subsubsection*{Boson-Fermion}
\[
\begin{split}
\Tmatrix \ket{Y_{\lAA\rAA} \Psi'_{\lBB\rbb}} = \ &
  + \half \tfrac{(p'-p)p'}{\cpp} \ket{Y_{\lAA\rAA} \Psi'_{\lBB\rbb}}
  + \ppcpp \ket{Y_{\lBB\rAA} \Psi'_{\lAA\rbb}} \\ &
  + \ppcpp \cosh\tfrac{\theta-\theta'}{2} \ket{\Psi_{\lAA\rbb} Y'_{\lBB\rAA}}
  - \ppcpp \sinh\tfrac{\theta-\theta'}{2} \levi_{\lAA\lBB} \levi^{\laa\lbb} \ket{\Bsi_{\laa\rAA} Z'_{\lbb\rbb}} \\[1mm]
\Tmatrix \ket{Y_{\lAA\rAA} \Bsi'_{\lbb\rBB}} = \ &
  + \half \tfrac{(p'-p)p'}{\cpp} \ket{Y_{\lAA\rAA} \Bsi'_{\lbb\rBB}}
  + \ppcpp \ket{Y_{\lAA\rBB} \Bsi'_{\lbb\rAA}} \\ &
  + \ppcpp \cosh\tfrac{\theta-\theta'}{2} \ket{\Bsi_{\lbb\rAA} Y'_{\lAA\rBB}}
  + \ppcpp \sinh\tfrac{\theta-\theta'}{2} \levi_{\rAA\rBB} \levi^{\raa\rbb} \ket{\Psi_{\lAA\raa} Z'_{\lbb\rbb}} \\[1mm]
\Tmatrix \ket{\Psi_{\lAA\raa} Y'_{\lBB\rBB}} = \ &
  + \half \tfrac{(p-p')p}{\cpp} \ket{\Psi_{\lAA\raa} Y'_{\lBB\rBB}}
  + \ppcpp \ket{\Psi_{\lBB\raa} Y'_{\lAA\rBB}} \\ &
  + \ppcpp \cosh\tfrac{\theta-\theta'}{2} \ket{Y_{\lAA\rBB} \Psi'_{\lBB\raa}}
  + \ppcpp \sinh\tfrac{\theta-\theta'}{2} \levi_{\lAA\lBB} \levi^{\laa\lbb} \ket{Z_{\laa\raa} \Bsi'_{\lbb\rBB}} \\[1mm]
\Tmatrix \ket{\Bsi_{\laa\rAA} Y'_{\lBB\rBB}} = \ &
  + \half \tfrac{(p-p')p}{\cpp} \ket{\Bsi_{\laa\rAA} Y'_{\lBB\rBB}}
  + \ppcpp \ket{\Bsi_{\laa\rBB} Y'_{\lBB\rAA}} \\ &
  + \ppcpp \cosh\tfrac{\theta-\theta'}{2} \ket{Y_{\lBB\rAA} \Bsi'_{\laa\rBB}}
  - \ppcpp \sinh\tfrac{\theta-\theta'}{2} \levi_{\rAA\rBB} \levi^{\raa\rbb} \ket{Z_{\laa\raa} \Psi'_{\lBB\rbb}}
\end{split}
\]

\[
\begin{split}
\Tmatrix \ket{Z_{\laa\raa} \Psi'_{\lBB\rbb}} = \ &
  - \half \tfrac{(p'-p)p'}{\cpp} \ket{Z_{\laa\raa} \Psi'_{\lBB\rbb}}
  - \ppcpp \ket{Z_{\laa\rbb} \Psi'_{\lBB\raa}} \\ &
  - \ppcpp \cosh\tfrac{\theta-\theta'}{2} \ket{\Psi_{\lBB\raa} Z'_{\laa\rbb}}
  - \ppcpp \sinh\tfrac{\theta-\theta'}{2} \levi_{\raa\rbb} \levi^{\rAA\rBB} \ket{\Bsi_{\laa\rAA} Y'_{\lBB\rBB}} \\[3mm]
\Tmatrix \ket{Z_{\laa\raa} \Bsi'_{\lbb\rBB}} = \ &
  - \half \tfrac{(p'-p)p'}{\cpp} \ket{Z_{\laa\raa} \Bsi'_{\lbb\rBB}}
  - \ppcpp \ket{Z_{\lbb\raa} \Bsi'_{\laa\rBB}} \\ &
  - \ppcpp \cosh\tfrac{\theta-\theta'}{2} \ket{\Bsi_{\laa\rBB} Z'_{\lbb\raa}}
  + \ppcpp \sinh\tfrac{\theta-\theta'}{2} \levi_{\laa\lbb} \levi^{\lAA\lBB} \ket{\Psi_{\lAA\raa} Y'_{\lBB\rBB}} \\[3mm]
\Tmatrix \ket{\Psi_{\lAA\raa} Z'_{\lbb\rbb}} = \ &
  - \half \tfrac{(p-p')p}{\cpp} \ket{\Psi_{\lAA\raa} Z'_{\lbb\rbb}}
  - \ppcpp \ket{\Psi_{\lAA\rbb} Z'_{\lbb\raa}} \\ &
  - \ppcpp \cosh\tfrac{\theta-\theta'}{2} \ket{Z_{\lbb\raa} \Psi'_{\lAA\rbb}}
  + \ppcpp \sinh\tfrac{\theta-\theta'}{2} \levi_{\raa\rbb} \levi^{\rAA\rBB} \ket{Y_{\lAA\rAA} \Bsi'_{\lbb\rBB}} \\[3mm]
\Tmatrix \ket{\Bsi_{\laa\rAA} Z'_{\lbb\rbb}} = \ &
  - \half \tfrac{(p-p')p}{\cpp} \ket{\Bsi_{\laa\rAA} Z'_{\lbb\rbb}}
  - \ppcpp \ket{\Bsi_{\lbb\rAA} Z'_{\laa\rbb}} \\ &
  - \ppcpp \cosh\tfrac{\theta-\theta'}{2} \ket{Z_{\laa\rbb} \Bsi'_{\lbb\rAA}}
  - \ppcpp \sinh\tfrac{\theta-\theta'}{2} \levi_{\laa\lbb} \levi^{\lAA\lBB} \ket{Y_{\lAA\rAA} \Psi'_{\lBB\rbb}}
\end{split}
\]


\end{document}